\documentclass[a4paper]{article} 
\usepackage{amssymb,amscd}
\usepackage[all]{xy}

\def\goth{\mathfrak}          
\def\double{\mathbb}

\def\cc{{\double C}}     
\def\nn{{\double N}}       
\def\zz{{\double Z}}

\def\rr{{\double R}}

\newtheorem{theorem}{Theorem}[section]
\newtheorem{lemma}[theorem]{Lemma}

\newtheorem{definition}[theorem]{Definition}
\newtheorem{proposition}[theorem]{Proposition}
\newtheorem{remark}[theorem]{Remark}

\newtheorem{assumption}[theorem]{Assumption}

\def\limind{\mathop{\mathrm{lim}}\limits_{\longrightarrow}}
\def\limproj{\mathop{\mathrm{lim}}\limits_{\longleftarrow}}
\def\cp{\rtimes}
\def\si{\sigma}

\def\cinf{C^{\infty}}

\newcommand{\be}{\begin{equation}}
\newcommand{\ee}{\end{equation}}
\newcommand{\beq}{\begin{eqnarray}}
\newcommand{\eeq}{\end{eqnarray}}
\newcommand{\om}{\omega}
\newcommand{\Om}{\Omega}
\newcommand{\al}{\alpha}
\def\nat{\natural}
\def\id{\mathop{\mathrm{id}}}
\def\Esf{{\sf E}}

\newcommand{\la}{\lambda}
\newcommand{\Ec}{{\cal E}}
\newcommand{\Vc}{{\cal V}}
\newcommand{\Lc}{{\cal L}}
\newcommand{\non}{\nonumber}
\newcommand{\eps}{\varepsilon}

\newcommand{\Wc}{{\cal W}}
\newcommand{\Rc}{{\cal R}}
\newcommand{\Mc}{{\cal M}}
\newcommand{\Ic}{{\cal I}}
\newcommand{\Jc}{{\cal J}}

\newcommand{\ch}{{\mathop{\mathrm{ch}}}}
\newcommand{\chb}{{\mathop{\textup{\bf ch}}}}

\newcommand{\Tr}{{\mathop{\mathrm{Tr}}}}

\newcommand{\Ac}{{\cal A}}

\newcommand{\te}{\theta}

\newcommand{\Te}{\Theta}

\newcommand{\cqfd}{\hfill\rule{1ex}{1ex}}

\def\d{\partial}
\def\dd{\textup{\bf d}}
\def\db{\underline{d}}
\def\odob{\underline{\odot}}
\def\Hc{{\cal H}}

\def\Bc{{\cal B}}
\def\Cc{{\cal C}}
\def\Jc{{\cal J}}
\def\Kc{{\cal K}}
\def\Fc{{\cal F}}

\def\ker{\mbox{\textup{Ker}}}
\def\Bb{\overline{B}}
\def\Bbe{\overline{B}_{\epsilon}}
\def\bb{\overline{b}}
\def\phib{\overline{\phi}}
\def\iotab{\overline{\iota}}

\def\hom{{\mathop{\mathrm{Hom}}}}
\def\dom{{\mathop{\mathrm{Dom}}}}

\def\End{{\mathop{\mathrm{End}}}}

\def\hotimes{\hat{\otimes}}

\def\St{\widetilde{S}}

\def\Qt{\widetilde{Q}}
\def\Sg{{\goth S}}
\def\Act{\widetilde{\cal A}}
\def\Tct{\widetilde{\cal T}}
\def\Bct{\widetilde{\cal B}}

\def\at{\widetilde{a}}

\def\Sge{{\goth S_{\epsilon}}}
\def\Sgan{{\goth S_{\textup{\scriptsize an}}}}
\def\Sgd{{\goth S_{\delta}}}
\def\Omt{\widetilde{\Omega}}

\def\chih{\widehat{\chi}}
\def\Ome{\Omega_{\epsilon}}
\def\Omd{\Omega_{\delta}}
\def\Omtd{\widetilde{\Omega}_{\delta}}
\def\Oman{\Omega_{\textup{\scriptsize an}}}
\def\Omtan{\widetilde{\Omega}_{\textup{\scriptsize an}}}
\def\Xc{{\mathcal X}}
\def\Xch{\widehat{{\mathcal X}}}
\def\Tc{{\cal T}}
\def\Dc{{\cal D}}

\def\Kc{{\cal K}}
\def\rlarrows{\textup{$\rightleftarrows$}}

\begin{document}

\begin{center}

{ \Large RETRACTION OF THE BIVARIANT\\
\vskip 2mm
CHERN CHARACTER}
\vskip 1cm
{\bf Denis PERROT}
\vskip 0.5cm
SISSA, via Beirut 2-4, 34014 Trieste, Italy \\[2mm]
{\tt perrot@fm.sissa.it}\\[2mm]
\today 
\end{center}
\vskip 0.5cm
\begin{abstract} 
We show that the bivariant Chern character in entire cyclic cohomology constructed in a previous paper in terms of superconnections and heat kernel regularization, retracts on periodic cocycles under some finite summability conditions. The trick is a bivariant generalization of the Connes-Moscovici method for finitely summable $K$-cycles. This yields concrete formulas for the Chern character of $p$-summable quasihomomorphisms and invertible extensions, analogous to those of Nistor. The latter formulation is completely algebraic and based on the universal extensions of Cuntz and Zekri naturally appearing in the description of bivariant $K$-theory. 
\end{abstract}

\vskip 0.5cm


\section{Introduction}

In a previous paper \cite{P} we introduced a bivariant Chern character in entire cyclic cohomology. We were mainly motivated by the need for a noncommutative generalization of the Atiyah-Singer index theorem for families of Dirac operators, with potential applications to mathematical physics. The basic ingredient of the construction is the use of heat kernel regularization for infinite-dimensional traces, in the spirit of the Bismut-Quillen approach of the families index theorem \cite{B,Q1}. For the sake of definiteness, we have to work in the category of {\it bornological algebras}. Recall that a (convex) bornology on a vector space is a collection of subsets playing at an abstract level the same role as the bounded sets of a normed space \cite{HN}. A bornological algebra is a bornological space for which the multiplication is bounded. Given two complete bornological algebras $\Ac$ and $\Bc$, we define in \cite{P} the $\zz_2$-graded semigroup $\Psi_*(\Ac,\Bc)$ of unbounded $\Ac$-$\Bc$-bimodules, in full analogy with the Baaj-Julg description of Kasparov's $KK$-groups \cite{Bl}. The bivariant Chern character then takes the form of a natural map
\be
\ch: \Psi^{\te}_*(\Ac,\Bc)\to HE_*(\Ac,\Bc)\ ,\quad *=0,1\ ,\label{biv}
\ee
from a restricted subset of $\te$-summable bimodules, to the bivariant entire cyclic cohomology $HE_*(\Ac,\Bc)$. The latter is a generalization of the entire cyclic cohomology of Connes \cite{C2} and was introduced by Meyer in \cite{Me}. The interesting feature of the entire cyclic theory is that it contains infinite-dimensional cocycles, which are well-adapted to the heat kernel method. In the particular case where $\Bc=\cc$, our formula exactly reduces to the JLO cocycle representing the Chern character of $\te$-summable $K$-cycles over $\Ac$ \cite{JLO}. When $\Ac=\cc$ and $\Bc$ is arbitrary, then (\ref{biv}) is a noncommutative generalization of Bismut's formula for the Chern character of the $K$-theory element defined by a family of Dirac operators over the ``manifold'' $\Bc$. In both cases, it is well-known that asymptotic expansions of the heat kernel \cite{BGV,Gi} give rise to {\it local} expressions allowing to compute the Chern character in many interesting geometric situations, including noncommutative ones \cite{CM95}. Thus our construction of (\ref{biv}) allows, in principle, to extend these methods to the bivariant case.\\
However, when dealing with unbounded bimodules, it is not easy to describe the most interesting aspect of bivariant $K$-theory, namely the Kasparov intersection product \cite{Bl}, as far as the latter is well-defined in our bornological context. Consequently, it is doubtful that we can check directly the compatibility of the bivariant Chern character with such a product, that is, the multiplicativity of the map (\ref{biv}).\\

To this classical ``geometric'' picture of the bivariant theory, we would like to oppose the purely algebraic approach of Cuntz \cite{Cu1,Cu2}. In this formulation, $KK$-theory is obtained via universal algebras and extensions. The algebraic nature of these objects together with the {\it excision} property of cyclic cohomology allows to construct a bivariant Chern character compatible with the Kasparov product, for various categories of algebras \cite{Cu2,Pu1,Pu2}. It is therefore tempting to compare this approach with our geometric method involving families of Dirac operators and heat kernel regularization. A compatibility between both constructions would automatically imply that the map (\ref{biv}) is a correct bivariant Chern character.\\

The aim of this paper is twofold. First, we show that under finite summability conditions, the entire Chern character retracts on a collection of {\it periodic} (i.e. finite-dimensional) cocycles. This is a bivariant generalization of the Connes-Moscovici retraction of the Chern character for finite-dimensional $K$-cycles over $\Ac$ \cite{CM86,CM93}. Second, these cocycles are well-adapted to the quasihomomorphism description of $KK$-groups, which connects our construction with the algebraic approach. Excision therefore implies the compatibility with the Kasparov product. In fact, one can show that our bivariant periodic cocycles are equivalent to Nistor's results \cite{Ni1,Ni2}.\\

The paper is organized as follows. In section \ref{sprel} we recall the basic notions concerning bornological spaces, universal algebras and extensions, and the different versions of bivariant cyclic cohomology we need, namely the entire and periodic ones. We then recall in section \ref{sbim} the construction of the entire bivariant Chern character of \cite{P} with some details. This is necessary in order to fix the fundamental objects and notations of the theory. This construction is performed in two steps. First, given an unbounded $\Ac$-$\Bc$-bimodule, we construct a chain map $\chi:\Ome\Ac\to X(\Bc)$ from the $(b+B)$-complex of entire chains over $\Ac$ \cite{C0,C1}, to the $X$-complex of $\Bc$ \cite{CQ1,Me}. This morphism is basically obtained from the exponential of the curvature of a Quillen superconnection. The requirement of {\it boundedness} for this exponential constitutes the $\te$-summability condition needed for the definiteness of $\chi$. This automatically incorporates the heat kernel regularization associated to the laplacian of the Dirac operator. The second step consists in lifting the whole construction of $\chi$, simply upon replacing $\Ac$ and $\Bc$ by their analytic tensor algebras $\Tc\Ac$ and $\Tc\Bc$ \cite{Me}, which yields, up to a Goodwillie equivalence \cite{CQ1,G,Me}, the bivariant Chern character in the entire cyclic cohomology $HE_*(\Ac,\Bc)$.\\
After these preparatory matters, we show in section \ref{sretr} that under $p$-summability hypotheses, the entire chain map $\chi$ retracts on a collection periodic cocycles. This retraction is exactly analogous to the procedure of Connes-Moscovici for $p$-summable $K$-cycles \cite{CM86,CM93}, the only difference is the use of the complex $X(\Bc)$ (or $X(\Tc\Bc)$ in the lifted case) instead of $\cc$ as the target of $\chi$. The net result is that we get explicit formulas for the bivariant Chern character of $p$-summable {\it bounded} Kasparov $\Ac$-$\Bc$-modules.\\
In section \ref{suniv}, we arrange these periodic cocycles into simpler expressions and relate them to the universal algebras of Cuntz and Zekri \cite{Cu1,Z}. This allows to provide a purely algebraic construction of a periodic Chern character on $KK$-groups in section \ref{sKK}. In the even case, we define the algebraic version of $KK_0(\Ac,\Bc)$ as the set of finitely summable quasihomomorphisms from $\Ac$ to $\Bc$ modulo homotopy and addition of degenerate elements. That is, if $\Lc$ and $\ell^p$ denote respectively the algebra of bounded operators and the Schatten ideal of $p$-summable operators on an infinite-dimensional separable Hilbert space, then a $p$-summable quasihomomorphism is given by a pair of homomorphisms $\rho_{\pm}:\Ac\to \Lc\hotimes\Bc$ whose difference $\rho_+-\rho_-$ maps to $\ell^p\hotimes\Bc$. The universal cocycles of the preceding section then yield a collection of cohomology classes in the bivariant groups $HC_{2n}(\Ac,\Bc)$ for $2n\ge p$, whose images in the periodic theory $HP_0(\Ac,\Bc)$ all coincide. Since we work fundamentally with the Cuntz algebra, the proof of excision in periodic cyclic cohomology implies that the bivariant Chern character is compatible with the Kasparov product whenever the latter is defined. There is also an analogous construction in the odd case, where we define the algebraic version of $KK_1(\Ac,\Bc)$ as the set of invertible extensions of $\Ac$ by $\ell^p\hotimes\Bc$ modulo homotopy and addition of degenerates. Again, we get a collection of classes in the odd groups $HC_{2n+1}(\Ac,\Bc)$ for $n$ sufficiently large, having the same image in the periodic theory $HP_1(\Ac,\Bc)$.\\

We insist on the fact that, though all the computations performed in this paper may look somewhat complicated, the explicit formulas of the periodic Chern character
\be
\ch: KK_*(\Ac,\Bc)\to HP_*(\Ac,\Bc)\ ,\quad *=0,1\label{per}
\ee
are really given by simple and useful expressions (see propositions \ref{uni0}, \ref{uni1} and section \ref{sKK}). Moreover, it should be understood that the construction of (\ref{per}) is by no means subordinate to the existence of the entire Chern character (\ref{biv}). The point is that both constructions, based on very different grounds, essentially coincide under suitable summability conditions.

\section{Preliminaries}\label{sprel}

Here we recall briefly the basic notions used throughout the paper. Most of this material is developed in great detail in the reference documents \cite{Cu2,CQ1,CQ2,HN,Me}. \\

\noindent {\bf Bornological algebras.} A bornological vector space $\Vc$ is a vector space endowed with a collection $\Sg(\Vc)$ of subsets $S\subset \Vc$, called the bornology of $\Vc$, satisfying certain axioms reminiscent of the properties fulfilled by the bounded sets of a normed space \cite{HN, Me}:
\begin{itemize}
\item $\{x\}\in\Sg(\Vc)$ for any vector $x\in\Vc$.
\item $S_1+S_2\in\Sg(\Vc)$ for any $S_1,S_2\in\Sg(\Vc)$.
\item If $S\in \Sg(\Vc)$, then $T\in\Sg(\Vc)$ for any $T\subset S$.
\item $S^{\diamond}\in\Sg(\Vc)$ for any $S\in\Sg(\Vc)$,
\end{itemize}
where $S^{\diamond}$ is the circled convex hull of the subset $S$. The elements of $\Sg(\Vc)$ are called the {\it small} subsets of $\Vc$. If a small subset $S$ is a disk (i.e. circled and convex), then its linear span $\Vc_S$ has a unique seminorm for which the closure of $S$ is the unit ball; $S$ is called completant iff $\Vc_S$ is a Banach space. $\Vc$ is a {\it complete bornological vector space} iff any small subset $T$ is contained in some completant small disk $S\in\Sg(\Vc)$. Typical examples of complete bornological spaces are provided by Banach or complete locally convex spaces, endowed with the bornology corresponding to the collection of all bounded subsets.\\
The interesting linear maps between two bornological spaces $\Vc$ and $\Wc$ are those which send $\Sg(\Vc)$ to $\Sg(\Wc)$. Such maps are called {\it bounded}. The vector space of bounded linear maps $\hom(\Vc,\Wc)$ is itself a bornological space, the small subsets corresponding to equibounded maps (see \cite{HN,Me}). It is complete if $\Wc$ is complete. Similarly, an $n$-linear map $\Vc_1\times\ldots\Vc_n\to\Wc$ is bounded if it sends an $n$-tuple of small sets $(S_1,\ldots, S_n)$ to a small set of $\Wc$. A {\it bornological algebra} $\Ac$ is a bornological vector space together with a bounded bilinear map $\Ac\times\Ac\to\Ac$. We will be concerned only with associative bornological algebras. \\
Given a bornological space $\Vc$, its {\it completion} $\Vc^c$ is a complete bornological space defined as the solution of a universal problem concerning the factorization of bounded maps $\Vc\to\Wc$ with complete target $\Wc$. This completion always exists \cite{HN,Me}. If $\Vc$ is a normed space endowed with the bornology of bounded subsets, then its bornological completion coincides with its Hausdorff completion. The {\it completed tensor product} $\Vc_1\hotimes\Vc_2$ of two bornological spaces $\Vc_1$ and $\Vc_2$ is the completion of their algebraic tensor product for the bornology generated by the bismall sets $S_1\otimes S_2$, $\forall S_i\in \Sg(\Vc_i)$. The completed tensor product is associative, whence the definition of the $n$-fold completed tensor product $\Vc_1\hotimes\ldots\hotimes\Vc_n$. For Fr\'echet spaces, this coincides with the usual projective tensor product. If $\Ac$ and $\Bc$ are bornological algebras, their completed tensor product $\Ac\hotimes\Bc$ is a complete bornological algebra.\\

\noindent {\bf Non-commutative differential forms.} Let $\Ac$ be a complete bornological algebra. The algebra of non-commutative differential forms over $\Ac$ is the direct sum $\Om\Ac=\bigoplus_{n\ge 0}\Om^n\Ac$ of the $n$-dimensional subspaces $\Om^n\Ac=\Act\hotimes\Ac^{\hotimes n}$ for $n\ge 1$ and $\Om^0\Ac=\Ac$, where $\Act=\Ac\oplus \cc$ is the unitalization of $\Ac$. It is customary to use the differential notation $a_0da_1\ldots da_n$ (resp. $da_1\ldots da_n$) for the string $a_0\otimes a_1\ldots\otimes a_n$ (resp. $1\otimes a_1\ldots\otimes a_n)$. The differential $d: \Om^n\Ac\to\Om^{n+1}\Ac$ is uniquely specified by $d(a_0da_1\ldots da_n)=da_0da_1\ldots da_n$ and $d^2=0$. The multiplication in $\Om\Ac$ is defined as usual and fulfills the Leibniz rule $d(\om_1\om_2)=d\om_1\om_2 +(-)^{|\om_1|}\om_1d\om_2$, where $|\om_1|$ is the degree of $\om_1$. Each $\Om^n\Ac$ is a complete bornological space by construction, and we endow $\Om\Ac$ with the direct sum bornology. This turns $\Om\Ac$ into a complete bornological differential graded (DG) algebra, i.e. the multiplication map and $d$ are bounded. It is the universal complete bornological DG algebra generated by $\Ac$. \\
On $\Om\Ac$ are defined various operators. First of all, the Hochschild boundary $b: \Om^{n+1}\Ac\to \Om^n\Ac$ is $b(\om da)=(-)^n[\om,a]$ for $\om\in\Om^n\Ac$, and $b=0$ on $\Om^0\Ac=\Ac$. One easily shows that $b$ is bounded and $b^2=0$. Then the Karoubi operator $\kappa:\Om^n\Ac\to \Om^n\Ac$ is defined by $1-\kappa=db+bd$. Therefore $\kappa$ is bounded and commutes with $b$ and $d$. One has $\kappa(\om\, da)=(-)^n da\,\om$ for any $\om\in \Om^n\Ac$ and $a\in\Ac$. The last operator is Connes' $B:\Om^n\Ac\to \Om^{n+1}\Ac$, equal to $(1+\kappa+\ldots+\kappa^n)d$ on $\Om^n\Ac$. It is bounded and verifies $B^2=0=Bb+bB$ and $B\kappa=\kappa B=B$. Thus $\Om\Ac$ endowed with the two anticommuting differentials $(b,B)$ is a complete bornological bicomplex.\\
We also define three other bornologies on $\Om\Ac$, leading to the notion of {\it entire} cyclic cohomology:
\begin{itemize}
\item The {\bf entire bornology} $\Sge(\Om\Ac)$ is generated by the sets
\be
\bigcup_{n\ge 0} [n/2]!\,\St (dS)^n\quad ,\ S\in\Sg(\Ac)\ ,\label{sge}
\ee 
where $[n/2]=k$ if $n=2k$ or $n=2k+1$, and $\St=S+\cc$. That is, a subset of $\Om\Ac$ is small iff it is contained in the circled convex hull of a set like (\ref{sge}). We write $\Ome\Ac$ for the completion of $\Om\Ac$ with respect to this bornology. $\Ome\Ac$ will give rise to the $(b,B)$-complex of entire chains.
\item  The {\bf analytic bornology} $\Sgan(\Om\Ac)$ is generated by the sets $\bigcup_{n\ge 0} \St(dS)^n$, $S\in\Sg(\Ac)$. The corresponding completion of $\Om \Ac$ is $\Oman\Ac$. It is related to the $X$-complex description of entire cyclic homology.
\item The {\bf de Rham-Karoubi bornology} $\Sgd(\Om\Ac)$ is generated by the collection of sets $\bigcup_{n\ge 0} \frac{1}{[n/2]!}\,\St(dS)^n$,  $S\in\Sg(\Ac)$, with completion $\Omd\Ac$. It gives rise to the correct analytic completion of the complex introduced by Karoubi for the construction of characteristic classes in algebraic $K$-theory \cite{Kar}.
\end{itemize}
The multiplication in $\Om\Ac$ is bounded for the three bornologies above, as well as all the operators $d,b,\kappa,B$. Moreover, the $\zz_2$-graduation of $\Om\Ac$ given by even and odd forms is preserved by the completion process, so that $\Ome\Ac, \Oman\Ac$ and $\Omd\Ac$ are $\zz_2$-graded differential algebras, endowed with the operators $b,\kappa,B$ fulfilling the usual relations. In particular, $\Ome\Ac$ is called the $(b,B)$-complex of {\it entire chains}. Note also that the rescaling $c:\Oman\Ac\to \Ome\Ac$ induced by its action on $n$-forms
\be
c(a_0da_1\ldots da_n)=(-)^{[n/2]}[n/2]!\, a_0da_1\ldots da_n\quad \forall n\in\nn\ ,\label{resc}
\ee
obviously provides a linear bornological isomorphism.\\

\noindent {\bf The tensor algebra.} Let $\Ac$ be a complete bornological algebra. The non-unital tensor algebra over $\Ac$ is the direct sum $T\Ac=\bigoplus_{n\ge 1}\Ac^{\hotimes n}$. It is a complete bornological algebra for the direct sum bornology. Let $m:T\Ac\to\Ac$ be the (bounded) multiplication map, sending $a_1\otimes\ldots\otimes a_n$ to the product $a_1\ldots a_n$. The inclusion $\si_{\Ac}:\Ac\to T\Ac$ is a bounded linear lift of $m$, hence the kernel $J\Ac$ of $m$ is a direct summand in $T\Ac$. This implies a linearly split exact sequence of complete bornological algebras $0\to J\Ac\to T\Ac \to \Ac \to 0$. As a bornological algebra, $T\Ac$ is isomorphic to the even part of $\Om\Ac$ endowed with the {\it Fedosov product} \cite{CQ1}
\be
\om_1\odot\om_2 :=\om_1\om_2 -d\om_1d\om_2\ ,\quad \om_i\in\Om^+\Ac\ .
\ee
The correspondence between the elements of $(\Om^+\Ac,\odot)$ and $T\Ac$ is given by
\be
\Om^+\Ac\ni a_0da_1\ldots da_{2n} \longleftrightarrow a_0\otimes\om(a_1,a_2)\otimes\ldots \otimes\om(a_{2n-1},a_{2n}) \in T\Ac\ ,\label{cor}
\ee
where $\om(a_i,a_j):=a_ia_j- a_i\otimes a_j \in \Ac\oplus \Ac^{\hotimes 2}$ is the {\it curvature} of $(a_i,a_j)$. The ideal $J\Ac$ is then isomorphic to the Fedosov subalgebra of differential forms of degree $\ge 2$:
\be
J\Ac=\bigoplus_{k\ge 1}\Om^{2k}\Ac\ .
\ee
It turns out that the Fedosov product $\odot$ is also bounded for the bornology $\Sgan$ restricted to the even part $\Om^+\Ac$, and thus extends to the analytic completion $\Oman^+\Ac$. The complete bornological algebra $(\Oman^+\Ac,\odot)$ is also denoted by $\Tc\Ac$ and called the {\it analytic tensor algebra} of $\Ac$ in \cite{Me}. Again, let $\si_{\Ac}$ be the bounded inclusion $\Ac\to \Tc\Ac$. Then there is a unique bounded homomorphism $v_{\Ac}:\Tc\Ac\to \Tc\Tc\Ac$ such that $v_{\Ac}\circ\si_{\Ac}=\si_{\Tc\Ac}\circ\si_{\Ac}$, see \cite{Me}. We denote by $\Jc\Ac$ the kernel of the multiplication map $\Tc\Ac\to \Ac$. It is a direct summand of $\Tc\Ac$, whence an exact sequence of complete bornological algebras
\be
0\to \Jc\Ac\to \Tc\Ac \to \Ac \to 0  \label{ext}
\ee
split by the bounded linear map $\si_{\Ac}$.\\

\noindent {\bf The Cuntz and Zekri algebras.} Let $\Ac$ and $\Bc$ be two complete bornological algebras. The free product $\Ac *\Bc$ is the (non-unital) complete bornological algebra generated by $\Ac$ and $\Bc$. We can write
\be
\Ac*\Bc = \Ac\oplus\Bc \oplus (\Ac\hotimes\Bc)\oplus (\Bc\hotimes\Ac)\oplus (\Ac\hotimes\Bc\hotimes\Ac)\oplus (\Bc\hotimes\Ac\hotimes\Bc)\oplus\ldots\ ,
\ee
where only alternate tensor products of $\Ac$ and $\Bc$ appear. $\Ac*\Bc$ is endowed with the direct sum bornology, and the product is such that whenever two elements of $\Ac$ meet, we take the multiplication in $\Ac$ instead of the tensor product, and similarly for the elements of $\Bc$. There are two canonical inclusions $\al:\Ac\to \Ac*\Bc$ and $\beta:\Bc\to\Ac*\Bc$ which are bounded algebra homomorphisms. The free product enjoys the following universal property: given two bounded homomorphisms $\phi:\Ac\to \Cc$ and $\psi:\Bc\to\Cc$ with target a complete bornological algebra $\Cc$, then there is a unique bounded homomorphism $\phi*\psi: \Ac*\Bc\to \Cc$ such that $\phi=(\phi*\psi)\circ \al$ and $\psi=(\phi*\psi)\circ \beta$. \\
Given a complete bornological algebra $\Ac$, we let $Q\Ac=\Ac*\Ac$ be the free product by itself. The two injections of $\Ac$ into $Q\Ac$ are denoted by $\iota$ and $\iotab$. Thus the universal property of $Q\Ac$ is that it factors pairs of homomorphisms $\Ac\rightrightarrows\Cc$ into an algebra $\Cc$. In particular \cite{Cu1}, the Cuntz algebra $q\Ac$ is the kernel of the folding map $\id*\id: Q\Ac\to\Ac$ sending both copies $\iota(\Ac)$ and $\iotab(\Ac)$ into $\Ac$. Since $\id*\id$ is split by the bounded homomorphism $\iota:\Ac\to Q\Ac$, the ideal $q\Ac$ is a direct summand and we have an exact sequence of complete bornological algebras
\be
0\to q\Ac\to Q\Ac \to \Ac \to 0\ ,
\ee
split by both homomorphisms $\iota$ and $\iotab$. This sequence is universal among the extensions split in two different ways: given any extension of complete bornological algebras $0\to \Ic\to \Ec \to \Ac \to 0$ split by two bounded homomorphisms $\phi,\phib:\Ac\rightrightarrows \Ec$, one has the commutative diagram
\be
\xymatrix{
0 \ar[r] & q\Ac \ar[r] \ar[d]_{\phi*\phib} & Q\Ac \ar[r] \ar[d]_{\phi*\phib} & \Ac \ar[r] \ar@{=}[d] & 0 \\
0 \ar[r] & \Ic \ar[r] & \Ec \ar[r] & \Ac \ar[r] \ar@/_1pc/[l]_{\phi,\phib} & 0 }
\ee
In fact $Q\Ac$ is isomorphic to the Fedosov algebra $(\Om\Ac,\odot)$ of {\it all} differential forms (not only even), for the product
\be
\om_1\odot\om_2 :=\om_1\om_2 -(-)^{|\om_1|}d\om_1d\om_2\ ,\quad \om_i\in\Om\Ac\ ,
\ee
where $|\om_1|$ is the degree of the (homogeneous) form $\om_1$. The correspondence $Q\Ac\leftrightarrow \Om\Ac$ sends $\iota(a)$ to $a+da$ and $\iotab(a)$ to $a-da$. Since $\Om\Ac$ is naturally $\zz_2$-graded by the degree of forms, $Q\Ac$ becomes a $\zz_2$-graded algebra. The action of the generator of $\zz_2$ exchanges the roles of $\iota$ and $\iotab$. We will denote by $Q^s\Ac$ this algebra {\it considered in the supersymmetric category}. $Q^s\Ac$ has the following universal property: given any bounded homomorphism $\rho$ from $\Ac$ to a unital $\zz_2$-graded complete bornological algebra $\Bc$ containing a symmetry $X\in\Bc$, $X^2=1$, implementing the grading on $\Bc$ (i.e. $b\to XbX$ generates the action of $\zz_2$), then there is a unique bounded homomorphism of even degree $\rho^s:Q^s\Ac\to\Bc$ mapping $\iota(a)$ to $\rho(a)$ and $\iotab(a)$ to $X\rho(a)X$. $Q^s\Ac$ is also closely related to the Zekri algebra $E\Ac=\Qt\Ac\cp\zz_2$, verifying a similar universal property \cite{Z} (here $\Qt\Ac$ is the unitalization of $Q\Ac$).\\

\noindent {\bf $X$-complex.} Let $\Ac$ be a complete bornological algebra. The space of one-forms $\Om^1\Ac$ is an $\Ac$-bimodule, hence we can take its quotient $\Om^1\Ac_{\nat}$ by the subspace of commutators $[\Ac,\Om^1\Ac]=b\Om^2\Ac$. It may fail to be complete in general. In order to avoid confusions in the subsequent notations, we always write a one-form $a_0\dd a_1\in\Om^1\Ac$ with a bold $\dd$ when dealing with the $X$-complex of $\Ac$. The latter is the supercomplex \cite{CQ1}
\be
X(\Ac):\quad \Ac\ \xymatrix@1{\ar@<0.5ex>[r]^{\nat \textup{\scriptsize \bf d}} &  \ar@<0.5ex>[l]^{\bb}}\ \Om^1\Ac_{\nat}\ ,
\ee
where $\Ac$ is located in degree zero and $\Om^1\Ac_{\nat}$ in degree one. The class of the generic element $(a_0\dd a_1\, \mbox{mod}\, [,])\in \Om^1\Ac_{\nat}$ is usually denoted by $\nat a_0\dd a_1$. The map $\nat \dd:\Ac\to \Om^1\Ac_{\nat}$ thus sends $a\in\Ac$ to $\nat \dd a$. Also, the Hochschild boundary $b:\Om^1\Ac\to \Ac$ vanishes on the commutator subspace $[\Ac,\Om^1\Ac]$, hence passes to a well-defined map $\bb:\Om^1\Ac_{\nat}\to\Ac$. Explicitly the image of $\nat a_0\dd a_1$ by $\bb$ is the commutator $[a_0,a_1]$. These maps are bounded and satisfy $\nat \dd\circ \bb=0$ and $\bb\circ\nat \dd=0$, so that $(X(\Ac),\nat\dd\oplus\bb)$ indeed defines a $\zz_2$-graded bornological complex. This construction is obviously functorial: if $\rho:\Ac\to\Bc$ is a bounded homomorphism, it induces a bounded chain map of even degree $X(\rho):X(\Ac)\to X(\Bc)$, by setting $X(\rho)(a)=\rho(a)$ and $X(\rho)(\nat a_0\dd a_1)=\nat \, \rho(a_0)\dd\rho(a_1)$.\\
Now let $\Ic\in\Ac$ be a two-sided ideal. By convention, $\Ic^n$ will always denote the {\it completion} of the $n$th power of $\Ic$ in $\Ac$ for any $n\in\zz$, with $\Ic^n=\Ac$ if $n\le 0$. As in \cite{CQ1}, we introduce the adic filtration of $X(\Ac)$ by the subcomplexes
\beq 
F_{\Ic}^{2n}X(\Ac) &:& \Ic^{n+1}+[\Ic^n,\Ac] \ \rightleftarrows \ \nat \Ic^n\dd\, \Ac \non\\
F_{\Ic}^{2n+1}X(\Ac) &:& \Ic^{n+1}\ \rightleftarrows\ \nat(\Ic^{n+1}\dd\, \Ac + \Ic^n\dd\, \Ic)\ .\label{filtration}
\eeq
This is a decreasing filtration: $F^{n+1}_{\Ic}X(\Ac)\subset F^{n}_{\Ic}X(\Ac)$, and $F^{n}_{\Ic}X(\Ac)=X(\Ac)$ for $n<0$. Let $\Xc^n(\Ac,\Ic)=X(\Ac)/F^{n}_{\Ic}X(\Ac)$ be the quotient complex. We thus get an inverse system $(\Xc^n(\Ac,\Ic))_{n\in\zz}$ of bornological complexes with surjective maps
\be
\ldots \to \Xc^{n}(\Ac,\Ic) \to \ldots \to \Xc^{1}(\Ac,\Ic) = X(\Ac/\Ic)\to \Xc^0(\Ac,\Ic)=\Ac/(\Ic+[\Ac,\Ac])\to 0\ ,
\ee
whose projective limit is the complex 
\be
\Xch(\Ac,\Ic)= \mathop{\limproj}\limits_n \Xc^n(\Ac,\Ic)\ .
\ee
$\Xch(\Ac,\Ic)$ is naturally filtered by the subcomplexes $F^n\Xch(\Ac,\Ic)=\ker(\Xch\to \Xc^n)$. If $\Bc$ is another complete bornological algebra and $\Jc\subset\Bc$ a two-sided ideal, then the space of linear maps between the two projective systems $(\Xc^n(\Ac,\Ic))_{n\in\zz}$ and $(\Xc^n(\Bc,\Jc))_{n\in\zz}$, or between $\Xch$ and $\Xch'$ for short, is given by
\be
\hom(\Xch,\Xch')=\mathop{\limproj}\limits_m\Big(\mathop{\limind}\limits_n \hom(\Xc^n,{\Xc'}^m)\Big)\ ,\label{hom}
\ee
where $\hom(\Xc^n,{\Xc'}^m)$ is the space of bounded linear maps between the $\zz_2$-graded bornological complexes $\Xc^n$ and ${\Xc'}^m$. Thus $\hom(\Xch,\Xch')$ is a $\zz_2$-graded complex. It corresponds also to the space of linear maps $\{f:\Xch\to\Xch'\ |\ \forall k,\,\exists n:\ f(F^n\Xch)\subset F^k\Xch'\}$. Sitting inside the latter are the subcomplexes of linear maps of order $\le n$:
\be
\hom^n(\Xch,\Xch')=\{ f:\Xch\to\Xch'\ |\ \forall k,\ f(F^{k+n}\Xch)\subset F^k\Xch'\}\ .\label{homn}
\ee

\noindent {\bf Entire and periodic cyclic cohomologies.} Let $\Ac$ be a complete bornological algebra. We focus on the $X$-complex of its analytic tensor algebra $\Tc\Ac$. In that case, the quotient $\Om^1\Tc\Ac_{\nat}=\Om^1\Tc\Ac/[\Tc\Ac,\Om^1\Tc\Ac]$ is always complete, and as a bornological vector space $X(\Tc\Ac)$ is canonically isomorphic to the analytic completion $\Oman\Ac$. The correspondence  goes as follows \cite{CQ1,Me}: first, one has a $\Tc\Ac$-bimodule isomorphism 
\beq
\Om^1\Tc\Ac &\cong& \Tct\Ac\hotimes \Ac\hotimes\Tct\Ac\\
x\, \dd a\, y &\leftrightarrow& x\otimes a\otimes y\qquad \mbox{for}\ a\in\Ac\ ,\ x,y\in\Tct\Ac\ , \non
\eeq
where $\Tct\Ac:=\cc\oplus\Tc\Ac$ is the unitalization of $\Tc\Ac$. This implies that the bornological space $\Om^1\Tc\Ac_{\nat}$ is isomorphic to $\Tct\Ac\hotimes\Ac$, which can further be identified with the analytic completion of odd forms $\Oman^-\Ac$, through the correspondence $x\otimes a\leftrightarrow xda$, $\forall a\in \Ac, x\in \Tct\Ac$. Thus collecting the even part $X_0(\Tc\Ac)=\Tc\Ac$ and the odd part $X_1(\Tc\Ac)=\Om^1\Tc\Ac_{\nat}$ together, yields a linear bornological isomorphism $X(\Tc\Ac)\cong \Oman\Ac$. The key point \cite{CQ1,Me} is that the complex $X(\Tc\Ac)$ is homotopy equivalent to the complex of entire chains $\Ome\Ac$ endowed with the differential $(b+B)$. Concretely, the square of the Karoubi operator $\kappa^2$ has discrete spectrum, hence the space of differential forms $\Om\Ac$ decomposes into a direct sum of generalized eigenspaces. One of the eigenvalues is 1. Let $P:\Om\Ac\to\Om\Ac$ denote the projection onto the corresponding eigenspace, annihilating the other ones. One shows that $P$ is bounded for the analytic and entire bornologies, hence extends to $\Oman\Ac$ and $\Ome\Ac$. The point is that the complexes $(X(\Tc\Ac),\nat\dd,\bb)$ and $(\Ome\Ac,b+B)$ retract on $PX(\Tc\Ac)$ and $P\Ome\Ac$ respectively \cite{Me}. Moreover, if $c:\Oman\Ac=X(\Tc\Ac)\to \Ome\Ac$ is the bornological isomorphism (\ref{resc}) induced by the rescaling $(-)^{[n/2]}[n/2]!$ on $n$-forms, then the restriction to 1-eigenspaces
\be
c: (PX(\Tc\Ac),\nat\dd\oplus\bb)\to (P\Ome\Ac,b+B)
\ee
is an isomorphism of bornological complexes, i.e. the differential $\nat\dd\oplus \bb$ corresponds exactly to $b+B$. In the following, $P\Om\Ac$ will be called the subspace of {\it cyclic forms}. From this we get the definition of bivariant entire cyclic cohomology \cite{Me}:
\begin{definition}
Let $\Ac$ and $\Bc$ be complete bornological algebras. Then the bivariant entire cyclic cohomology of $\Ac$ and $\Bc$ is the homology of the $\zz_2$-graded complex of bounded linear maps from $X(\Tc\Ac)$ to $X(\Tc\Bc)$:
\be
HE_*(\Ac,\Bc)=H_*\big(\hom(X(\Tc\Ac),X(\Tc\Bc))\big)\ , \quad *=0,1\ .
\ee
\end{definition}
Thus the bivariant cocycles are the bounded chain maps between $X(\Tc\Ac)$ and $X(\Tc\Bc)$. One can show that the $X$-complex of $\Tc\cc$ is homotopy equivalent to $X(\cc):\cc\rightleftarrows 0$, hence for any algebra $\Ac$, $HE_*(\cc,\Ac)$ is isomorphic to the entire cyclic homology $HE_*(\Ac)=H_*(X(\Tc\Ac))$, and $HE_*(\Ac,\cc)$ is isomorphic to the entire cyclic cohomology $HE^*(\Ac)=H_*(\hom(X(\Tc\Ac),\cc))$. The composition of bounded chain maps being bounded, one has an obvious composition product $HE_i(\Ac,\Bc)\times HE_j(\Bc,\Cc)\to HE_{i+j +2\zz}(\Ac,\Cc)$ for three complete bornological algebras $\Ac,\Bc,\Cc$. Any bounded homomorphism $\rho:\Ac\to\Bc$ extends uniquely to a bounded homomorphism $\Tc\rho:\Tc\Ac\to \Tc\Bc$ by letting $\Tc\rho(a)=\rho(a)$, $\forall a\in\Ac$, whence a bivariant cyclic cohomology class represented by the induced chain map $X(\Tc\rho): X(\Tc\Ac)\to X(\Tc\Bc)$. This gives the functoriality of entire cyclic cohomology.\\
Let us now turn to the periodic theory. It is based on the adic filtration (\ref{filtration}) for the complex $X(\Tc\Ac)$ with respect to the ideal $\Jc\Ac\subset \Tc\Ac$ associated to the universal extension $0\to\Jc\Ac\to\Tc\Ac\to\Ac\to 0$. Under the isomorphism of bornological algebras $\Tc\Ac\cong(\Oman^+\Ac,\odot)$, $\Jc\Ac$ is isomorphic to the completion of the space of differential forms of degree $\ge 2$ for the analytic bornology. Consequently, the {\it completion} $(\Jc\Ac)^n$ of the $n$th power of $\Jc\Ac$ in $\Tc\Ac$ is 
\be
(\Jc\Ac)^n= \Big( \bigoplus_{k\ge n}\Om^{2k}\Ac\Big)^c\ ,
\ee
and under the vector space isomorphism $X(\Tc\Ac)\cong \Oman\Ac$, the filtration $F^n_{\Jc\Ac}X(\Tc\Ac)$ corresponds to the {\it Hodge filtration} of $\Oman\Ac$ \cite{CQ1,Me}:
\be
F^n\Oman\Ac:= F^n_{\Jc\Ac}X(\Tc\Ac)= b(\Om^{n+1}\Ac)\oplus\Big( \bigoplus_{k\ge n+1}\Om^{k}\Ac\Big)^c\ .\label{hodge}
\ee
This allows to compute explicitly the associated $\zz_2$-graded bornological complex $\Xc^n(\Tc\Ac,\Jc\Ac)$:
\be
\Xc^n(\Tc\Ac,\Jc\Ac)=\bigoplus_{k=0}^{n-1}\Om^k\Ac\oplus \Om^n\Ac/b(\Om^{n+1}\Ac)\ .\label{xn}
\ee
The homology in degree $n\ \textup{mod}\ 2$ of this complex is the $n$th cyclic homology group $HC_n(\Ac)$. The projective system $(\Xc^n(\Tc\Ac,\Jc\Ac))_{n\in\zz}$ and its limit $\Xch(\Tc\Ac,\Jc\Ac)$ lead to the definition of periodic cyclic cohomology:
\begin{definition}
Let $\Ac$ and $\Bc$ be complete bornological algebras. The bivariant periodic cyclic cohomology of $\Ac$ and $\Bc$ is the homology of the $\zz_2$-graded complex (\ref{hom}) of linear maps between the projective limits $\Xch(\Tc\Ac,\Jc\Ac)$ and $\Xch(\Tc\Bc,\Jc\Bc)$:
\be
HP_*(\Ac,\Bc)=H_*\big(\hom(\Xch(\Tc\Ac,\Jc\Ac),\Xch(\Tc\Bc,\Jc\Bc))\big)\ ,\quad *=0,1\ .
\ee
For any $n\in\nn$, the non-periodic cyclic cohomology group $HC_n(\Ac,\Bc)$ of degree $n$ is the homology, in degree $n$ \textup{mod} $2$, of the $\zz_2$-graded subcomplex (\ref{homn}) of linear maps of order $\le n$:
\be
HC_n(\Ac,\Bc)=H_{n+2\zz}\big(\hom^n(\Xch(\Tc\Ac,\Jc\Ac),\Xch(\Tc\Bc,\Jc\Bc))\big)\ .
\ee
The embedding $\hom^n\hookrightarrow \hom^{n+2}$ induces, for any $n$, the $S$-operation in bivariant cyclic cohomology $S:HC_{n}(\Ac,\Bc)\to HC_{n+2}(\Ac,\Bc)$, and $\hom^n\hookrightarrow \hom$ yields a natural map $HC_n(\Ac,\Bc)\to HP_{n+2\zz}(\Ac,\Bc)$.
\end{definition}
As in the case of entire cyclic cohomology, the groups $HP_*(\cc,\Ac)$ and $HP_*(\Ac,\cc)$ compute respectively the periodic cyclic homology $HP_*(\Ac)=H_*(\Xch(\Tc\Ac,\Jc\Ac))$ and the periodic cyclic cohomology $HP^*(\Ac)=H_*(\hom(\Xch(\Tc\Ac,\Jc\Ac),\cc))$ of $\Ac$. For the non-periodic theory, $HC_n(\Ac,\cc)$ computes the $n$th cyclic cohomology group $HC^n(\Ac)=H_{n+2\zz}(\hom(\Xc^n(\Tc\Ac,\Jc\Ac),\cc))$. One has also the composition products $HC_n(\Ac,\Bc)\times HC_m(\Bc,\Cc)\to HC_{n+m}(\Ac,\Cc)$ and $HP_i(\Ac,\Bc)\times HP_j(\Bc,\Cc)\to HP_{i+j+2\zz}(\Ac,\Cc)$.

\section{Unbounded bimodules}\label{sbim}

Let $\Ac$ and $\Bc$ be two complete bornological algebras. In this section we recall the construction of the bivariant Chern character 
\be
\ch: \Psi_*^{\te}(\Ac,\Bct)\to HE_*(\Ac,\Bc)\ ,\qquad *=0,1\label{ch}
\ee
performed in \cite{P}. This map associates a bivariant entire cyclic cohomology class to any $\te$-summable unbounded $\Ac$-$\Bct$-bimodule (where $\Bct$ is the unitalization of $\Bc$, see the definitions below). First one has
\begin{definition}\label{dbim}
Let $\Ac$ and $\Bc$ be complete bornological algebras. We assume that $\Ac$ is trivially graded and $\Bc$ is $\zz_2$-graded. Then a (Hilbert) family of spectral triples over $\Bc$, or an unbounded $\Ac$-$\Bc$-bimodule, is a triple $(\Ec,\rho,D)$ corresponding to the following data:
\begin{itemize}
\item A $\zz_2$-graded separable Hilbert space $\Hc$ and the obvious right $\Bc$-module $\Ec=\Hc\hotimes\Bc$.
\item A bounded homomorphism $\rho:\Ac\to \End_{\Bc}(\Ec)$ sending $\Ac$ to the \emph{even degree} endomorphisms of $\Ec$. Hence $\Ec$ is a bornological left $\Ac$-module.
\item An unbounded endomorphism $D:\dom(D)\subset \Ec\to\Ec$ of odd degree, defined on a dense domain of $\Ec$ and commuting with the right action of $\Bc$:
\be
D\cdot(\xi b)=(D\cdot \xi)b\quad \forall \xi\in\Ec\ ,\ b\in\Bc\ .
\ee
We call $D$ a \emph{Dirac operator}.
\item The commutator $[D,\rho(a)]$ extends to an element of $\End_{\Bc}(\Ec)$ for any $a\in\Ac$.
\item For any $t\in \rr_+$, the \emph{heat kernel} $\exp(-t D^2)$ is densely defined and extends to a bounded endomorphism of $\Ec$.
\end{itemize}
We denote by $\Psi(\Ac,\Bc)$ the set of such unbounded bimodules.
\end{definition}
This definition differs slightly from \cite{P}, where $\Hc$ could be any complete bornological vector space. We restrict ourselves to separable Hilbert spaces here in order to introduce the Schatten ideals $\ell^p(\Hc)$ in the subsequent sections. Thus any triple $(\Ec,\rho,D)$ should be viewed as a Hilbert bundle $\Ec$ over the noncommutative ``space'' $\Bc$, endowed with a representation of $\Ac$ as bounded endomorphisms, together with a family of Dirac operators ``over $\Bc$''. The geometric meaning of such objects should be clear.\\

If $\Ac$ and $\Bc$ are both trivially graded, we introduce the higher unbounded bimodules. Let $C_1=\cc\oplus\eps\cc$, be the first complex Clifford algebra. It is the $\zz_2$-graded algebra generated by the unit $1$ in degree zero and $\eps$ in degree one, with $\eps^2=1$. For any trivially graded complete bornological algebra $\Bc$, the tensor product $C_1\hotimes \Bc$ is thus a $\zz_2$-graded complete bornological algebra.
\begin{definition}
Let $\Ac$ and $\Bc$ be trivially graded complete bornological algebras. We set $\Psi_0(\Ac,\Bc):=\Psi(\Ac,\Bc)$ and $\Psi_1(\Ac,\Bc):=\Psi(\Ac,C_1\hotimes \Bc)$. These are respectively the sets of \emph{even} and \emph{odd} $\Ac$-$\Bc$-bimodules.
\end{definition}
One has the following descriptions of $\Psi_*(\Ac,\Bc)$ for $*=0,1$ \cite{P}:
\vskip 2mm
\noindent $\bullet$ $\Psi_0(\Ac,\Bc)$: Let $(\Ec,\rho,D)$ be such a bimodule, with $\Ec=\Hc\hotimes \Bc$. The Hilbert space $\Hc$ is $\zz_2$-graded, hence it comes equipped with an involutive operator $\Gamma$, $\Gamma^2=1$, which splits $\Hc$ into two eigenspaces $\Hc_+$ and $\Hc_-$ of even and odd vectors respectively. Also the right $\Bc$-module $\Ec$ splits into two eigenspaces $\Ec_{\pm}=\Hc_{\pm}\hotimes \Bc$. We adopt the usual $2\times 2$ matrix notation
\be
\Ec=\left( \begin{array}{c}
          \Ec_+ \\
          \Ec_- \\
     \end{array} \right) \qquad
\Gamma=\left( \begin{array}{cc}
          1 & 0 \\
          0 & -1 \\
     \end{array} \right)\ .
\ee
The even (resp. odd) part of the $\zz_2$-graded algebra $\End_{\Bc}(\Ec)$ is represented by diagonal (resp. off-diagonal) matrices. By definition the homomorphism $\rho\to \End_{\Bc}(\Ec)$ commutes with $\Gamma$, whereas $D$ anticommutes. In matricial notations one thus has 
\be
\rho(a)=\left( \begin{array}{cc}
          \rho_+(a) & 0 \\
          0 & \rho_-(a) \\
     \end{array} \right)\qquad
D=\left( \begin{array}{cc}
          0 & D_- \\
          D_+ & 0 \\
     \end{array} \right) \label{dec0}
\ee
for any $a\in\Ac$.
\vskip 2mm
\noindent $\bullet$ $\Psi_1(\Ac,\Bc)$: Let $(\Ec,\rho,D)$ be an odd bimodule. Then $\Ec$ is the direct sum of two copies of $\Kc\hotimes\Bc$ for a given {\it trivially graded} Hilbert space $\Kc$:
\be
\Ec=\left( \begin{array}{c}
          \Kc\hotimes\Bc \\
          \Kc\hotimes\Bc \\
     \end{array} \right) \ ,
\ee
and the right action of $C_1\hotimes\Bc$ on $\Ec$ is such that $\eps$ flips the two factors. It follows that any endomorphism $z\in\End_{C_1\hotimes\Bc}(\Ec)$ reads
\be
z=\left( \begin{array}{cc}
          x & y \\
          y & x \\
     \end{array} \right)\ ,
\ee
with $x,y\in \End_{\Bc}(\Kc\hotimes\Bc)$. As a consequence, there is a bounded homomorphism $\al:\Ac\to \End_{\Bc}(\Kc\hotimes\Bc)$ and an unbounded endomorphism $Q:\Kc\hotimes\Bc\to \Kc\hotimes\Bc$ such that
\be
\rho(a)=\left( \begin{array}{cc}
          \al(a) & 0 \\
          0 & \al(a) \\
     \end{array} \right)\qquad
D=\left( \begin{array}{cc}
          0 & Q \\
          Q & 0 \\
     \end{array} \right)\ .\label{dec1}
\ee
This is the general matricial form for an element of $\Psi_1(\Ac,\Bc)$.\\

The construction of the bivariant Chern character (\ref{ch}) for a bimodule $(\Ec,\rho,D)$ carrying suitable $\te$-summability conditions is then performed in two steps:
\vskip 2mm
\noindent a) We work in Quillen's algebra cochains formalism \cite{Q2}. Using the exponential of the curvature of a superconnection, we introduce a bounded chain map $\chi(\Ec,\rho,D)$ from the $(b+B)$-complex of entire chains $\Ome\Ac$ to the $X$-complex of $\Bc$.
\vskip 2mm
\noindent b) $(\Ec,\rho,D)$ is lifted to a $\Tc\Ac$-$\Tct\Bc$-bimodule. Step a) thus yields a chain map from $\Ome\Tc\Ac$ to $X(\Tc\Bc)$. The bivariant Chern character corresponds to the left composition of the latter map with the homotopy equivalence $X(\Tc\Ac)\stackrel{\sim}{\longrightarrow}\Ome\Tc\Ac$ given by the Goodwillie theorem \cite{Me,P}.\\

For our purpose we need essentially to recall the detailed construction of step a). For this, we use the formalism of algebra cochains \cite{Q2} adapted to the bornological framework \cite{P}. Let $\Ac$ be a complete bornological algebra, $\Om \Ac=\bigoplus_{n\ge 0}\Om^n\Ac$ the $(b,B)$-bicomplex of noncommutative forms over $\Ac$, with $\Om^n\Ac=\Act\hotimes\Ac^{\hotimes n}$. Introduce the \emph{bar complex} 
\be
\Bb(\Ac)=\bigoplus_{n\ge 0}\Bb_n(\Ac)\ ,
\ee
where $\Bb_n(\Ac)=\Ac^{\hotimes n}$ and $\Bb_0(\Ac)=\cc$. Endow $\Bb(\Ac)$ with the {\it entire bornology} $\Sge(\Bb(\Ac))$ generated by the sets $\bigcup_{n\ge 0}[n/2]!\, S^{\otimes n}$ for any $S\in\Sg(\Ac)$. We denote by $\Bbe(\Ac)$ the completion of $\Bb(\Ac)$ with respect to this bornology. $\Bbe(\Ac)$ is a coassociative coalgebra, the (bounded) coproduct $\Delta:\Bbe(\Ac)\to\Bbe(\Ac)\hotimes\Bbe(\Ac)$ is induced by
\be
\Delta(a_1\otimes\ldots\otimes a_n)=\sum_{i=0}^n (a_1\otimes\ldots\otimes a_i)\otimes(a_{i+1}\otimes\ldots\otimes a_n)\ ,
\ee
for any $a_j\in\Ac$. There is a natural $\zz_2$-graduation on $\Bbe(\Ac)$, coming from the parity of $n$ in the string $a_1\otimes\ldots \otimes a_n$. Furthermore, one has a (bounded) boundary map $b':\Bbe(\Ac)\to\Bbe(\Ac)$ of odd degree:
\be
b'(a_1\otimes\ldots\otimes a_n)=\sum_{i=1}^{n-1}(-)^{i+1}a_1\otimes\ldots\otimes a_ia_{i+1}\otimes\ldots\otimes a_n
\ee
verifying ${b'}^2=0$ and $\Delta b'=(b'\otimes\id+\id\otimes b')\Delta$. Thus $\Bbe(\Ac)$ is a complete $\zz_2$-graded differential coalgebra. There is an associated free bicomodule $\Om_1\Bbe(\Ac)=\Bbe(\Ac)\hotimes\Ac\hotimes\Bbe(\Ac)$, with bounded left and right bicomodule maps
\beq
\Delta_l=\Delta\otimes\id\otimes\id &:& \Om_1\Bbe(\Ac)\to \Bbe(\Ac)\hotimes\Om_1\Bbe(\Ac)\ ,\non\\
\Delta_r=\id\otimes\id\otimes\Delta &:& \Om_1\Bbe(\Ac)\to \Om_1\Bbe(\Ac)\hotimes\Bbe(\Ac)\ .
\eeq
It is gifted with the natural $\zz_2$-graduation coming from the parity of $n$ in the string $(a_1\otimes\ldots\otimes a_{i-1})\otimes a_i\otimes(a_{i+1}\otimes\ldots\otimes a_n)$. $\Om_1\Bbe(\Ac)$ has a differential $b'':\Om_1\Bbe(\Ac)\to\Om_1\Bbe(\Ac)$ of odd degree,
\beq
\lefteqn{ b''\left((a_1\otimes\ldots\otimes a_{i-1})\otimes a_i\otimes (a_{i+1}\otimes \ldots\otimes a_n)\right)=}\non\\
&&\qquad\qquad b'(a_1,...,a_{i-1})\otimes a_i\otimes(a_{i+1}\otimes\ldots\otimes a_n)\\
&&\qquad\qquad +(-)^i(a_1\otimes\ldots\otimes a_{i-2})\otimes a_{i-1}a_i\otimes(a_{i+1}\otimes\ldots\otimes a_n)\non\\
&&\qquad\qquad +(-)^{i+1}(a_1\otimes\ldots\otimes a_{i-1})\otimes a_ia_{i+1}\otimes(a_{i+2}\otimes\ldots\otimes a_n)\non\\
&&\qquad\qquad +(-)^i(a_1\otimes\ldots\otimes a_{i-1})\otimes a_i\otimes b'(a_{i+1}\otimes\ldots\otimes a_n)\ ,\non
\eeq
verifying ${b''}^2=0$. $\Delta_{l,r}$ are then morphisms of graded complexes, i.e. $\Delta_l b'' =(b'\otimes 1 + 1\otimes b'')\Delta_l$ and similarly for $\Delta_r$. There is also a bounded projection of even degree $\partial:\Om_1\Bbe(\Ac)\to\Bbe(\Ac)$ defined by
\be
\partial\big((a_1\otimes\ldots\otimes a_{i-1})\otimes a_i\otimes(a_{i+1}\otimes\ldots\otimes a_n)\big)=a_1\otimes\ldots\otimes a_n\ .
\ee
It is a coderivation ($\Delta\d=(\id\otimes\d)\Delta_l+(\d\otimes\id)\Delta_r$) and a morphism of complexes ($\d b''=b'\d$). The last operator we need is the injection $\nat: \Ome\Ac\to \Om_1\Bbe(\Act)$ defined on the space of entire chains over $\Ac$:
\be
\nat(\at_0da_1\ldots da_n)=\sum_{i=0}^n(-)^{n(i+1)}(a_{i+1}\otimes\ldots\otimes a_n)\otimes \at_0\otimes(a_1\otimes\ldots\otimes a_i)\ ,
\ee
for any $\at_0$ in the unitalization $\Act=\cc\oplus\Ac$ and $a_j\in\Ac$. Then $\nat$ is a \emph{cotrace}, i.e. if $\si:\Om_1\Bbe(\Act)\hotimes \Bbe(\Act)\rightleftarrows \Bbe(\Act)\hotimes \Om_1\Bbe(\Act)$ denotes the map which permutes the two factors (with signs according to the degrees of the elements), then one has $\Delta_l\nat = \si \Delta_r\nat$ and $\si\Delta_l\nat =\Delta_r\nat$.\\

Consider now a trivially graded complete bornological algebra $\Bc$, and let $(\Ec,\rho,D)\in\Psi_*(\Ac,\Bct)$ be an unbounded $\Ac$-$\Bct$-bimodule, with $\Ec=\Hc\hotimes\Bct$ for some Hilbert space $\Hc$. The unitalized de Rham-Karoubi completion  $\Omtd\Bc$ of differential forms over $\Bc$ is naturally a left $\Bct$-module, so that we can form the bornological tensor product over $\Bct$
\be
\Omd\Ec := \Ec\hotimes_{\Bct}\Omtd\Bc= \Hc\hotimes \Omtd\Bc\ .
\ee
It is a complete bornological right $\Omtd\Bc$-module, endowed with the differential induced by $\Omtd\Bc$. For the compatibility with the $X$-complex notations, we write $\dd$ for the differential of $\Omtd\Bc$, so that a string like $b_0\dd b_1\ldots\dd b_n$ is a $n$-form over $\Bc$. Then the induced map $\dd:\Omd\Ec\to\Omd\Ec$ is a derivation of right $\Omtd\Bc$-modules.\\
Let $\Lc=\End_{\Omtd\Bc}(\Omd\Ec)$ be the unital algebra of bounded endomorphisms of $\Omd\Ec$, commuting with the right action of $\Omtd\Bc$. It is a complete DG algebra. Its differential is induced by the action of $\dd$ on $\Omd\Ec$, and the unit element $1\in\Lc$ verifies $\dd 1=0$. We denote by $m:\Lc\hotimes\Lc\to\Lc$ the associative product of endomorphisms, and by $m':\Lc\hotimes\Omd\Ec\to\Omd\Ec$ the left action.\\
Introduce now the complete bornological space of bounded linear maps
\be
\Rc=\hom(\Bbe(\Act),\Lc)\ .
\ee
It is a $\zz_2$-graded associative algebra for the convolution product $fg=m\circ(f\otimes g)\circ \Delta$, $\forall f,g\in\Rc$. It is endowed with two anticommuting differentials $\delta$ and $\dd$, coming respectively from the boundaries $b'$ on $\Bbe(\Act)$ and $\dd$ on $\Lc$:
\be
\dd f=\dd\circ f\ ,\quad \delta f=-(-)^{|f|}f\circ b'\ ,\quad \forall f\in\Rc\ ,
\ee
which moreover satisfy the Leibniz rule for the convolution product. Associated to the $\Bbe(\Act)$-bicomodule $\Om_1\Bbe(\Act)$ is the $\zz_2$-graded $\Rc$-bimodule
\be
\Mc=\hom(\Om_1\Bbe(\Act),\Lc)\ ,
\ee
the left and right multiplication maps being respectively given by $f\gamma=m\circ(f\otimes\gamma)\circ\Delta_l$ and $\gamma f=m\circ(\gamma\otimes f)\circ\Delta_r$, for any $f\in\Rc$, $\gamma\in\Mc$. $\Mc$ is also endowed with two anticommuting differentials $\dd\gamma=\dd\circ\gamma$ and $\delta\gamma=-(-)^{|\gamma|}\gamma\circ b''$, compatible with the $\Rc$-bimodule maps in the sense that they fulfill the Leibniz rule. Also, the transposed of the canonical coderivation $\d:\Om_1\Bbe(\Act)\to\Bbe(\Act)$ yields a bounded derivation $\d:\Rc\to\Mc$, commuting with $\dd$ and $\delta$. Finally, the left $\Lc$-module $\Omd\Ec$ yields a left $\Rc$-module
\be
\Fc=\hom(\Bbe(\Act),\Omd\Ec)\ .
\ee
The module map $\Rc\times\Fc\to\Fc$ comes from the convolution product $f\cdot\xi=m'\circ(f\otimes\xi)\circ\Delta$, $\forall f\in\Rc$, $\xi\in\Fc$. It is immediate to check the compatibility of this action with the product on $\Rc$: the coassociativity of $\Bbe(\Act)$ implies $f\cdot(g\cdot\xi)=(fg)\cdot\xi$. The differentials $\dd$ on $\Omd\Ec$ and $b'$ on $\Bbe(\Act)$ imply as above that $\Fc$ is a bidifferential $\Rc$-bimodule: one has $\dd\xi=\dd\circ\xi$ and $\delta\xi=-(-)^{|\xi|}\xi\circ b'$ for any $\xi\in\Fc$, and these differentials are compatible with the ones on $\Rc$, i.e. $\dd(f\cdot\xi)=\dd f\cdot\xi +(-)^{|f|}f\cdot \dd\xi$ and $\delta(f\cdot\xi)=\delta f\cdot\xi +(-)^{|f|}f\cdot \delta\xi$ for any $f\in\Rc$.\\

We now turn to the construction of a {\it superconnection} on $\Fc$. The bounded homomorphism $\rho:\Ac\to\End_{\Bct}(\Ec)$ yields an homomorphism from $\Ac$ to $\Lc$, which we extend to a unital homomorphism from $\Act$ to $\Lc$ still denoted by $\rho$. The thus defined map can be viewed as a cochain on $\Bbe(\Act)$ with values in $\Lc$, so that $\rho$ is an odd degree element of $\Rc$. Moreover, the unbounded endomorphism $D$ on $\Ec$ determines an unbounded operator of odd degree on $\Fc$. Then the superconnection $\Dc:\Fc\to\Fc$ is defined as
\be
\Dc=\delta -\dd+\rho+D\ .
\ee
Its curvature $\Dc^2=-\dd (\rho+D)+[D,\rho]+D^2$ is an unbounded endomorphism of $\Fc$. Since the Chern character involves the exponential of $\Dc^2$, it is necessary to impose the following $\te$-summability condition:
\begin{definition}[\cite{P}]\label{dwte}
The bimodule $(\Ec,\rho,D)\in\Psi_*(\Ac,\Bct)$, $\Ec=\Hc\hotimes\Bct$, is called (weakly) $\te$-summable iff the heat kernel associated to $\Dc^2$, defined by the power series
\be
\exp(-t\Dc^2):= \sum_{n\ge 0}(-t)^n\int_{\Delta_n}ds_1\ldots ds_n\, e^{-ts_0D^2}\Te e^{-ts_1D^2}\ldots \Te e^{-ts_nD^2}\ ,\label{du}
\ee
with $\Te=-\dd(\rho+D)+[D,\rho]$, converges to an element of $\Rc$ for any $t\ge 0$. Here $\Delta_n=\{(s_0,\ldots,s_n)\in [0,1]^{n+1}\ |\ \sum_i s_i=1\}$ is the fundamental $n$-simplex. Moreover for $t>0$, $\exp(-t\Dc^2)$ must define a \emph{trace-class map} $\Bbe(\Act)\to \ell^1(\Hc)\hotimes\Omtd\Bc$, where $\ell^1(\Hc)$ is the ideal of trace-class operators on $\Hc$.
\end{definition}
Recall that we have a derivation of degree zero $\d:\Rc\to\Mc$. Thus the derivative $\d\rho$ of the homomorphism $\rho\in\Rc$ is an odd element of $\Mc$. Assuming the $\te$-summability condition \ref{dwte}, we form the following map defining an odd element of $\Mc$:
\be
\mu= \int_0^1 dt\, e^{-t\Dc^2}\partial\rho \,e^{(t-1)\Dc^2}\ :\ \Om_1\Bbe(\Act)\to\ell^1(\Hc)\hotimes\Omtd\Bc\ .\label{mu}
\ee
$\mu$ should be viewed as the component of the formal exponential $\exp(-\Dc^2+\d\rho)$ of degree one with respect to $\d\rho$. Then composing $\mu$ by the cotrace $\nat:\Ome \Ac\to \Om_1\Bbe(\Act)$ yields an entire cochain on the $(b+B)$-complex of $\Ac$, namely $\mu\nat \in \hom(\Ome \Ac, \ell^1(\Hc)\hotimes\Omtd\Bc)$. The next step is to compose $\mu\nat$ with a partial supertrace $\tau:\ell^1(\Hc)\hotimes\Omtd\Bc\to\Omtd\Bc$. This depends on the parity of the bimodule $(\Ec,\rho,D)$:
\vskip 2mm
\noindent $\bullet$ $(\Ec, \rho,D)\in\Psi_0(\Ac,\Bct)$: The Hilbert space $\Hc=\Hc_+\oplus\Hc_-$ is $\zz_2$-graded, and we denote by $\Tr_s$ the supertrace of operators on $\Hc$. Then $\tau$ is by definition the \emph{even} partial supertrace
\be
\tau=\Tr_s\otimes\id\ :\ \ell^1(\Hc)\hotimes\Omtd\Bc\to\Omtd\Bc\ .\label{tau0}
\ee
The composite $\tau\mu\nat$ is therefore an even entire cochain in $\hom(\Ome\Ac,\Omtd\Bc)$. Remark that $\tau$ commutes with the differential $\dd$ on $\ell^1(\Hc)\hotimes\Omtd\Bc\subset \Lc$ and $\Omtd\Bc$ respectively.
\vskip 2mm
\noindent $\bullet$ $(\Ec, \rho,D)\in\Psi_1(\Ac,\Bct)$: Then $\Ec=\Hc\hotimes\Bct$, where $\Hc$ is the direct sum of two copies of a trivially graded Hilbert space $\Kc$. The algebra of endomorphisms of $\Ec$ corresponds to the tensor product $\End_{\Bct}(\Kc\hotimes\Bct)\hotimes C_1$, where $C_1$ is the Clifford algebra generated by the odd element $\eps$. Since $\mu$ is made out of $\rho$ and $D$, one sees that $\mu$ takes its values in the algebra $\ell^1(\Kc)\hotimes\Omtd\Bc\hotimes C_1$. Then $\tau$ corresponds to the \emph{odd} partial supertrace
\be
\tau=\Tr\otimes\id\otimes\zeta\ :\ \ell^1(\Kc)\hotimes\Omtd\Bc\hotimes C_1\to\Omtd\Bc\ ,\label{tau1}
\ee
where $\Tr$ is the usual trace of operators on the trivially graded algebra $\ell^1(\Kc)$ and $\zeta$ is the unique odd supertrace on $C_1$, verifying $\zeta(1)=0$ and $\zeta(\eps)=\sqrt{2i}$. The latter choice is the only normalization compatible with Bott periodicity and the bivariant Chern character, see \cite{P}. The composite $\tau\mu\nat$ is then an odd entire cochain in $\hom(\Ome\Ac,\Omtd\Bc)$. Remark that $\tau$ {\it anticommutes} with the differential $\dd$ on $\ell^1(\Kc)\hotimes\Omtd\Bc\hotimes C_1\subset \Lc$ and $\Omtd\Bc$ respectively.\\

Let now $p_X:\Omtd\Bc\to X(\Bc)$ be the natural projection. Then the composite $\chi(\Ec,\rho,D)=p_X\tau\mu\nat$ defines a bounded entire cochain on $\Ac$ with values in the $X$-complex of $\Bc$.
\begin{proposition}[\cite{P}]
For any $\te$-summable bimodule $(\Ec,\rho,D)\in\Psi_*(\Ac,\Bct)$, $*=0,1$, the map $\chi(\Ec,\rho,D)\in \hom(\Ome\Ac,X(\Bc))$ is a cocycle of parity equal to $*$, i.e.
\be
(\bb,\nat\dd)\circ\chi(\Ec,\rho,D)-(-)^{*}\chi(\Ec,\rho,D)\circ(b+B)=0\ ,
\ee
where $*=0,1$ is the degree of the bimodule.
\end{proposition}
{\it Proof:} See \cite{P} proposition 6.5.\cqfd\\

The fundamental property of $\chi(\Ec,\rho,D)$ is its homotopy invariance with respect to $\rho$ and $D$. This requires to introduce the Chern-Simons transgression, obtained by a straightforward modification of the construction above. Let $\rho_t$ and $D_t$ depend smoothly on a real parameter $t\in [0,1]$ (for a precise statement of all this, see \cite{P}). Then we replace the previous module $\Fc$ by the new one $\Fc=\hom(\Bbe(\Act),\Omd\Ec\hotimes\Om^*[0,1])$, where $\Om^*[0,1]$ is the de Rham complex of differential forms over $[0,1]$ endowed with its locally convex topology. We denote by $d_t$ the de Rham coboundary with respect to the variable $t$. The new superconnection $\Dc:\Fc\to\Fc$ is then 
\be
\Dc=\delta -(\dd+d_t)+\rho_t+D_t\ ,\label{supercon}
\ee
whose curvature reads $\Dc^2=-(\dd+d_t)(\rho_t+D_t)+[D_t,\rho_t]+D_t^2$. This gives rise to the trace-class map
\be
\mu= \int_0^1 ds\, e^{-s\Dc^2}\partial\rho_t \,e^{(s-1)\Dc^2}\ :\ \Om_1\Bbe(\Act)\to\ell^1(\Hc)\hotimes\Omtd\Bc\hotimes\Om^*[0,1]\ .
\ee
The composite $\tau\mu\nat$ therefore defines a bounded map from $\Ome\Ac$ to $\Omd\Bc\hotimes\Om^*[0,1]$. Let $p_{X,1}:\Omd\Bc\hotimes\Om^*[0,1]\to X(\Bc)\hotimes\Om^1[0,1]$ be the natural projection (for simplicity, we shall assume in the following that $X(\Bc)$ is complete, which is always the case for the universal algebras we will deal with). Then the entire chain
\be
cs(\Ec,\rho_t,D_t)=p_{X,1}\tau\mu\nat\in \hom(\Ome\Ac, X(\Bc)\hotimes\Om^1[0,1])
\ee
is by definition the Chern-Simons transgression associated to the families $\rho_t$ and $D_t$.
\begin{proposition}[\cite{P}]\label{phomo}
Let $(\Ec,\rho_t,D_t)\in\Psi_*(\Ac,\Bct)$, $t\in [0,1]$,  be a differentiable family of $\te$-summable bimodules. Then the cohomology class of the cocycles $\chi(\Ec,\rho_t,D_t)$ is independent of $t$. One has
\be
d_t\chi(\Ec,\rho_t,D_t)=-(\bb,\nat\dd)\circ cs(\Ec,\rho_t,D_t)+(-)^{*}cs(\Ec,\rho_t,D_t)\circ(b+B)\ ,
\ee
where $*=0,1$ is the degree of the bimodule.
\end{proposition}
{\it Proof:} See \cite{P} proposition 6.7. \cqfd\\ 

We now discuss briefly step b), that is, the lifting of $(\Ec,\rho,D)\in \Psi_*(\Ac,\Bct)$ to a $\Tc\Ac$-$\Tct\Bc$-bimodule. Consider the analytic completion $\Oman\Bc$ (section \ref{sprel}) and its unitalization $\Omtan\Bc$. Thus $\Omtan\Bc$ is a unital DG algebra, whose differential is $d$. It is also a bornological left $\Bct$-module, so that we can form the space of {\it analytic} forms with values in $\Ec=\Hc\hotimes\Bct$:
\be
\Oman\Ec := \Ec\hotimes_{\Bct}\Omtan\Bc\ .
\ee
As a bornological vector space, $\Oman\Ec$ is isomorphic to $\Hc\hotimes\Omtan\Bc$ and is naturally a right $\Omtan\Bc$-module. Moreover, it is canonically endowed with the action of the differential $d$, by $d(h\otimes\om)=(-)^{|h|}h\otimes d\om$ for any $h\in\Hc$ and $\om\in\Omtan\Bc$. Let $\Oman^+\Ec=\Hc\hotimes\Omtan^+\Bc$ be the subspace of even degree forms. We define on $\Omtan^+\Ec$ a right action $\odot$ of the unitalized tensor algebra $\Tct\Bc\cong(\Omtan^+\Bc,\odot)$, by deforming the usual product of differential forms into a Fedosov one:
\beq
\odot&:& \Oman^+\Ec\times \Omtan^+\Bc \to \Oman^+\Ec \\
&& (\xi,\om) \mapsto \xi\odot\om:= \xi\om- (-)^{|\xi|}d\xi d\om\ ,\non
\eeq
where $d\xi\in \Oman\Ec$ and $d\om\in\Oman\Bc$ are odd degree differential forms. It is easy to check the right $\Tct\Bc$-module relation $(\xi\odot\om_1)\odot \om_2=\xi\odot(\om_1\odot\om_2)$ for any $\om_1,\om_2\in\Omtan^+\Bc=\Tct\Bc$. Hence, as a right $\Tct\Bc$-module, $\Oman^+\Ec$ is exactly isomorphic to $\Hc\hotimes\Tct\Bc$. \\
Our goal now is to define a left representation of $\Tc\Ac$ as a bounded homomorphism $\rho_*:\Tc\Ac\to \End_{\Tct\Bc}(\Oman^+\Ec)$. This is also given by a Fedosov-type deformation of the usual product. The homomorphism $\rho:\Ac\to \End_{\Bct}(\Ec)$ provides a bounded linear map $\rho_*:\Ac\to \End_{\Tct\Bc}(\Oman^+\Ec)$ by
\be
\rho_*(a)\odot\xi:= \rho(a)\xi- d\rho(a) d\xi\ ,\quad \forall a\in\Ac\ ,\ \xi\in\Oman^+\Ec\ ,
\ee
where $\rho(a)$ and $d\rho(a)$ are viewed as elements of the DG algebra $\End_{\Omtan\Bc}(\Oman\Ec)$, while $\xi,d\xi$ are elements of $\Oman\Ec$. One has $\rho_*(a)\odot(\xi\odot\om)=(\rho_*(a)\odot\xi)\odot\om$ for any $\om\in\Tct\Bc$, hence $\rho_*(a)$ is indeed an endomorphism of $\Oman^+\Ec$. This induces a representation of the analytic tensor algebra $\rho_*:\Tc\Ac\to\End_{\Tct\Bc}(\Oman^+\Ec)$ by
\be
\rho_*(a_1\otimes\ldots\otimes a_n)\odot\xi=\rho_*(a_1)\odot\ldots\odot\rho_*(a_n)\odot\xi\ ,\quad \forall a_1\otimes\ldots\otimes a_n\in \Tc\Ac\ ,\ \xi\in\Oman^+\Ec\ .
\ee
Under the identification of algebras $\Tc\Ac\simeq(\Oman^+\Ac,\odot)$, the above action reads
\beq
\lefteqn{\rho_*(a_0da_1\ldots da_{2n})\odot\xi = (\rho(a_0)d\rho(a_1)\ldots d\rho(a_{2n}))\odot\xi}\\
&&\qquad\qquad= \rho(a_0)d\rho(a_1)\ldots d\rho(a_{2n})\xi -d\rho(a_0)d\rho(a_1)\ldots d\rho(a_{2n})d\xi\ ,\non
\eeq
for any $a_0da_1\ldots da_{2n}\in\Oman^+\Ac$. Hence $\Oman^+\Ec$ is a $\Tc\Ac$-$\Tct\Bc$-bimodule.\\
It remains to lift the Dirac operator $D$ to an unbounded endomorphism of $\Oman^+\Ec$. Once again, we define the Fedosov action
\be
D\odot\xi:= D\xi+dD d\xi\ ,\quad \forall \xi\in\Oman^+\Ec\ ,
\ee
so that $D\odot(\xi\odot\om)=(D\odot\xi)\odot\om$ for any $\om\in\Omtan^+\Bc$. The sign $+$ in front of $dD$ is due to the odd degree of the endomorphism $D$. In general, the thus obtained triple $(\Oman^+\Ec,\rho_*,D)$ does not determine an element of $\Psi_*(\Tc\Ac,\Tct\Bc)$ for two obvious reasons: first, the Fedosov commutator $[D,\rho_*(x)]_{\odot}:= D\odot x - x\odot D$ may not be a bounded endomorphism for any $x\in\Tc\Ac$, and second the heat kernel associated to the Laplacian of $D$ has to be defined carefully. Since we work with Fedosov products, note that the Laplacian is $D^{\odot 2}=D^2+dDdD$, and its exponential is given by the formal power series
\be
\exp_{\odot}(-t D^{\odot 2}):=\sum_{n\ge 0}\frac{(-t)^n}{n!}(D^{\odot 2})^{\odot n}\ .
\ee
We showed in \cite{P} lemma 7.1 that this {\it Fedosov exponential} can be interpreted as a Duhamel-type expansion
\be
\exp_{\odot}H= \sum_{n\ge 0}(-)^n\int_{\Delta_n}ds_1\ldots ds_n \, e^{s_0H}dHd(e^{s_1H})dH\ldots d(e^{s_{n-1}H})dHd(e^{s_nH})\ ,
\ee
with $H=-t(D^2+dDdD)$. Here $e^{s_iH}$ is the exponential for the usual product. A complete expansion of $\exp_{\odot}H$ in degree of differential forms yields a finite number of terms in each degree, automatically incorporating the usual heat kernel $e^{-s_iD^2}\in \End_{\Bct}(\Ec)$. This formula is really the definition of the Fedosov exponential.\\
Even if $(\Oman\Ec,\rho_*,D)$ is not in $\Psi_*(\Tc\Ac,\Tct\Bc)$, we can still perform the construction of the previous chain map 
\be
\chi(\Oman^+\Ec,\rho_*,D): \Ome\Tc\Ac\to X(\Tc\Bc)\ ,
\ee
provided appropriate $\te$-summability conditions are imposed on the lifted triple (this is called {\it strong $\te$-summability} in \cite{P}). The $(b+B)$-complex $\Ome\Tc\Ac$ calculates the entire cyclic homology of $\Tc\Ac$, which is known to be isomorphic to the entire cyclic homology of $\Ac$ by the generalized Goodwillie theorem \cite{Me}. We constructed in \cite{P} an explicit bounded chain map $\gamma: X(\Tc\Ac)\to \Ome\Tc\Ac$ realizing this equivalence. Hence the composition with $\chi(\Oman^+\Ec,\rho_*,D)$ yields an entire cocycle in $\hom(X(\Tc\Ac),X(\Tc\Bc))$ whose class 
\be
\ch(\Ec,\rho,D)\in HE_*(\Ac,\Bc)
\ee 
is the bivariant Chern character of the unbounded bimodule $(\Ec,\rho,D)\in \Psi_*(\Ac,\Bc)$. It is important to note that, under the homotopy equivalence $(\Ome\Tc\Ac,b+B)\sim (X(\Tc\Tc\Ac),\nat\dd\oplus\bb)$ of section \ref{sprel}, the map $\gamma: X(\Tc\Ac)\to \Ome\Tc\Ac$ of \cite{P} is homotopic to the chain map $X(v_{\Ac}):X(\Tc\Ac)\to X(\Tc\Tc\Ac)$ induced by the canonical bounded homomorphism $v_{\Ac}:\Tc\Ac\to \Tc\Tc\Ac$ (see \cite{P} corollary 4.5).

\section{Retraction of the entire Chern character}\label{sretr}

Let $(\Ec,\rho,D)$ be an unbounded $\Ac$-$\Bct$-bimodule. Under some $p$-summability assumptions, we shall retract the entire cocycle $\chi(\Ec,\rho,D)\in \hom(\Ome\Ac, X(\Bc))$ into a collection of finite-dimensional (periodic) cocycles $\chih^n_{\infty}(\Ec,\rho,D)$ for any $n>p-1$. This will allow us to obtain a bivariant Chern character for $p$-summable {\it bounded} Kasparov bimodules. The retraction process presented here is exactly the bivariant analogue of the Connes-Moscovici retraction \cite{CM93} of the Chern character for finitely summable $K$-cycles.\\

Recall that in the preceding section we constructed for any $\te$-summable bimodule $(\Ec,\rho,D)\in\Psi_*(\Ac,\Bct)$, $*=0,1$, an entire cocycle $\chi(\Ec,\rho,D)$ with values in $X(\Bc)$ and with parity $*$. Furthermore, if $(\Ec,\rho_t,D_t)$ is a differentiable family of $\te$-summable bimodules, for a real parameter $t$ running in some interval $[0,T]$, then we associate the Chern-Simons transgression $cs\in\hom(\Ome\Ac,X(\Bc)\hotimes\Om^1[0,T])$. Then proposition \ref{phomo} shows that, upon integration of $cs$ over $[0,T]$, the difference $\chi(\Ec,\rho_0,D_0)-\chi(\Ec,\rho_T,D_T)$ is a coboundary. One can use this property applied to the family $(\Ec,\rho,D_t=tD)$ to cut the tail of the entire cocycle $\chi(\Ec,\rho,D)$ as in \cite{CM93}. This requires to impose some finite-summability conditions. For any $n\in\nn$, let $\chi^n(\Ec,\rho,D)$ be the restriction of $\chi(\Ec,\rho,D)$ to the space of $n$-forms $\Om^n\Ac$. In other words, $\chi^n$ is the composition of $\chi$ with the bounded canonical map $\Om^n\Ac\to\Ome\Ac$. Similarly, we denote by $cs^n$ the restriction of $cs$ to $\Om^n\Ac$.
\begin{definition}\label{dpsum}
Let $p\in\rr_+$. A $\te$-summable bimodule $(\Ec,\rho,D)\in\Psi_*(\Ac,\Bct)$ is $p$-summable iff for any $t>0$, the bimodule $(\Ec,\rho,tD)$ is $\te$-summable, and for any integer $n>p$ one has\\
\noindent i) $\lim_{t\to 0} \chi^{n+1}(\Ec,\rho,tD)=0$;\\
\noindent ii) $cs^n(\Ec,\rho,tD)$, as a one-form with respect to $dt$, is integrable over $[0,T]$, $\forall T>0$.
\end{definition}
Then given a $p$-summable bimodule, the entire cocycle $\chi$ retracts onto a collection of periodic cocycles, i.e. chain maps from the $(b+B)$-complex $\Ome\Ac$ to $X(\Bc)$ which vanish on the subspaces $\Om^k\Ac$ for $k$ sufficiently large:
\begin{proposition}\label{pper}
Let $(\Ec,\rho,D)\in\Psi_*(\Ac,\Bct)$ be a $p$-summable bimodule. Then for any integer $n>p-1$ and $T\in\rr^*_+$, the formula
\beq
\lefteqn{\chih_T^n(\Ec,\rho,D)=\sum_{k=0}^{n+1}\chi^n(\Ec,\rho,TD)+}\\
&&+ \int_{[0,T]}\Big((\bb,\nat\dd)\circ cs^{n+1}(tD)-(-)^*(cs^{n+1}(tD)+cs^{n+2}(tD))\circ B\Big)\non
\eeq
defines a cocycle of degree $*$ in $\hom(\Ome\Ac,X(\Bc))$. It vanishes on $\Om^k\Ac$, $\forall k>n+1$, hence is in fact a \emph{periodic} cocycle. The cohomology class of $\chih_T^n(\Ec,\rho,D)$ inside the complex of periodic cochains is independent of $n>p-1$ and $T\in\rr^*_+$. Moreover, $\chih_T^n(\Ec,\rho,D)$ is cohomologous to $\chi(\Ec,\rho,D)$ in the complex of entire cochains $\hom(\Ome\Ac,X(\Bc))$.
\end{proposition}
{\it Proof:} From proposition \ref{phomo}, one has (with $\chi^n(tD)=\chi^n(\Ec,\rho,tD)$)
\be
d_t\chi^n(tD)=-(\bb,\nat\dd)\circ cs^n(tD)+(-)^{*}(cs^{n-1}(tD)\circ b+cs^{n+1}(tD)\circ B)\label{v1}
\ee
for any $n\in\nn$. By the $p$-summability assumption, one has $\lim_{t\to 0} \chi^{n}(tD)=0$ for any $n>p+1$, thus
\be
\chi^n(TD)=\int_{[0,T]} d_t\chi^n(tD)\qquad \forall n>p+1\ ,\ T>0\ .\label{v2}
\ee
Now let $n>p-1$. The infinite sum $\sum_{k=n+1}^{\infty}cs^k(tD)$ is an entire cochain over $\Ac$, thus the entire cocycle
$$
\chih_T^n(D):= \chi(TD)+ \int_{[0,T]}\sum_{k=n+1}^{\infty}\Big( (\bb,\nat\dd)\circ cs^k(tD)-(-)^{*}cs^k(tD)\circ (b+B)\Big)
$$
is cohomologous to $\chi(TD)$ for any $T>0$, thus in particular cohomologous to $\chi(\Ec,\rho,D)$ by homotopy invariance with respect to $T$. Since $\chi=\sum_{k=0}^{\infty}\chi^n$, we may write 
\beq
&&\chih_T^n(D)=\sum_{k=0}^{n+1}\chi^n(TD)\non\\
&&+\int_{[0,T]}\Big((\bb,\nat\dd) cs^{n+1}(tD)-(-)^*(cs^{n+1}(tD)+cs^{n+2}(tD)) B\Big)\non\\
&&+\sum_{k=n+2}^{\infty}\Big( \chi^k(TD)+\int_{[0,T]}\big[ (\bb,\nat\dd) cs^k(tD)-(-)^{*}(cs^{k-1}(tD) b+cs^{k+1}(tD) B)\big]\Big)\ .\non
\eeq
The infinite sum vanishes by virtue of equations (\ref{v1}) and (\ref{v2}), whence the formula given for $\chih_T^n(D)$. Next, from the definition one sees that the difference
$$
\chih_T^{n+1}(D)-\chih_T^n(D)=\int_{[0,T]}\Big( (\bb,\nat\dd) cs^{n+1}(tD)-(-)^{*}cs^{n+1}(tD) (b+B)\Big)
$$
is a coboundary in the complex of periodic cochains. Moreover, the derivative of $\chih_T^n(D)$ with respect to $T$ reads
\beq
\frac{d}{dT}\chih_T^n(D)&=&\frac{d}{dT}\chi(TD)+\sum_{k=n+1}^{\infty}\Big( (\bb,\nat\dd) cs^k(TD)-(-)^{*}cs^k(TD) (b+B)\Big)\non\\
&=& -(\bb,\nat\dd) cs(TD)+(-)^{*}cs(TD) (b+B)\non\\
&&\qquad\qquad+ \sum_{k=n+1}^{\infty}\Big( (\bb,\nat\dd) cs^k(TD)-(-)^{*}cs^k(TD) (b+B)\Big)\non\\
&=&\sum_{k=0}^{n}\Big( -(\bb,\nat\dd) cs^k(TD)+(-)^{*}cs^k(TD) (b+B)\Big)\ ,\non
\eeq
which proves that the periodic cyclic cohomology class of $\chih_T^n(D)$ is independent of $T$.\cqfd\\

We now take a zero-temperature limit of the retracted cocycles. In order to avoid analytical difficulties with the kernel of the Dirac operator, we shall assume from now on that $D^2$ is an invertible (unbounded) endomorphism of the Hilbert module $\Ec$. This will allow to derive the concrete formulas for a bivariant Chern character on quasihomomorphisms and invertible extensions in section \ref{sKK}. We first have to assume the following stronger conditions:
\begin{assumption}\label{ass}
The $p$-summable bimodule $(\Ec,\rho,D)\in\Psi_*(\Ac,\Bct)$ is such that:\\
\noindent i) $\la+D^2$ is invertible for any $\la\in [0,\infty)$ and the function $\la\to \la^{-u}(\la+D^2)^{-1}$ is integrable over $[0,\infty)$ for any real $u>0$.\\
\noindent ii) For any integer $n\ge 0$, one has $\lim_{t\to\infty}\chi^n(\Ec,\rho,tD)=0$ and the one-form $cs^n(tD)$ is integrable over $[T,\infty)$ for any $T>0$.
\end{assumption}
Condition ii) implies that the limit $T\to\infty$ of the cocycle $\chih^n_T(\Ec,\rho,D)$ exists for all $n>p-1$. Thus the finite-dimensional cocycles
\be
\chih_{\infty}^n(\Ec,\rho,D)= \int_0^{\infty}\Big((\bb,\nat\dd)\circ cs^{n+1}(tD)-(-)^*(cs^{n+1}(tD)+cs^{n+2}(tD))\circ B\Big)\label{chih}
\ee
represent the same periodic cohomology class for any $n>p-1$. With condition i), it is possible to retract $(\Ec,\rho,D)$ onto a bounded Fredholm bimodule \cite{Bl}. For any parameter $u\in (0,1]$, we define
\be
|D|^{-u}=C(u/2)\int_0^{\infty} d\la\, \frac{\la^{-u/2}}{\la+D^2}\ ,
\ee
where $C(u)^{-1}:=(1-u)^{-1}\int_0^{\infty}d\la/(1+\la^{1/(1-u)})$ is a normalization factor. For $u=0$ we set $|D|^0=1$. Then the family of endomorphisms $D_u=D|D|^{-u}$, $u\in [0,1]$, connects homotopically the Dirac operator $D$ to the bounded operator $F:=D_1$, verifying $F^2=1$. 
\begin{proposition}
Let $(\Ec,\rho,D)\in\Psi_*(\Ac,\Bct)$ be a $p$-summable bimodule. Under assumption \ref{ass}, we set $D_u=D/|D|^u$, $F=D/|D|$. Then for $n$ sufficiently large, the cohomology class of the periodic cocycle $\chih^n_{\infty}(\Ec,\rho,D_u)$ is homotopy invariant with respect to $u\in[0,1]$. Hence $\chih^n_{\infty}(\Ec,\rho,F)$ is a periodic cocycle cohomologous to $\chih^n_{\infty}(D)$.
\end{proposition}
{\it Proof:} Now we have two real parameters $t\in\rr_+$ and $u\in[0,1]$. Therefore we need to perform a two-parameters transgression. Consider the Fr\'echet algebra $\Om^*\rr_+$ of rapidly decreasing smooth de Rham forms on $\rr_+$, with differential $d_t$, and the algebra $\Om^*[0,1]$ with differential $d_u$. These are complete bornological algebras for the bounded bornology associated to the locally convex topology. We form the graded tensor product of DG algebras $\Omtd\Bc\hotimes\Om^*\rr_+\hotimes\Om^*[0,1]$, endowed with the total differential $\dd+d_t+d_u$. Recall that in section \ref{sbim} we introduced the right $\Omtd\Bc$-module $\Omd\Ec=\Ec\hotimes_{\Bct}\Omtd\Bc$. We shall mimic the construction of the map $\mu$ (\ref{mu}), with $\Omd\Ec$ replaced by $\Omd\Ec\hotimes\Om^*\rr_+\hotimes\Om^*[0,1]$, as a right module for the algebra $\Omtd\Bc\hotimes\Om^*\rr_+\hotimes\Om^*[0,1]$. Let $\Fc$ be the space of bounded linear maps
$$
\Fc=\hom(\Bbe(\Act),\Omd\Ec\hotimes\Om^*\rr_+\hotimes\Om^*[0,1])\ ,
$$
and let $\Dc:\Fc\to\Fc$ be the superconnection
$$
\Dc=\delta -(\dd+d_t+d_u)+\rho+tD_u\ .
$$
Then with the notations of section \ref{sbim}, the following bounded map (here we have to replace $\Om^*\rr_+$ by $\Om^*[\eta,\infty)$ with $\eta>0$ in order to avoid divergencies when $t\to 0$)
$$
\mu= \int_0^1 ds\, e^{-s\Dc^2}\partial\rho_t \,e^{(s-1)\Dc^2}\ :\ \Om_1\Bbe(\Act)\to\ell^1(\Hc)\hotimes\Omtd\Bc\hotimes\Om^*[\eta,\infty)\hotimes\Om^*[0,1]
$$
verifies the Bianchi identity of \cite{P}, proposition 6.4:
$$
(\dd+d_t+d_u)\mu\nat =\mu\nat(b+B)+[\mu,\rho+tD_u]\nat\ ,
$$
where $\nat:\Ome\Ac\to\Om_1\Bbe(\Act)$ is the cotrace. Let $\tau$ be the partial trace (\ref{tau0},\ref{tau1}), and $p_X:\Omtd\Bc\to X(\Bc)$ be the natural projection. Then $p_X\tau\mu\nat$ is a bounded map from $\Ome(\Ac)$ to $X(\Bc)\hotimes\Om^*[\eta,\infty)\hotimes\Om^*[0,1]$ satisfying the cocycle condition (see \cite{P} propositions 6.5 and 6.7)
$$
(d_t+d_u)p_X\tau\mu\nat=-(\bb,\nat\dd)p_X\tau\mu\nat+(-)^*p_X\tau\mu\nat(b+B)\ ,
$$
where $*=|\tau|$ is the degree of the bimodule $(\Ec,\rho,D)$. We may decompose the map $p_X\tau\mu\nat$ into its components corresponding to different powers of $dt,du$:
\beq
\chi=&\textup{component of $p_X\tau\mu\nat$}&\textup{involving no $dt,du$;}\non\\
cs=&\textup{''}&\textup{ containing only $dt$;}\non\\
\beta=&\textup{''}&\textup{ containing only $du$;}\non\\
\al=&\textup{''}&\textup{ containing $dt\,du$.}\non
\eeq
Also, we denote by $\chi^n,cs^n,\beta^n$ and $\al^n$ their restrictions to the space of $n$-forms $\Om^n\Ac$. Taking the $dt\,du$ component of the cocycle condition above yields
$$
d_t\beta+ d_ucs=-(\bb,\nat\dd)\al +(-)^*\al(b+B)\ ,
$$
or equivalently for any $n\in\nn$:
$$
d_t\beta^n+ d_ucs^n=-(\bb,\nat\dd)\al^n +(-)^*(\al^{n-1}b+\al^{n+1}B)\ .
$$
Let $\int_0^{\infty}$ denote the integration of differential forms over $[0,\infty)$ with respect to the parameter $t$. For any $n\in\nn$, one has $\lim_{t\to\infty}\beta^n=0$. For $n$ sufficiently large, the limit $\lim_{t\to 0}\beta^n$ exists, and it vanishes because the curvature $\Dc^2$ involves no term proportional to $du$. One thus has $\int_0^{\infty} d_t\beta^n=0$ for $n\gg 0$. This implies 
$$
\int_0^{\infty}d_ucs^n=\int_0^{\infty}\Big(-(\bb,\nat\dd)\al^n +(-)^*(\al^{n-1}b+\al^{n+1}B)\Big)\ ,\quad n\gg 0\ .
$$
Therefore the derivative of the periodic cocycle $\chih_{\infty}^n(D_u)$ with respect to $u$ reads (remark that $d_u$ anticommutes with the one-dimensional current $\int_0^{\infty}$)
\beq
d_u\chih_{\infty}^n(D_u) &=& d_u\int_0^{\infty}\Big((\bb,\nat\dd)cs^{n+1} -(-)^*(cs^{n+1}+cs^{n+2})B\Big)\non\\
& =& \int_0^{\infty}\Big((\bb,\nat\dd)d_ucs^{n+1} +(-)^*(d_ucs^{n+1}+d_ucs^{n+2})B\Big)\non\\
&=& -(\bb,\nat\dd)\int_0^{\infty}\Big(-(\bb,\nat\dd)\al^{n+1} +(-)^*(\al^{n}b+\al^{n+2}B)\Big)\non\\
&&\qquad + (-)^*\int_0^{\infty}\Big(-(\bb,\nat\dd)\al^{n+1} +(-)^*\al^{n}b\Big)B\non\\
&&\qquad +(-)^*\int_0^{\infty}\Big(-(\bb,\nat\dd)\al^{n+2} +(-)^*\al^{n+1}b\Big)B\non\\
&=& \int_0^{\infty}\Big((-)^*(\bb,\nat\dd)\al^{n}b - (-)^*(\bb,\nat\dd)\al^{n+1}B +  \al^{n}bB +\al^{n+1}bB \Big)\non\\
&=& \int_0^{\infty}\Big((-)^*(\bb,\nat\dd)\al^{n}b +\al^{n}b(b+B)\non\\
&&\qquad -(-)^*(\bb,\nat\dd)\al^{n+1}B -\al^{n+1}B(b+B)\Big)\ ,\non
\eeq
and since the cochains $\al^nb$ and $\al^{n+1}B$ have degree $(*+1)$ mod 2, one sees that $d_u\chih_{\infty}^n(D_u)$ is a periodic coboundary.\cqfd \\

The retracted cocycles $\chih^n_{\infty}(\Ec,\rho,F)$ are fundamentally associated to the Chern character for $p$-summable bounded Kasparov bimodules according to the definition below. Recall that for any $p\ge 1$, the Schatten ideal $\ell^p(\Hc)$ is the set of bounded operators $x$ on the Hilbert space $\Hc$ such that $\Tr(|x|^{p}) <\infty$. Then $\ell^p(\Hc)$ is a Banach algebra for the norm $||x||_p=(\Tr\,|x|^{p})^{1/p}$.
\begin{definition}
Let $\Ac$ and $\Bc$ be complete bornological algebras, with $\Bc$ $\zz_2$-graded, and let $p\in [1,\infty)$. A $p$-summable bounded Fredholm $\Ac$-$\Bc$-module is a triple $(\Ec,\rho,F)$ where
\begin{itemize}
\item $\Ec=\Hc\hotimes\Bc$ is a right $\Bc$-module for a given separable $\zz_2$-graded Hilbert space $\Hc$.
\item  $\rho:\Ac\to \End_{\Bc}(\Ec)$ is a bounded homomorphism sending $\Ac$ to the \emph{even degree} endomorphisms of $\Ec$. Hence $\Ec$ is a bornological left $\Ac$-module.
\item $F\in\End_{\Bc}(\Ec)$ is a bounded endomorphism of odd degree, such that $F^2=1$.
\item $[F,\rho(a)]\in \ell^p(\Hc)\hotimes\Bc$ for any $a\in\Ac$, where $\ell^p(\Hc)$ is the Schatten ideal of $p$-summable operators on $\Hc$.
\end{itemize}
$(\Ec,\rho,F)$ is called \emph{degenerate} if $[F,\rho(a)]=0$ for all $a\in\Ac$. We denote by $\Esf(\Ac,\Bc)$ the union, for all $p$, of the $p$-summable bounded Fredholm bimodules. If $\Bc$ is trivially graded, then we define $\Esf_0(\Ac,\Bc)=\Esf(\Ac,\Bc)$ and $\Esf_1(\Ac,\Bc)=\Esf(\Ac,C_1\hotimes\Bc)$ in complete analogy with the unbounded case.
\end{definition}
A smooth homotopy between two finitely summable Fredholm bimodules is provided by an interpolating element of $\Esf_*(\Ac,\Bc\hotimes\cinf[0,1])$, where $\cinf[0,1]$ is the Fr\'echet algebra of smooth functions on $[0,1]$. Smooth homotopy is an equivalence relation \cite{Cu2}. The following proposition gives the explicit form of the cocycle $\chih^n_{\infty}(\Ec,\rho,F)$ associated to a $p$-summable Fredholm bimodule.
\begin{proposition}\label{pfred}
Let $\Ac,\Bc$ be trivially graded complete bornological algebras and  $(\Ec,\rho,F)\in\Esf_*(\Ac,\Bct)$ be a $p$-summable Fredholm bimodule, with $\Ec=\Hc\hotimes\Bct$. Then for any integer $n>p-1$, such that $*=n\ \textup{mod}\ 2$, the periodic cocycle $\chih^n_{\infty}(\Ec,\rho,F)\in\hom(\Ome\Ac,X(\Bc))$ vanishes on $\Om^k\Ac$ for any $k\neq n,n+1$, and for any elements $\at_0\in\Act$, $a_1,...,a_{n+1}\in\Ac$,
\beq
\lefteqn{\chih^n_{\infty}(\Ec,\rho,F)(\at_0da_1\ldots da_n)=}\label{35}\\
&&\qquad\qquad (-)^n\frac{\Gamma(1+\frac{n}{2})}{(n+1)!}\, \frac{1}{2}\sum_{\la\in S_{n+1}}\eps(\la)\tau(F[F,\rho(a_{\la(0)})]\ldots[F,\rho(a_{\la(n)})])\ ,\non
\eeq
\beq
\lefteqn{\chih^n_{\infty}(\Ec,\rho,F)(\at_0da_1\ldots da_{n+1})=}\label{36} \\
&&\qquad (-)^n\frac{\Gamma(1+\frac{n}{2})}{(n+1)!}\, \frac{1}{2}\nat\tau\Big( \dd(\rho(a_0)F[F,\rho(a_{1})]\ldots[F,\rho(a_{n+1})])\non\\
&&\qquad\qquad\qquad\qquad +\sum_{\la\in S_{n+2}}\eps(\la) F[F,\rho(a_{\la(0)})]\ldots[F,\rho(a_{\la(n)})]\dd \rho(a_{\la(n+1)})\non\\
&&\qquad\qquad\qquad\qquad -\frac{1}{2}\sum_{\la\in S_{n+2}}\eps(\la) F[F,\rho(a_{\la(0)})]\ldots[F,\rho(a_{\la(n+1)})]\dd F\Big)\ ,\non
\eeq
where $S_k$ denotes the cyclic permutation group of $k$ elements and $\tau$ is the canonical supertrace of degree $*$ on $\ell^1(\Hc)$. Moreover, $\chih^n_{\infty}(\Ec,\rho,F)$ is a cyclic cocycle: if $\kappa$ denotes the Karoubi operator on $\Om\Ac$, then $\chih^n_{\infty}(\Ec,\rho,F)\circ\kappa=\chih^n_{\infty}(\Ec,\rho,F)$. The periodic cohomology class of $\chih^n_{\infty}(\Ec,\rho,F)$ does not depend on $n$ and is homotopy invariant with respect to $\rho$ and $F$. Furthermore, $\chih^n_{\infty}(\Ec,\rho,F)$ vanishes on the degenerate elements of $\Esf_*(\Ac,\Bct)$. If $(\Ec,\rho,F)$ is the retraction of an unbounded bimodule $(\Ec,\rho,D)\in\Psi_*(\Ac,\Bct)$, then $\chih^n_{\infty}(\Ec,\rho,F)$ and $\chi(\Ec,\rho,D)$ define the same entire cyclic cohomology class in $\hom(\Ome\Ac,X(\Bc))$.
\end{proposition}
{\it Proof:} We have to compute the cocycle (\ref{chih}) where $D$ is replaced by $F$, with $F^2=1$. Recall that the Chern-Simons form $cs^n(tF)$ defines a map from $\Om^n\Ac$ to $X(\Bc)\hotimes\Om^1[0,\infty)$, for $n$ sufficiently large. We shall decompose $cs^n(tF)$ into the two components corresponding to the projections
$X(\Bc)\to \Bc$ and $X(\Bc)\to\Om^1\Bc_{\nat}$:
$$
cs^n_0(tF):\Om^n\Ac\to \Bc\hotimes\Om^1[0,\infty)\ ,\quad cs^n_1(tF):\Om^n\Ac\to \Om^1\Bc_{\nat}\hotimes\Om^1[0,\infty)\ .
$$
Consider the superconnection (\ref{supercon}) with $D_t=tF$, and $\rho_t=\rho$ is constant. Thus
$$
\Dc=\delta-(\dd+d_t)+\rho+tF\ ,
$$
and its curvature reads
$$
\Dc^2=-\dd(\rho+tF)-dt\, F+t[F,\rho]+t^2\ .
$$
Let $p_{X,1}$ be the projection of $\Omtd\Bc\hotimes\Om^*[0,\infty)$ onto $X(\Bc)\hotimes\Om^1[0,\infty)$. Then $cs(tF)$ is given by the exponential
$$
cs(tF)= p_{X,1}\tau \int_0^1ds\,  e^{-s\Dc^2}\d\rho\, e^{(s-1)\Dc^2}\nat\ ,
$$
and thus $cs^n(tF)$ is the restriction of this map to $\Om^n\Ac$. Let us first compute $cs^{n+1}_0(tF)B$ on $\Om^n\Ac$. Setting $\te=t[F,\rho]+t^2$, we have
$$
cs_0(tF)= \tau\int_0^1ds\, e^{s(-\te+dtF)}\d\rho e^{(1-s)(-\te+dtF)}\nat_{dt}\ ,
$$
where we retain only the component of degree 1 with respect to $dt$. Using \cite{P} lemma A.2, one thus has
$$
cs^{n+1}_0(tF)B=\tau \d(e^{-\te+dtF})\nat_{dt,\Om^n\Ac}\ .
$$
Since we take the restriction to the subspace of $n$-forms and retain only the degree one in $dt$, the exponential reduces to
\beq
&&cs^{n+1}_0(tF)B = \frac{1}{(n+2)!}e^{-t^2}\tau\d\big((-t[F,\rho]+dtF)^{n+2}\big)\nat_{dt}\non\\
&&= \frac{(-t)^{n+1}}{(n+2)!}e^{-t^2}\tau\d(dtF[F,\rho]^{n+1}+[F,\rho]dtF[F,\rho]^n+\ldots +[F,\rho]^{n+1}dtF)\nat\ .\non
\eeq
Next, one has $[F,\rho]dtF=dtF[F,\rho]$, so that
$$
cs^{n+1}_0(tF)B = (-)^{|\tau|}\frac{(-t)^{n+1}}{(n+1)!}e^{-t^2}dt\tau\d(F[F,\rho]^{n+1})\nat\ ,
$$
where the sign $(-)^{|\tau|}=(-)^*$ comes from the permutation of $dt$ with the graded trace $\tau$, of the same parity as the bimodule $(\Ec,\rho,F)$. Recall that $\d$ is a derivation of degree zero, and since $F$ is a 0-cochain on $\Bbe(\Act)$, one has $\d F=0$. Hence
\beq
\lefteqn{cs^{n+1}_0(tF)B = (-)^{*}\frac{(-t)^{n+1}}{(n+1)!}e^{-t^2}dt\times}\non\\
&&\times \tau\big( F([F,\d\rho][F,\rho]^{n}+[F,\rho][F,\d\rho][F,\rho]^{n-1}+\ldots+ [F,\rho]^n[F,\d\rho])\big)\nat\ .\non
\eeq
Next, using the integral
$$
\int_0^{\infty}t^ne^{-t^2}dt=\frac{1}{2}\,\Gamma\left(\frac{n+1}{2}\right)
$$
and evaluating on a $n$-chain $\at_0da_1\ldots da_n\in\Om^n\Ac$ yields
\beq
\lefteqn{\int_0^{\infty}(-)^*cs^{n+1}_0(tF)B(\at_0da_1\ldots da_n)=}\non\\
&&(-)^{n+1}\frac{\Gamma(1+\frac{n}{2})}{(n+1)!}\,\frac{1}{2}\sum_{\la\in S_{n+1}}\eps(\la)\tau\big(F[F,\rho(a_{\la(0)})]\ldots[F,\rho(a_{\la(n)})]\big)\ .\non
\eeq
Remark that this expression contains $n+2$ powers of $F$. If the parity $*$ is even, then the trace $\tau$ selects only even powers of $F$, whereas if $*$ is odd, then $\tau$ selects only odd powers of $F$ (see \cite{P}). Therefore, the expression above vanishes whenever the parity of $n$ differs from $*$.\\
Let us now compute $cs_1^{n+1}(tF)B$. The trace property of $\nat\tau \cdot \nat$ implies
$$
cs_1(tF)=\nat\tau\int_0^1ds\, e^{s(-\te+dtF)}\d\rho\, e^{(1-s)(-\te+dtF)}\dd (\rho+tF)\nat_{dt}\ .
$$
Thus by \cite{P} lemma A.2, 
$$
cs_1^{n+1}(tF)B=\nat\tau\d(e^{-\te+dtF}\dd(\rho+tF))\nat_{dt}\ ,
$$
again with $\te=t[F,\rho]+t^2$, and
\beq
cs^{n+1}_1(tF)B &=& e^{-t^2}\nat\tau\d\Big( \frac{1}{(n+1)!}(-t[F,\rho]+dtF)^{n+1}\dd\rho +\non\\
&&\qquad\qquad\qquad\qquad+ \frac{t}{(n+2)!}(-t[F,\rho]+dtF)^{n+2}\dd F\Big)\nat_{dt}\ .\non
\eeq
Using the same tricks as before, this leads to
\beq
\lefteqn{\int_0^{\infty}(-)^*cs^{n+1}_1(tF)B(\at_0da_1\ldots da_n)=}\non\\
&& (-)^n\frac{\Gamma(\frac{n+1}{2})}{n!}\, \frac{1}{2}\sum_{\la\in S_{n+1}}\eps(\la)\nat\tau\, F[F,\rho(a_{\la(0)})]\ldots[F,\rho(a_{\la(n-1)})]\dd \rho(a_{\la(n)})\non\\
&&-(-)^n\frac{\Gamma(\frac{n+3}{2})}{(n+1)!}\, \frac{1}{2}\sum_{\la\in S_{n+1}}\eps(\la)\nat\tau\, F[F,\rho(a_{\la(0)})]\ldots[F,\rho(a_{\la(n)})]\dd F\ .\non
\eeq
Again, the properties of $\tau$ imply that this expression vanishes whenever the parity of $n$ is equal to $*$.\\
We now turn to the computation of $\nat\dd\, cs^n_0(tF)$. The trace property of $\nat\dd\tau\cdot\nat$ implies 
$$
\nat\dd\tau\int_0^1ds\, e^{s(-\te+dtF)}\d\rho\, e^{(1-s)(-\te+dtF)}\nat = \nat\dd\tau\big( \d\rho\, e^{-\te+dtF}\big)\nat\ ,
$$
and $\nat\dd\, cs^n_0(tF)$ is the $dt$ component of this expression restricted to $\Om^n\Ac$. Thus
\beq
\nat\dd\, cs^n_0(tF)&=& \frac{1}{(n+1)!}e^{-t^2}\nat\dd\tau\Big( \d\rho(-t[F,\rho]+dtF)^{n+1}\Big)\nat_{dt}\non\\
&=& (-)^*\frac{(-t)^n}{n!}e^{-t^2}dt\,\nat\dd\tau(\d\rho F[F,\rho]^n)\nat\ ,\non
\eeq
where we took care of the fact that $dt$ anticommutes with $\d\rho,\dd$ and with $\tau$ when the latter is odd. Thus one has
\beq
\lefteqn{\int_0^{\infty}\nat\dd cs^n_0(tF)(\at_0da_1\ldots da_n)=}\non\\
&& (-)^*(-)^{n+1}\frac{\Gamma(\frac{n+1}{2})}{n!}\, \frac{1}{2}\nat\dd \tau(\rho(\at_0)F[F,\rho(a_1)]\ldots [F,\rho(a_n)])\ .\non
\eeq
This expression vanishes if $n$ has parity $*$. Now, we want to compute only the cocycles $\chih^n_{\infty}(\Ec,\rho,F)$ where $n$ has parity $*$. In this case, the expression (\ref{chih}) evaluated on $\Om^n\Ac$ and $\Om^{n+1}\Ac$ reduces to
\beq
\chih^n_{\infty}(\Ec,\rho,F)|_{\Om^n\Ac}&=& -\int_0^{\infty}(-)^*cs^{n+1}_0(tF)B\ ,\non\\
\chih^n_{\infty}(\Ec,\rho,F)|_{\Om^{n+1}\Ac}&=& \int_0^{\infty}\nat\dd cs^{n+1}_0(tF) -\int_0^{\infty}(-)^*cs^{n+2}_1(tF)B\ ,\non
\eeq
because of the properties of $\tau$, whence equations (\ref{35}, \ref{36}). For the cyclicity of $\chih^n_{\infty}(\Ec,\rho,F)$, remark that $B\kappa=B$, and a direct computation shows that $\nat\dd\, cs^n_0(tF)\kappa=\nat\dd\, cs^n_0(tF)$. Finally, $\chih^n_{\infty}(\Ec,\rho,F)$ vanishes on degenerates because it involves the commutators $[F,\rho]$. The other statements are a consequence of the preceding propositions. \cqfd\\

At this point we would like to discuss the lifting of the retracted cocycles to the tensor algebras $\Tc\Ac$ and $\Tc\Bc$ in order to get the full bivariant Chern character in cyclic cohomology. For doing this, we first have to lift a $p$-summable bounded Fredholm bimodule $(\Ec,\rho,F)\in \Esf_*(\Ac,\Bct)$ to a $\Tc\Ac$-$\Tct\Bc$-bimodule. Exactly as in the case of unbounded bimodules (section \ref{sbim}), we replace $\Ec$ by the right $\Tct\Bc$-module $\Oman^+\Ec=\Ec\hotimes_{\Bct}\Omtan^+\Bc$, and $\rho$ extends to a left representation $\rho_*:\Tc\Ac\to \End_{\Tct\Bc}(\Oman^+\Ec)$. It remains to treat the Fredholm operator $F$. Once again, we define a Fedosov deformation of its action on $\Oman^+\Ec$ by setting
\be
F\odot\xi:= F\xi+dF d\xi\ ,\quad \forall \xi\in\Oman^+\Ec\ .
\ee
One has $F\odot(\xi\odot\om)=(F\odot\xi)\odot\om$ for any $\om\in\Omtan^+\Bc$, so that $F$ becomes an odd element of the algebra of endomorphisms $\End_{\Tct\Bc}(\Oman^+\Ec)$. However, one must take care of the fact that while $F^2=1$ as an endomorphism of $\Ec$, its lift does not fulfill this property because the endomorphisms of $\Oman^+\Ec$ are endowed with the Fedosov product. One thus has $F^{\odot 2}=1 +dFdF$, where the new term $dFdF$ is the {\it curvature} of $F$. Another difficulty comes from $p$-summability: one can not assume that the Fedosov commutator $[F,\rho_*(x)]_{\odot}$ lies in $\ell^p(\Hc)\hotimes \Tct\Bc$ for any $x\in \Tc\Ac$, because it contains the possibly non-compact term $dF d\rho_*(x)+d\rho_*(x) dF$. Since these new terms always arise together with some powers of the differential $d$ on $\Om\Bc$, we can control them as follows. Instead of working in $\hom(\Ome\Tc\Ac,X(\Tc\Bc))$, we consider the adic filtration of $\Tc\Bc$ by the ideal $\Jc\Bc$ and, for any $m\in\zz$, the graded complex
\be
\hom(\Ome\Tc\Ac,\Xc^m(\Tc\Bc,\Jc\Bc))\ ,
\ee
where $\Xc^m(\Tc\Bc,\Jc\Bc)$ is the quotient complex $X(\Tc\Bc)/F^m_{\Jc\Bc}X(\Tc\Bc,\Jc\Bc)$. Equation (\ref{xn}) shows that it only contains differential forms over $\Bc$ of degree less than $m$. Hence for $n$ sufficiently large, expression (\ref{chih}) applied to the lifted bimodule $(\Oman^+\Ec, \rho_*, F)$, 
\be
\chih_{\infty}^n(\Oman^+\Ec, \rho_*, F)= \int_0^{\infty}\Big((\bb,\nat\dd)\circ cs^{n+1}(tF)-(-)^*(cs^{n+1}(tF)+cs^{n+2}(tF))\circ B\Big)
\ee
yields a well-defined cocycle
\be
\chih_{\infty}^{n,m}(\Oman^+\Ec,\rho_*,F):\Ome\Tc\Ac\to \Xc^m(\Tc\Bc,\Jc\Bc)\ .\label{cm}
\ee 
It is indeed sufficient to take $n\ge p+m/2$, so that the commutators $[F,\rho_*(x)]_{\odot}$ bring at least $p$ compact terms in $\ell^p(\Hc)\hotimes \Tct\Bc$ whose product is trace-class. Hence the Chern-Simons forms are well-defined. Because of the presence of the curvature $dFdF$, the cocycles $\chih_{\infty}^{n,m}(\Oman^+\Ec,\rho_*,F)$ are no longer given by the simple formulas of proposition \ref{pfred}. If we follow the proof of the proposition, we see that in many places a Fedosov exponential of $dFdF$ must be inserted, coming from the new value of $\te=t[F,\rho_*]_{\odot}+t^2+t^2dFdF$. For any integer $k\ge 0$ and $n+1$ elements $x_0, \ldots, x_n$ of $\Tc\Ac$, we introduce the following polynomial,
\beq
\lefteqn{P^n_k(x_0,\ldots, x_n)= \sum_{k_0+\ldots+k_n=k} \sum_{i=0}^{n+1} (-)^i (dFdF)^{k_0} [F,\rho_*(x_0)]_{\odot}(dFdF)^{k_1}\ldots}\non\\
&&\qquad\ldots [F,\rho_*(x_{i-1})]_{\odot}(dFdF)^{k_i}\odot F\odot (dFdF)^{k_{i+1}}[F,\rho_*(x_i)]_{\odot}\ldots\non\\
&&\qquad\ldots (dFdF)^{k_{n+1}} [F,\rho_*(x_n)]_{\odot}(dFdF)^{k_{n+2}}\ ,
\eeq
where all products are taken in $\End_{\Tct\Bc}(\Oman^+\Ec)$. Then the chain map (\ref{cm}) is explicitly obtained as follows. It vanishes on the components of $\Om\Tc\Ac$ of degree different from $n$ or $n+1$, and one finds
\beq
\lefteqn{\chih_{\infty}^{n,m}(\Oman^+\Ec,\rho_*,F)(\widetilde{x}_0dx_1\ldots dx_n)=}\label{f1}\\
&&\qquad \sum_{k\ge 0}(-)^{n+k}\frac{\Gamma(k+1+\frac{n}{2})}{(n+k+2)!}\, \frac{1}{2}\sum_{\la\in S_{n+1}} \eps(\la)\tau P^n_k(x_{\la(0)},\ldots, x_{\la(n)})\ ,\non
\eeq
for any $\widetilde{x}_0dx_1\ldots dx_n\in \Om^n\Tc\Ac$. Similarly on $\Om^{n+1}\Tc\Ac$:
\beq
\lefteqn{\chih_{\infty}^{n,m}(\Oman^+\Ec,\rho_*,F)(\widetilde{x}_0dx_1\ldots dx_{n+1})=}\label{f2}\\
&&\qquad \sum_{k\ge 0}(-)^{n+k}\frac{\Gamma(k+1+\frac{n}{2})}{(n+k+2)!}\, \frac{1}{2}\nat\tau\Big(\dd(\rho_*(\widetilde{x}_0)P^n_k(x_1,\ldots, x_{n+1}))\non\\
&&\qquad\qquad +\sum_{\la\in S_{n+2}} \eps(\la) P^n_k(x_{\la(0)},\ldots, x_{\la(n)})\dd\rho_*(x_{\la(n+1)})\non\\
&&\qquad\qquad - \frac{k+1+\frac{n}{2}}{n+k+3} \sum_{\la\in S_{n+2}} \eps(\la) P^{n+1}_k(x_{\la(0)},\ldots, x_{\la(n+1)})\dd F\Big)\ .\non
\eeq
Since the target complex is $\Xc^m(\Tc\Bc,\Jc\Bc)$ and the polynomial $P^n_k(x_0,\ldots, x_n)$ involves at least $2k$ powers of $dF$, the sum over $k$ in formulas (\ref{f1}, \ref{f2}) is actually finite.\\
If $(\Ec,\rho,F)\in \Esf_*(\Ac,\Bct)$ comes from an unbounded bimodule $(\Ec,\rho,D)\in\Psi_*(\Ac,\Bct)$, i.e. if $F=D/|D|$, we would like to know if the periodic cocycles $\chih_{\infty}^{n,m}(\Oman^+\Ec,\rho_*,F)$ are actually retractions of the entire cocycle $\chi(\Oman^+\Ec,\rho_*,D)$. Since we work in the complex $\hom(\Ome\Tc\Ac,\Xc^m(\Tc\Bc,\Jc\Bc))$, for fixed $m$, we are reduced to the study of the projections
\beq
\chi^{n,m}(\Oman^+\Ec,\rho_*,tD) &:& \Om^n\Tc\Ac\to \Xc^m(\Tc\Bc,\Jc\Bc)\ ,\label{coch}\\
cs^{n,m}(\Oman^+\Ec,\rho_*,tD) &:& \Om^n\Tc\Ac\to \Xc^m(\Tc\Bc,\Jc\Bc)\hotimes \Om^1\rr_+^*\ ,\non
\eeq
In order to perform a retraction process, we must impose a finite-summability condition as stated in definition \ref{dpsum}, with $\Ac$ and $\Bc$ replaced by $\Tc\Ac$ and $\Tc\Bc$ respectively. Note however that due to the presence of an unlimited number of unbounded operators like $dDd\rho_*+d\rho_*dD$ coming from the Fedosov commutator $[D,\rho_*]_{\odot}$, the degree $p$ appearing in definition \ref{dpsum} cannot be taken fixed, but must depend on $m$ (and will generally increase with $m$). We call this property a {\it sequential} finite summability for the cochains (\ref{coch}). Together with the analogue of assumption \ref{ass} {\it ii)}, sequential finite summability implies that for $n$ sufficiently large (depending on $m$), the entire cocycle $\chi(\Oman^+\Ec,\rho_*,D)$ rectracts onto the periodic cocycle $\chih_{\infty}^{n,m}(\Oman^+\Ec,\rho_*,F)\in \hom(\Ome\Tc\Ac,\Xc^m(\Tc\Bc,\Jc\Bc))$. Therefore, the composition with the Goodwillie map $\gamma: X(\Tc\Ac)\to \Ome\Tc\Ac$ yields a cohomology class independent of the choice of $n$, 
\be
[\chih_{\infty}^{n,m}(\Oman^+\Ec,\rho_*,F)\circ\gamma]\in H_*(\hom(X(\Tc\Ac),\Xc^m(\Tc\Bc,\Jc\Bc)))\ ,
\ee
which coincides with the image of the entire Chern character $\ch(\Ec,\rho,D)$ in the cohomology of $\hom(X(\Tc\Ac),\Xc^m(\Tc\Bc,\Jc\Bc))$. These classes are compatible with the projections $\Xc^{m+1}(\Tc\Bc,\Jc\Bc)\to \Xc^m(\Tc\Bc,\Jc\Bc)$, hence by taking their union for all $m\in\zz$ we obtain a projective system. Put in another way, the periodic and entire Chern characters induce the same map
\be
\big([\chih_{\infty}^{n,m}(\Oman^+\Ec,\rho_*,F)\circ\gamma]\big)_{m\in\zz}= \ch(\Ec,\rho,D) \ :\ HE_*(\Ac) \to \mathop{\limproj}\limits_m HC_m(\Bc)\ .
\ee
Note that in general the projective limit $\mathop{\limproj} HC_m(\Bc)$ does not strictly coincide with the periodic cyclic homology $HP_*(\Bc)$, but both are related by a ${\limproj}^1$-exact sequence (see for instance \cite{L}).\\

Formulas (\ref{f1}, \ref{f2}) are general but not very tractable because of their complexity. In what follows, we will suppose that the Fredholm bimodule $(\Ec,\rho,F)$ is put into a {\it canonical form}, for which $dF=0$. This means that $F=G\otimes 1_{\Bct}$ for some bounded operator $G\in \End(\Hc)$. $F$ is therefore a ``constant'' field of Fredholm operators over the space $\Bc$. When dealing with ordinary Kasparov theory \cite{Bl}, it is always possible to represent a $KK$-element by this kind of reduced bimodule. Since by construction $d1_{\Bct}=0$, one automatically has $F^{\odot 2}=1$ and $[F,\rho_*(x)]_{\odot}\in \ell^p(\Hc)\hotimes \Tct\Bc$, where $p$ is the summability degree of $(\Ec, \rho,F)$. Hence it is no longer necessary to project $\chih_{\infty}^n(\Oman^+\Ec,\rho_*,F)$ in the quotient complexes $\hom(\Ome\Tc\Ac,\Xc^m(\Tc\Bc,\Jc\Bc))$, and (\ref{f1}, \ref{f2}) exactly reduce to the formulas of proposition \ref{pfred}, with $\rho$ replaced by $\rho_*$ and all products are replaced by the Fedosov ones. The bivariant Chern character therefore is the composition of the Goodwillie equivalence $\gamma\in\hom(X(\Tc\Ac), \Ome\Tc\Ac)$ with $\chih_{\infty}^n(\Oman^+\Ec,\rho_*,F)\in \hom(\Ome\Tc\Ac,X(\Tc\Bc))$. In the following sections, we shall relate this construction to a universal language appropriate to quasihomomorphisms and invertible extensions, for which $F$ is always a constant field. We will show that the retracted bivariant Chern character is in fact a bivariant periodic cocycle, that is, an element of the $\zz_2$-graded complex $\hom(\Xch(\Tc\Ac,\Jc\Ac),\Xch(\Tc\Bc,\Jc\Bc))$ of section \ref{sprel}.

\section{Universal cocycles}\label{suniv}

Let $\Ac$ be a complete bornological algebra. We shall construct a collection of bounded chain maps of even degree $\chb^{2n}:X(\Tc\Ac) \to F^{4n}_{q\Ac}X(Q\Ac)$, and of odd degree $\chb^{2n+1}: X(\Tc\Ac)\to F^{4n+2}_{q^s\Ac}X(Q^s\Ac)$ for any $n$, where $Q^s\Ac$ is the supersymmetric Fedosov algebra of differential forms over $\Ac$. These universal cocycles are cohomologous to the retractions $\chih^n_{\infty}$ introduced in the last section, associated to {\it universal} Fredholm bimodules. They will play a central role in the bivariant Chern character for $KK$-groups of section \ref{sKK}.\\

Consider the universal free product $Q\Ac=\Ac * \Ac$ of section \ref{sprel}. The $X$-complex
\be
X(Q\Ac)\ :\  Q\Ac\ \xymatrix@1{\ar@<0.5ex>[r]^{\nat \textup{\scriptsize{\bf d}}} &  \ar@<0.5ex>[l]^{\bb}}\ \Om^1Q\Ac_{\nat}
\ee
is a $\zz_2$-graded bornological complex. Since $Q\Ac$ is generated by the range of the two injections $\iota,\iotab:\Ac\to Q\Ac$, the odd part of the $X$-complex $\Om^1Q\Ac_{\nat}$ is isomorphic, as a vector space, to $\Qt\Ac\hotimes (\Ac\oplus\Ac)$, where $\Qt\Ac$ is the unitalization of $Q\Ac$. Indeed, any element of $\Om^1Q\Ac_{\nat}$ may be decomposed into a sum of terms like $\nat\,x\dd \iota(a)$ or $\nat\,x\dd\iotab(a)$ for $x\in \Qt\Ac$ and $a\in\Ac$. As usual, let $q\Ac$ be the kernel of the folding map $\id * \id : Q\Ac\to\Ac$ sending both $\iota(a)$ and $\iotab(a)$ to $a\in\Ac$, i.e. one has an exact sequence of algebras 
\be
0 \to q\Ac \to Q\Ac \to \Ac \to 0\ .
\ee
We recall that $q\Ac$ is the two-sided ideal of $Q\Ac$ generated by the elements $q(a)=\iota(a)-\iotab(a)$, $\forall a\in\Ac$. As in section \ref{sprel}, we will consider the adic filtration of $X(Q\Ac)$ by the subcomplexes 
\beq
F_{q\Ac}^{2n}X(Q\Ac) &:& (q\Ac)^{n+1}+[(q\Ac)^n,Q\Ac] \ \rlarrows \ \nat (q\Ac)^n\dd\, Q\Ac \non\\
F_{q\Ac}^{2n+1}X(Q\Ac) &:& (q\Ac)^{n+1}\ \rightleftarrows\ \nat((q\Ac)^{n+1}\dd\, Q\Ac + (q\Ac)^n\dd\, q\Ac)\label{filt}
\eeq
for any $n\ge 0$. The following proposition introduces a collection of chain maps form the $X$-complex of the analytic tensor algebra $\Tc\Ac$ to the subcomplexes $F^{4n}_{q\Ac}X(Q\Ac)$, i.e. a collection of cocycles in the $\zz_2$-graded complexes $\hom(X(\Tc\Ac),F^{4n}_{q\Ac}X(Q\Ac))$. By convention we extend the two injections $\iota,\iotab$ to unital homomorphisms $\iota,\iotab:\Act\to\Qt\Ac$ by setting $\iota(1)=\iotab(1)=1$, whence $q(1)=0$.
\begin{proposition}\label{uni0}
Let $\Ac$ be a complete bornological algebra. We identify $\Tc\Ac$ with the analytic completion $(\Oman^+\Ac,\odot)$ of the Fedosov algebra of non-commutative differential forms with even degree over $\Ac$. For any $n\ge 0$, a linear map
\be
\chb^{2n}: X(\Tc\Ac)\to F^{4n}_{q\Ac}X(Q\Ac)
\ee
is constructed as follows. For any $x\in \Om^{2k}\Ac$ with $k\ne n$ and $a\in\Ac$, we set $\chb^{2n}(x)=0$ and $\chb^{2n}(\nat x\dd a)=0$; whereas for any $2n$-chain $\at_0da_1\ldots da_{2n}\in \Om^{2n}\Ac$ and $a_{2n+1}\in\Ac$ we set
\beq
&&\chb^{2n}(\at_0da_1\ldots da_{2n}) = \frac{(n!)^2}{(2n)!}\, q(\at_0)q(a_1)\ldots q(a_{2n})\label{ga}\\
&&\chb^{2n}(\nat \at_0da_1\ldots da_{2n}\dd a_{2n+1}) = \frac{(n!)^2}{(2n)!} \, \nat( \iota(\at_0)q(a_1)\ldots q(a_{2n})\dd \iota(a_{2n+1}) -\non\\
 && \hspace{6cm}  -\iotab(\at_0)q(a_1)\ldots q(a_{2n})\dd \iotab(a_{2n+1}))\ .\non
\eeq
Then $\chb^{2n}$ is a bounded chain map of even degree from $X(\Tc\Ac)$ to $F^{4n}_{q\Ac}X(Q\Ac)$, vanishing on the subcomplex $F^{2n+1}_{\Jc\Ac}X(\Tc\Ac)$. In other words, it defines an even \emph{periodic} cocycle in the $\zz_2$-graded complex $\hom(X(\Tc\Ac),F^{4n}_{q\Ac}X(Q\Ac))$. Moreover, if $\kappa$ denotes the Karoubi operator acting on $X(\Tc\Ac)$ identified with $\Oman\Ac$, then this universal cocycle fulfills the following cyclic property:
\be
\chb^{2n}\circ\kappa^{2n+1}=\chb^{2n}\ ,\qquad \forall n\ge 0\ .
\ee
\end{proposition}
{\it Proof:} By direct computation, using repeatedly the identity $q(ab)=\iota(a)q(b)+q(a)\iotab(b)=\iotab(a)q(b)+q(a)\iota(b)$ for any $a,b\in\Ac$. This is tedious but rather straightforward, and we omit the details. \cqfd\\

These cocycles are related to the character $\chih^{2n}_{\infty}$ (proposition \ref{pfred}) of the {\it universal Fredholm bimodule} $u_0\in\Esf_0(\Ac,\Qt\Ac)$ constructed as follows: we set $u_0=(\Ec,\rho,F)$, where $\Ec=\Qt\Ac\oplus\Qt\Ac$ is the sum of two copies of $\Qt\Ac$ with canonical $\zz_2$-graduation; the homomorphism $\rho:\Ac\to M_2(\Qt\Ac)\cong\End_{\Qt\Ac}(\Ec)$ and the odd Fredholm endomorphism $F\in\End_{\Qt\Ac}(\Ec) $ are given by the matrices
\be
\rho(a)=\left( \begin{array}{cc}
          \iota(a) & 0 \\
          0 & \iotab(a) \\
     \end{array} \right)\ ,\qquad
F=\left( \begin{array}{cc}
          0 & 1 \\
          1 & 0 \\
     \end{array} \right)\ .
\ee
\begin{proposition}\label{pcycl0}
Let $u_0\in \Esf_0(\Ac,\Qt\Ac)$ be the universal even bimodule. If $c:X(\Tc\Ac)\to \Ome\Ac$ denotes the rescaling (\ref{resc}) of $n$-forms by $(-)^{[n/2]}[n/2]!$, then for any $n\ge 0$ the $\kappa$-invariant cochain $\chih^{2n}_{\infty}(u_0)\circ c$ is a cocycle in the complex $\hom(X(\Tc\Ac),F^{4n}_{q\Ac}X(Q\Ac))$, corresponding to the cyclic projection of $\chb^{2n}$:
\be
\chih^{2n}_{\infty}(u_0)\circ c=\frac{1}{2n+1}\, \chb^{2n}\circ(1+\kappa+\ldots +\kappa^{2n})\ .
\ee
Consequently, for any $n\ge 0$, $\chb^{2n+2}$ and $\chb^{2n}$ are cohomologous as \emph{periodic cocycles}, i.e. the difference $\chb^{2n+2}-\chb^{2n}$ is the coboundary of a cochain in $\hom(X(\Tc\Ac),F^{4n}_{q\Ac}X(Q\Ac))$ vanishing on $F^k_{\Jc\Ac}X(\Tc\Ac)$ for $k$ sufficiently large.
\end{proposition}
{\it Proof:} Since $q(a)=\iota(a)-\iotab(a)$, one has
$$
[F,\rho(a)]=\left(\begin{array}{cc}
		0 & -q(a) \\
		q(a) & 0
		\end{array} \right)\ ,\qquad
F[F,\rho(a)]=\left(\begin{array}{cc}
		q(a) & 0 \\
		0 & -q(a)
		\end{array} \right)\ ,
$$
so that with the supertrace $\Tr_s$ on supersymmetric $2\times 2$ matrices:
$$
\frac{1}{2}\Tr_s F[F,\rho(a_0)]\ldots [F,\rho(a_{2n})]= (-)^n q(a_0)\ldots q(a_{2n})\ .
$$ 
Taking into account the rescaling $c:X(\Tc\Ac)\cong \Oman\Ac\to \Ome\Ac$ by a factor $(-)^nn!$ on a $2n$-form, we deduce from proposition \ref{pfred}
\beq
\chih^{2n}_{\infty}(u_0)\circ c (\at_0da_1\ldots da_{2n})&=& n!\,\frac{\Gamma(1+n)}{(2n+1)!}\sum_{\la\in S_{2n+1}} q(a_{\la(0)})\ldots q(a_{\la(2n+1)})\non\\
&=& \frac{(n!)^2}{(2n+1)!}\sum_{\la\in S_{2n+1}} q(a_{\la(0)})\ldots q(a_{\la(2n+1)})\ .\non
\eeq
A direct computation shows that it coincides with the evaluation of the $\kappa$-invariant cocycle $\frac{1}{2n+1}\chb^{2n}\circ (1+\kappa+\ldots+\kappa^{2n})$ on $\at_0da_1\ldots da_{2n}$.\\
Now we consider the component of odd degree. Since $\dd 1=0$, the term containing $\dd F$ disappears. One has 
\beq
\lefteqn{\frac{1}{2}\nat\Tr_s\big( \dd(\rho(\at_0) F[F,\rho(a_1)]\ldots [F,\rho(a_{2n+1})])\big)=}\non\\
&&\qquad\qquad (-)^n\frac{1}{2}\nat \dd\big((\iota(a_0)+\iotab(a_0))q(a_1)\ldots q(a_{2n+1})\big)\ ,\non
\eeq
and also
\beq
\lefteqn{\frac{1}{2}\nat\Tr_s \big( F[F,\rho(\at_0)]\ldots [F,\rho(a_{2n})]\dd \rho(a_{2n+1})\big)=}\non\\
&&\qquad\qquad (-)^n\frac{1}{2}\nat\, q(\at_0)\ldots q(a_{2n})\dd(\iota(a_{2n+1})+\iotab(a_{2n+1}))\ .\non
\eeq
Consequently, proposition \ref{pfred} implies
\beq
\lefteqn{\chih^{2n}_{\infty}(u_0)\circ c (\nat\,\at_0da_1\ldots da_{2n}\dd a_{2n+1}) =}\non\\
&&\qquad\qquad n!\,\frac{\Gamma(1+n)}{(2n+1)!}\,\frac{1}{2}\nat\big( \dd((\iota(\at_0)+\iotab(\at_0))q(a_1)\ldots q(a_{2n+1})) +\non\\
&&\qquad\qquad +\sum_{\la\in S_{2n+2}}\eps(\la)q(a_{\la(0)})\ldots q(a_{\la(2n)})\dd(\iota(a_{\la(2n+1)})+\iotab(a_{\la(2n+1)}))\big)\ ,\non
\eeq
and a direct computation shows that it corresponds to $\frac{1}{2n+1}\, \chb^{2n}\circ(1+\kappa+\ldots +\kappa^{2n})$ evaluated on the odd chain $\nat\,\at_0da_1\ldots da_{2n}\dd a_{2n+1}$.\\
To show that $\chb^{2n+2}$ and $\chb^{2n}$ are cohomologous as periodic cocycles, we first consider the cocycles $\chih_{\infty}^{2n+2}(u_0)$ and $\chih_{\infty}^{2n}(u_0)$. From the proof of proposition \ref{pper}, one knows that their difference is a periodic coboundary:
\beq
&&\chih_{\infty}^{2n+2}(u_0)-\chih_{\infty}^{2n}(u_0)=(\nat\dd,\bb) I+ I(b+B)\ ,\non\\
&&I=-\int_{[0,\infty)}(cs^{2n+2}(tF)+cs^{2n+1}(tF))\ .\non
\eeq
Indeed by construction, the Chern-Simons transgressions $cs^{2n+2}$ and $cs^{2n+1}$ vanish on $\Om^k\Ac$ for $k>2n+2$, hence are periodic cochains. A direct computation as in the proof of proposition \ref{pfred} shows that they involve sufficiently many commutators $[F,\rho]$ so that their range lies in the subcomplex $F^{4n}_{q\Ac}X(Q\Ac)$. This shows that $I$ is a periodic cochain in $\hom(\Ome\Ac,F^{4n}_{q\Ac}X(Q\Ac))$. As in section \ref{sprel}, let $P:\Ome\Ac\to\Ome\Ac$ be the spectral projection onto the 1-eigenspace for the square of the Karoubi operator $\kappa^2$, and $P^{\bot}=1-P$ be the orthogonal projection. The identities $(b+B)\circ P\circ c=P\circ c\circ(\nat\dd\oplus\bb)$ and $\chih_{\infty}=\chih_{\infty}\circ P$ imply
$$
\chih_{\infty}^{2n+2}(u_0)c-\chih_{\infty}^{2n}(u_0)c=(\nat\dd,\bb)IPc+IPc(\nat\dd,\bb)\ ,
$$
hence the cyclic projection of $\chb^{2n+2}-\chb^{2n}$ is a periodic coboundary in the complex $\hom(X(\Tc\Ac),F^{4n}_{q\Ac}X(Q\Ac))$. The conclusion follows from the fact that $P^{\bot}X(\Tc\Ac)$ is a contractible subcomplex as well as the inverse system of complexes $(P^{\bot}\Xc^n(\Tc\Ac,\Jc\Ac))_{n\in\nn}$ assocated to the Hodge filtration \cite{CQ1}. \cqfd\\

Let us now turn to the construction of odd cocycles. Recall that the algebra $Q\Ac$ is isomorphic to the Fedosov algebra $(\Om\Ac,\odot)$ of {\it all} differential forms over $\Ac$, for the product
\be
\om_1\odot\om_2 = \om_1\om_2 -(-)^{|\om_1|}d\om_1 d\om_2\ ,\qquad \forall \om_i\in\Om\Ac\ .
\ee
The correspondence sends $\iota(a)$ to $a+da$ and $\iotab(a)$ to $a-da$. Since $(\Om\Ac, \odot)$ is naturally $\zz_2$-graded by the parity of differential forms, $Q\Ac$ inherits this graduation. Under this correspondence, the action of the generator of $\zz_2$ exchanges $\iota(a)$ and $\iotab(a)$ for any $a\in\Ac$, and $Q^s\Ac$ denotes this algebra $(\Om\Ac,\odot)$ {\it considered in the $\zz_2$-graded category} (section \ref{sprel}). Now we have to define carefully what is the $X$-complex of a superalgebra. As in the case of $Q\Ac$, the $X$-complex of $Q^s\Ac$ is the supercomplex
\be
X(Q^s\Ac)\ :\  Q^s\Ac\ \xymatrix@1{\ar@<0.5ex>[r]^{\nat \textup{\scriptsize{\bf d}}} &  \ar@<0.5ex>[l]^{\bb}}\ \Om^1Q^s\Ac_{\nat}\cong \Qt^s\Ac\hotimes(\Ac\oplus\Ac)\ ,
\ee
where $\Om^1Q^s\Ac_{\nat}$ is the quotient of $\Om^1Q^s\Ac$ by the subspace of {\it supercommutators} $[\om_1,\om_2\dd\om_3]_{\odot}= (\om_1\odot \om_2)\dd\om_3-(-)^{|\om_1|(|\om_2|+|\om_3|)}\om_2\dd(\om_3)\om_1$, for any $\om_i\in\Om\Ac\cong Q^s\Ac$. The map $\bb$ on the generic element of $\Om^1Q^s\Ac_{\nat}$ reads 
\be
\bb \nat \om_1\dd\om_2= [\om_1,\om_2]_{\odot}=\om_1\odot\om_2- (-)^{|\om_1||\om_2|}\om_2\odot\om_1\ , \qquad \forall \om_i\in\Om\Ac\ .
\ee
Furthermore, the differential $d$ on forms induces a {\it chain map} $d:X(Q^s\Ac)\to X(Q^s\Ac)$ by setting $d(\om)=d\om$ and $d(\nat\om_1\dd\om_2)= \nat(d\om_1\dd\om_2 +(-)^{|\om_1|}\om_1\dd(d\om_2))$. It is easy to show that it is indeed a chain map, i.e. $d$ commutes with the boundaries $\nat\dd$ and $\bb$.\\
As before, denote by $q^s\Ac$ the kernel of the folding map $\id * \id :Q^s\Ac\to \Ac$. Then, $q^s\Ac$ is the two-sided ideal of $Q^s\Ac$ generated by the one-forms $q(a):=\iota(a)-\iotab(a)=2da\in\Om^1\Ac$. The $X$-complex of $Q^s\Ac$ admits a decreasing filtration $F^n_{q^s\Ac}X(Q^s\Ac)$ analogous to (\ref{filt}). This gives rise to the construction of universal odd cocycles in the complexes $\hom(X(\Tc\Ac),F^{4n+2}_{q^s\Ac}X(Q^s\Ac))$:
\begin{proposition}\label{uni1}
Let $\Ac$ be a complete bornological algebra. We identify $\Tc\Ac$ with the analytic completion $(\Oman^+\Ac,\odot)$ of the Fedosov algebra of even forms, and $Q^s\Ac$ with the (non-completed) Fedosov algebra $(\Om\Ac,\odot)$ of all forms. For any $n\ge 0$, a bounded linear map
\be
\chb^{2n+1}: X(\Tc\Ac)\to F^{4n+2}_{q^s\Ac}X(Q^s\Ac)
\ee
is constructed as follows. For any $x\in \Om^{2k}\Ac$ with $k\ne n+1$ we set $\chb^{2n+1}(x)=0$. For any $x\in \Om^{2k}\Ac$ with $k\ne n$ and $a\in\Ac$, we set $\chb^{2n+1}(\nat x\dd a)=0$. Otherwise, for any elements $\at_0\in\Act$ and $a_1\ldots ,a_{2n+2}\in\Ac$ we set
\beq
&&\chb^{2n+1}(\nat \at_0da_1\ldots da_{2n}\dd a_{2n+1}) = d(\at_0da_1\ldots da_{2n+1})\\
&&\chb^{2n+1}(\at_0da_1\ldots da_{2n+2}) = -d\sum_{i=1}^{n+1}\nat (\at_0da_1\ldots da_{2i-1}\dd(a_{2i})da_{2i+1}\ldots da_{2n+2})\ .\non
\eeq
Then the map $\chb^{2n+1}$ defines an odd \emph{periodic} cocycle in the $\zz_2$-graded complex $\hom(X(\Tc\Ac),F^{4n+2}_{q^s\Ac}X(Q^s\Ac))$. Moreover, this universal cocycle fulfills the following cyclic property:
\be
\chb^{2n+1}\circ\kappa^{2n+2}=\chb^{2n+1}\ , \qquad \forall n\ge 0\ .
\ee
In fact the restriction of $\chb^{2n+1}$ to the even part $\Tc\Ac$ is already $\kappa^2$-invariant: for any $x\in\Om^{2n+2}\Ac$, one has $\chb^{2n+1}\kappa^2(x)=\chb^{2n+1}(x)$.
\end{proposition}
{\it Proof:} By direct computation as in the even case. \cqfd\\

These odd cocycles are also related to the character $\chih^{2n+1}_{\infty}$ of an odd universal Fredholm bimodule. This requires to use the Zekri algebra $E\Ac=\Qt\Ac\cp\zz_2$. We identify $\Qt\Ac$ with the unitalization of the Fedosov algebra $(\Omt\Ac,\odot)$. Thus any element of $E\Ac$ is a linear combination of elements of the form $\om$ or $\om X$, for some $\om\in \Qt\Ac$, where $X$ denotes the generator of $\zz_2$ (i.e. $X^2=1$ and $\om\to X\om X$ implements the action of $\zz_2$ on $\Qt\Ac$). Since $E\Ac$ is already unital, we let $u_1\in\Esf_1(\Ac,E\Ac)$ be the triple $(\Ec,\rho,F)$ where $\Ec=E\Ac\oplus E\Ac$, and $\rho:\Ac\to M_2(E\Ac)$ and $F\in M_2(E\Ac)$ are given by the matrices
\be
\rho(a)=\left( \begin{array}{cc}
          a & 0 \\
          0 & a \\
     \end{array} \right)\ ,\qquad
F=\left( \begin{array}{cc}
          0 & X \\
          X & 0 \\
     \end{array} \right)\ ,\qquad F^2=1\ .
\ee
Next, we need a bounded chain map of even degree $\eta: X(E\Ac)\to X(Q^s\Ac)$. The latter is constructed as follows. First, remark that the even part $X_0(E\Ac)=E\Ac$ decomposes into elements corresponding to one of the six following possibilities: $1$, $X$, $\om_+$, $\om_-$, $\om_+X$ or $\om_-X$, for given differential forms of even/odd degree $\om_{\pm}\in\Om^{\pm}\Ac$. Then we define $\eta$ as
\be
\eta(1)=\eta(X)=\eta(\om_{\pm})=\eta(\om_-X)=0\ ,\qquad \eta(\om_+X)=\om_+\ .
\ee
For the odd part $X_1(E\Ac)$, a similar decomposition holds, and for any $\al_{\pm},\om_{\pm}\in\Omt^{\pm}\Ac$ we set
\beq	
&&\eta(\nat\,\al_+\dd(\om_+X))=\nat\,\al_+\dd\om_+\ ,\quad \eta(\nat\,(\al_+X)\dd\om_+)=\nat\,\al_+\dd\om_+\ ,\non\\
&&\eta(\nat\,\al_-\dd(\om_-X))=\nat\,\al_-\dd\om_-\ ,\quad \eta(\nat\,(\al_-X)\dd\om_-)=-\nat\,\al_-\dd\om_-\ ,
\eeq
and $\eta$ vanishes on all other possibilities. With these definitions, it is easy to show that $\eta$ is indeed a chain map.
\begin{proposition}
Let $u_1\in \Esf_1(\Ac,E\Ac)$ be the universal bimodule of odd degree, $c:X(\Tc\Ac)\to \Ome\Ac$ be the rescaling of $n$-forms by $(-)^{[n/2]}[n/2]!$, and $\eta:X(E\Ac)\to X(Q^s\Ac)$ be the canonical chain map. Then for any $n\ge 0$ the $\kappa$-invariant cochain $\eta\circ\chih^{2n+1}_{\infty}(u_1)\circ c$ is a cocycle in $\hom(X(\Tc\Ac),F^{4n+2}_{q^s\Ac}X(Q^s\Ac))$, corresponding to the cyclic projection of $\chb^{2n+1}$:
\be
\eta\circ \chih^{2n+1}_{\infty}(u_1)\circ c=\sqrt{2\pi i}\,\frac{1}{2n+2}\, \chb^{2n+1}\circ(1+\kappa+\ldots +\kappa^{2n+1})\ ,
\ee
where $\sqrt{2\pi i}$ is the normalization factor accounting for Bott periodicity. Consequently, for any $n\ge 0$, $\chb^{2n+3}$ and $\chb^{2n+1}$ are cohomologous as \emph{periodic cocycles} in $\hom(X(\Tc\Ac),F^{4n+2}_{q^s\Ac}X(Q^s\Ac))$.
\end{proposition}
{\it Proof:} Under the identification of $Q^s\Ac$ with $\Om\Ac$, one has $q(a)=X[X,a]=2da$ and $\iota(a)+\iotab(a)=a+XaX=2a$. We calculate, with the odd trace (\ref{tau1})
\beq
&& \tau(F[F,\rho(a_0)][F,\rho(a_1)]\ldots [F,\rho(a_{2n+1})]) = \sqrt{2i}\,X[X,a_0][X,a_1]\ldots [X,a_{2n+1}]\non\\
&&\qquad\qquad\qquad= (-)^{n+1}\sqrt{2i}\, X[X,a_0]X[X,a_1]X\ldots X[X,a_{2n+1}]X\non\\
&&\qquad\qquad\qquad = (-)^{n+1}\sqrt{2i}\, 2^{2n+2} da_0da_1\ldots da_{2n+1}X\ ,\non
\eeq
so that
$$
\eta\tau(F[F,\rho(a_0)]\ldots [F,\rho(a_{2n+1})])=(-)^{n+1}\sqrt{2i}\,2^{2n+2}da_0\ldots da_{2n+1}\ .
$$
Then proposition \ref{pfred} yields, taking into account the rescaling $c$
\beq
\lefteqn{\eta\chih^{2n+1}_{\infty}(u_1) c (\nat \at_0da_1\ldots da_{2n}\dd a_{2n+1})=}\non\\
&&\qquad\qquad \sqrt{2i}\, 2^{2n+1}n!\frac{\Gamma(n+\frac{3}{2})}{(2n+2)!}\sum_{\la\in S_{2n+2}}\eps(\la)da_{\la(0)}\ldots da_{\la(2n+1)}\ .\non
\eeq
The equality
$$
2^{2n+1}n!\frac{\Gamma(n+\frac{3}{2})}{(2n+2)!}=\frac{\sqrt{\pi}}{2n+2}
$$
implies
$$
\eta\chih^{2n+1}_{\infty}(u_1) c (\nat \at_0da_1\ldots da_{2n}\dd a_{2n+1})=\frac{\sqrt{2\pi i}}{2n+2}\sum_{\la\in S_{2n+2}}\eps(\la)da_{\la(0)}\ldots da_{\la(2n+1)}\ .
$$
Similarly, one has
$$
\nat\tau\dd(\rho(\at_0)F[F,\rho(a_1)]\ldots [F,\rho(a_{2n+2})])=-\sqrt{2i}\,\nat\dd(\at_0X[X,a_1]\ldots [X,a_{2n+2}])\ ,
$$
where the minus sign comes from the permutation of the odd trace $\tau$ with $\dd$, and
$$
\nat\tau\, F[F,\rho(\at_0)]\ldots [F,\rho(a_{2n+1})]\dd\rho(a_{2n+2})= \sqrt{2i}\,\nat\, X[X,\at_{0}]\ldots [X,a_{2n+1}]\dd a_{2n+2}\ .
$$
Thus proposition \ref{pfred} implies
\beq
\lefteqn{\eta\chih^{2n+1}_{\infty}(u_1) c ( \at_0da_1\ldots d a_{2n+2})=}\non\\
&&\qquad\qquad\sqrt{2\pi i}\,\frac{1}{2}\nat\big(\dd(\at_0da_1\ldots da_{2n+2})-\sum_{\la\in S_{2n+3}} da_{\la(0)}\ldots da_{\la(2n+2)}\big)\ ,\non
\eeq
and a direct computation shows that $\eta\chih^{2n+1}_{\infty}(u_1) c$ corresponds precisely to the cyclic cocycle $\frac{\sqrt{2\pi i}}{2n+2}\, \chb^{2n+1}(1+\kappa+\ldots +\kappa^{2n+1})$. One proves as in the even case that $\chb^{2n+3}$ and $\chb^{2n+1}$ are cohomologous as periodic cocycles. \cqfd\\

\section{Bivariant Chern character on $KK$-groups}\label{sKK}

The universal cocycles introduced in the preceding section will allow us to derive a bivariant Chern character for $p$-summable quasihomomorphisms and invertible extensions, by the usual lifting procedure to tensor algebras. We thus get a natural map from the algebraic version of Kasparov theory to bivariant periodic cyclic cohomology. Due to finite summability, the approach followed here is largely independent of bornology and in fact can be performed over arbitrary algebras, without additional structure. We remain in the bornological framework mainly for convenience and compatibility with the previous sections. \\

We start with the even $KK$-group. Let $\Ac$ and $\Bc$ be trivially graded complete bornological algebras. In the following, $\Hc$ will always denote a trivially graded infinite-dimensional separable Hilbert space, $\Lc=\Lc(\Hc)$ the algebra of bounded operators on $\Hc$, and for any $p\in [1,\infty)$, $\ell^p=\ell^p(\Hc)$ is the Schatten ideal of $p$-summable operators. We will assume that a basis of $\Hc$ has been chosen, in other words $\Hc=\ell^2(\nn)$. Furthermore, we fix a bijection $\nn\times \{1,2\}\to\nn$, so that the matrix algebra $M_2(\Lc)$ is isomorphic to $\Lc$. Also, we fix another bijection $\nn\times\nn\to\nn$ yielding an isomorphism between $\Hc$ and the Hilbert space completion of $\Hc\otimes\Hc$ (which does not agree with the bornological tensor product). This in turn yields a bounded homomorphism $\ell^p\hotimes\ell^q\to \ell^r$ for any $p,q\ge 1$ and $r=\max(p,q)$.\\ 
By definition, a $p$-summable quasihomomorphism from $\Ac$ to $\Bc$ is a $p$-summable bounded Fredholm bimodule $(\Ec,\rho,F)\in\Esf_0(\Ac,\Bct)$ whose matricial representation reads
\be
\Ec=\left( \begin{array}{c}
          \Hc\hotimes\Bct \\
          \Hc\hotimes\Bct \\
     \end{array} \right)\ ,\quad
\rho(a)=\left( \begin{array}{cc}
          \rho_+(a) & 0 \\
          0 & \rho_-(a) \\
     \end{array} \right)\in M_2(\Lc\hotimes\Bc)\ ,\quad
F=\left( \begin{array}{cc}
          0 & 1 \\
          1 & 0 \\
     \end{array} \right)\ .
\ee
The $p$-summability condition implies $\rho_+(a)-\rho_-(a)\in \ell^p\hotimes\Bc$ for any $a\in\Ac$. The quasihomomorphism is degenerate iff $\rho_+=\rho_-$. Remark that this definition is more restrictive than the usual one \cite{Bl,Cu1}, since the range of $\rho$ does not lie in the algebra of endomorphisms $\End_{\Bct}(\Ec)$, but in the subalgebra $M_2(\Lc\hotimes\Bc)$. This restriction can be removed only upon the use of excision, see remark \ref{rem} below.\\
Equivalently, we may describe a $p$-summable quasihomomorphism as the free product $\rho_+ * \rho_-: Q\Ac\to \Lc\hotimes\Bc$, sending the ideal $q\Ac$ to $\ell^p\hotimes\Bc$. Degenerate elements are such that $q\Ac$ is mapped to 0. As usual, a (differentiable) homotopy between $p$-summable quasihomomorphisms is an interpolating $p$-summable quasihomomorphism $Q\Ac\to \Lc\hotimes\Bc\hotimes\cinf[0,1]$. The set of quasihomomorphisms is a semigroup for the direct sum $\varphi_1\oplus\varphi_2: Q\Ac\to M_2(\Lc)\hotimes\Bc\cong \Lc\hotimes\Bc$.
\begin{definition}
Let $\Ac$ and $\Bc$ be complete bornological algebras. Then $KK_0(\Ac,\Bc)$ is the union, for all $p\ge 1$, of the classes of $p$-summable quasihomomorphisms $\varphi:Q\Ac\to \Lc\hotimes\Bc$ defined modulo homotopy and addition of degenerates. $KK_0(\Ac,\Bc)$ is an abelian group.
\end{definition}
The zero element of $KK_0(\Ac,\Bc)$ corresponds to the class of degenerates, and the inverse class of a quasihomomorphism $\rho_+ * \rho_-: Q\Ac\to \Lc\hotimes\Bc$ is represented by $\rho_-*\rho_+$.\\

Let us now turn to the construction of the bivariant Chern character. Let $v_\Ac:\Tc\Ac\to\Tc\Tc\Ac$ be the unique bounded homomorphism such that $v_{\Ac}\circ \si_{\Ac}=\si_{\Tc\Ac}\circ\si_{\Ac}$ (section \ref{sprel}). We know that it induces an homotopy equivalence $X(v_\Ac):X(\Tc\Ac)\to X(\Tc\Tc\Ac)$. On the other hand, the folding map $\id * \id: Q\Ac\to\Ac$ lifts to a bounded homomorphism on tensor algebras $\Tc(\id*\id): \Tc Q\Ac\to\Tc\Ac$. Let $\Ic$ be the its kernel. Then $\Tc(\id*\id)$ has two right inverses given by the bounded homomorphisms $\Tc\iota,\Tc\iotab:\Tc\Ac\to\Tc Q\Ac$, so that the universal propery of $Q\Tc\Ac$ implies the following commutative diagram:
\be
\xymatrix{
0 \ar[r] & q\Tc\Ac \ar[r] \ar[d]_{\phi} & Q\Tc\Ac \ar[r] \ar[d]_{\phi} & \Tc\Ac \ar[r] \ar@{=}[d] & 0 \\
0 \ar[r] & \Ic \ar[r] & \Tc Q\Ac \ar[r]_{\ {}^{\Tc(\id*\id)}} & \Tc\Ac \ar[r] \ar@/_1pc/[l]_{\Tc\iota,\Tc\iotab} & 0 }
\ee
The classifying map $\phi:Q\Tc\Ac\to \Tc Q\Ac$ can be explicitly described as follows. If $a_1\otimes\ldots \otimes a_n$ is an element of $\Tc\Ac$, then $\phi(\iota(a_1\otimes\ldots\otimes a_n))$ is the element $\iota(a_1)\otimes\ldots \otimes \iota(a_n)\in \Tc Q\Ac$, and similarly with $\iotab$. The bounded homomorphism $\phi$ induces a chain map $X(\phi):X(Q\Tc\Ac)\to X(\Tc Q\Ac)$, compatible with the adic filtrations by the powers of the ideals $q\Tc\Ac$ and $\Ic$. Then for any $n\in\nn$, we let $\gamma^{2n}_{\Ac}$ be the composite
\be
\begin{CD}
 X(\Tc\Ac) @>{X(v_{\Ac})}>> X(\Tc\Tc\Ac) @>{\chb^{2n}}>> F^{4n}_{q\Tc\Ac} X(Q\Tc\Ac) @>{X(\phi)}>> F^{4n}_{\Ic}X(\Tc Q\Ac)\ ,
\end{CD}
\ee
where $\chb^{2n}$ is the universal cocycle of proposition \ref{uni0}, with $\Ac$ replaced by $\Tc\Ac$.
\begin{proposition}\label{deg}
Let $\Ac$ be a complete bornological algebra. For any $n\ge 0$, the bounded chain map $\gamma^{2n}_{\Ac}$ defines a cocycle in the complex $\hom^{2n}(X(\Tc\Ac),X(\Tc Q\Ac))$ of maps of order $\le 2n$. Consequently it represents a cyclic cohomology class of degree $2n$:
\be
[\gamma^{2n}_{\Ac}] \in HC_{2n}(\Ac, Q\Ac)\ .
\ee
Moreover, all these cocycles have the same image in the bivariant periodic cyclic cohomology $HP_0(\Ac,Q\Ac)$.
\end{proposition}
{\it Proof:} We first need to fix the notations concerning the identifications of $\Tc\Ac$ and $\Tc\Tc\Ac$ with Fedosov algebras of differential forms. As before, $\Tc\Ac\cong(\Oman^+\Ac,\odot)$ will denote the analytic completion of even degree differential forms over $\Ac$, $\odot$ is the Fedosov product, and $d$ is the differential on $\Oman\Ac$. Thus $a_0da_1\ldots da_{2n}$ is a $2n$-form in $\Om^{2n}\Ac$. On the other hand, we have an isomorphism $\Tc\Tc\Ac\cong(\Oman^+\Tc\Ac,\odob)$, where $\odob$ is the Fedosov product for $\Oman^+\Tc\Ac$. We denote by $\db$ the differential on $\Oman\Tc\Ac$. Thus $x_0\db x_1\ldots\db x_{2n}\in\Om^{2n}\Tc\Ac$, $x_i\in\Tc\Ac$, denotes a $2n$-form over $\Tc\Ac$. The canonical homomorphism $v_{\Ac}:\Tc\Ac\to \Tc\Tc\Ac$ may be expressed in terms of differential forms as follows. By definition, an element $a\in\Ac=\Om^0\Ac$ maps to the element $a\in\Ac\subset\Tc\Ac=\Om^0\Tc\Ac$. Next, using the homomorphism property, for any element $da_1da_2\in\Om^2\Ac$ one has
\beq
&&\qquad v_{\Ac}(da_1da_2)=v_{\Ac}(a_1a_2-a_1\odot a_2)=a_1a_2-a_1\odob a_2\non\\
&&=a_1a_2-a_1\odot a_2 +a_1\odot a_2 -a_1\odob a_2 = da_1da_2+ \db a_1\db a_2\ \in \Om^0\Tc\Ac\oplus \Om^2\Tc\Ac\ ,\non
\eeq
so that the evaluation of $v_{\Ac}$ on a $2n$-form in $\Om^{2n}\Ac$ can be determined recursively using the homomorphism property. This leads to the adic behaviour of $v_{\Ac}$ with respect to the ideal $\Jc\Ac$ associated to the universal extension $0\to \Jc\Ac\to\Tc\Ac\to\Ac\to 0$. First, recall that $\Jc\Ac=(\bigoplus_{n\ge 1} \Om^{2n}\Ac)^c$ is the ideal of $\Tc\Ac$ generated by the $2$-forms $da_1da_2$. Similarly, $\Jc\Tc\Ac=(\bigoplus_{n\ge 1} \Om^{2n}\Tc\Ac)^c$ is the kernel of the extension $0\to \Jc\Tc\Ac \to \Tc\Tc\Ac\to\Tc\Ac\to 0$ for $\Tc\Ac$. Now let $\Jc\subset\Tc\Tc\Ac$ be the ideal generated by the linear embedding $\Jc\Ac\hookrightarrow\Tc\Ac=\Om^0\Tc\Ac$. The equation above shows that $v_{\Ac}(\Jc\Ac)\subset \Jc + \Jc\Tc\Ac$.\\ 
Let $n$ be a fixed integer. Since $\Jc$ and $\Jc\Tc\Ac$ are ideals in $\Tc\Tc\Ac$, we can write for any $k>n$
\be
v_{\Ac}((\Jc\Ac)^k)\subset (\Jc+\Jc\Tc\Ac)^k\subset \sum_{i=0}^n\Jc^{k-i}+\sum_{i=n+1}^k(\Jc\Tc\Ac)^i\subset \Jc^{k-n} + (\Jc\Tc\Ac)^{n+1}\ ,\label{spli}
\ee
and for $k\le n$ the same inclusion holds obviously, since in this case $\Jc^{k-n}=\Tc\Tc\Ac$. Now, denote by $\Kc$ the ideal in $Q\Tc\Ac$ generated by $\iota(\Jc\Ac)+\iotab(\Jc\Ac)$. We want to show that the composition $\chb^{2n}\circ X(v_{\Ac})$ maps the Hodge filtration $F^m_{\Jc\Ac}X(\Tc\Ac)$ to $F^{m-2n}_{\Kc}X(Q\Tc\Ac)$ for any $m$. Let us first investigate the case of $m=2k$, where $k$ is an arbitrary integer. Recall that the filtration of degree $2k$ is given by the subcomplex
$$
F_{\Jc\Ac}^{2k}X(\Tc\Ac) \ :\ (\Jc\Ac)^{k+1}+[(\Jc\Ac)^k,\Tc\Ac] \ \rightleftarrows \ \nat (\Jc\Ac)^k\dd\, \Tc\Ac \ .
$$
Since $v_{\Ac}((\Jc\Ac)^{k+1})\subset\Jc^{k-n+1} + (\Jc\Tc\Ac)^{n+1}$, we deduce 
$$
\chb^{2n}\circ v_{\Ac}((\Jc\Ac)^{k+1})\subset \chb^{2n}(\Jc^{k-n+1})\ ,
$$
because proposition \ref{uni0} states that $\chb^{2n}$ vanishes on the filtration
$$
F^{2n+1}_{\Jc\Tc\Ac}X(\Tc\Tc\Ac): (\Jc\Tc\Ac)^{n+1} \rightleftarrows \nat\,((\Jc\Tc\Ac)^{n+1}\dd\Tc\Tc\Ac + (\Jc\Tc\Ac)^n\dd\Jc\Tc\Ac)\ .
$$
Next, $\chb^{2n}$ maps a $2n$-form $x_0\db x_1\ldots \db x_{2n}\in \Om^{2n}\Tc\Ac$ to $q(x_0)q(x_1)\ldots q(x_{2n})\in Q\Tc\Ac$ up to a multiplicative factor, hence if $x_i\in (\Jc\Ac)^{l_i}$ one has
$$
\chb^{2n}(x_0\db x_1\ldots \db x_{2n})\in q((\Jc\Ac)^{l_0})q((\Jc\Ac)^{l_1})\ldots q((\Jc\Ac)^{l_{2n}})\subset \Kc^{\sum_i l_i}\ .
$$
This shows that $\chb^{2n}(\Jc^{k-n+1})\subset \Kc^{k-n+1}$ and
$$
\chb^{2n}\circ v_{\Ac}((\Jc\Ac)^{k+1})\subset \Kc^{k-n+1}\ .
$$
In the same way we get
$$
X(v_{\Ac})\, \nat (\Jc\Ac)^k\dd\, \Tc\Ac\subset \nat\, (\Jc^{k-n}+(\Jc\Tc\Ac)^{n+1}) \dd \Tc\Tc\Ac\ ,
$$
so that
$$
\chb^{2n}\circ X(v_{\Ac})(\nat (\Jc\Ac)^k\dd\, \Tc\Ac)\subset \nat \Kc^{k-n}\dd Q\Tc\Ac\ .
$$
We deal now with the commutator $[(\Jc\Ac)^k,\Tc\Ac]$. It corresponds to the image of $\nat (\Jc\Ac)^k\dd\, \Tc\Ac$ by $\bb$, hence we can use the cocycle property of $\chb^{2n}\circ X(v_{\Ac})$:
$$
\chb^{2n}\circ X(v_{\Ac})([(\Jc\Ac)^k,\Tc\Ac])= \bb\, \chb^{2n}\circ X(v_{\Ac})(\nat (\Jc\Ac)^k\dd\, \Tc\Ac)\subset [\Kc^{k-n}, Q\Tc\Ac]\ .
$$
This shows that $\chb^{2n}\circ X(v_{\Ac})$ maps $F_{\Jc\Ac}^{2k}X(\Tc\Ac)$ to $F_{\Kc}^{2k-2n}X(Q\Tc\Ac)$.\\
It remains to treat the odd case $m=2k+1$. One has
$$
F_{\Jc\Ac}^{2k+1}X(\Tc\Ac)\ :\ (\Jc\Ac)^{k+1}\ \rightleftarrows\ \nat((\Jc\Ac)^{k+1}\dd\, \Tc\Ac + (\Jc\Ac)^k\dd\, \Jc\Ac)\ .
$$
From the preceding computations, the cocycle $\chb^{2n}\circ X(v_{\Ac})$ sends the subspaces $(\Jc\Ac)^{k+1}$ and $\nat(\Jc\Ac)^{k+1}\dd\, \Tc\Ac$ to $\Kc^{k-n+1}$ and $\nat \Kc^{k-n+1}\dd Q\Tc\Ac$ respectively. The remaining term $\nat (\Jc\Ac)^k\dd \Jc\Ac$ requires more work. One has
$$
X(v_{\Ac})\, \nat (\Jc\Ac)^k\dd \Jc\Ac = \nat\, v_{\Ac}((\Jc\Ac)^k)\dd v_{\Ac}(\Jc\Ac) \subset \nat \, (\Jc+\Jc\Tc\Ac)^k\dd(\Jc+\Jc\Tc\Ac)\ .
$$
We compute separately
$$
\nat\, (\Jc+\Jc\Tc\Ac)^k\dd \Jc \subset \nat\, (\Jc^{k-n}+\Jc\Tc\Ac^{n+1})\dd\Jc\ ,
$$
and by choosing another splitting of the sum in (\ref{spli}),
$$
\nat\, (\Jc+\Jc\Tc\Ac)^k\dd \Jc\Tc\Ac \subset \nat\, (\Jc^{k-n+1}+\Jc\Tc\Ac^{n})\dd \Jc\Tc\Ac\ .
$$
Since $\chb^{2n}$ vanishes on the subspaces $\nat(\Jc\Tc\Ac)^{n+1}\dd\Tc\Tc\Ac\supset(\Jc\Tc\Ac)^{n+1}\dd\Jc$ and $\nat(\Jc\Tc\Ac)^n\dd\Jc\Tc\Ac$, we deduce
\beq
&&\chb^{2n}\circ X(v_{\Ac})(\nat (\Jc\Ac)^k\dd \Jc\Ac)\subset \chb^{2n}\,\nat(\Jc^{k-n}\dd\Jc+ \Jc^{k-n+1}\dd \Jc\Tc\Ac)\non\\
&&\qquad \subset   \nat(\Kc^{k-n}\dd\Kc+ \Kc^{k-n+1}\dd Q\Tc\Ac)\subset F_{\Kc}^{2k+1-2n}X(Q\Tc\Ac)\ .\non
\eeq
This proves that $\chb^{2n}\circ X(v_{\Ac})$ maps $F_{\Jc\Ac}^{2k+1}X(\Tc\Ac)$ to $F_{\Kc}^{2k+1-2n}X(Q\Tc\Ac)$ for any $k$.\\
Finally, let us investigate the classifying homomorphism $\phi: Q\Tc\Ac\to\Tc Q\Ac$. For any differential form $a_0da_1\ldots da_{2k}\in\Om^{2k}\Ac\subset \Tc\Ac$, one has
$$
\phi(\iota(a_oda_1\ldots da_{2k}))=\iota(a_0)d\iota(a_1)\ldots d\iota(a_{2k})\ ,
$$
and similarly for $\iotab$, so that $\phi$ obviously maps $\Kc$ to the ideal $\Jc Q\Ac$ associated to the universal extension $0\to \Jc Q\Ac\to \Tc Q\Ac\to Q\Ac\to 0$. This shows that the cocycle $\gamma^{2n}_{\Ac}$ maps $F^m_{\Jc\Ac}X(\Tc\Ac)$ to $F^{m-2n}_{\Jc Q\Ac}X(\Tc Q\Ac)$ for any $m$, hence by definition it is of order $\le 2n$ and its cohomology class is an element of $HC_{2n}(\Ac,Q\Ac)$.\\
By proposition \ref{pcycl0}, we know that the difference $\chb^{2n+2}-\chb^{2n}$ is the coboundary of a chain in $\hom(X(\Tc\Tc\Ac), X(Q\Tc\Ac))$ vanishing on $F^k_{\Jc\Tc\Ac}X(\Tc\Tc\Ac)$ for $k$ sufficiently large, and this implies as above that $\gamma^{2n+2}_{\Ac}-\gamma^{2n}_{\Ac}$ is the coboundary of a chain of finite order in $\hom(X(\Tc\Ac),X(\Tc Q\Ac))$. Hence the periodic cohomology class of $\gamma^{2n}_{\Ac}$ is independent of $n$. \cqfd\\

Let now $\varphi:Q\Ac\to \Lc\hotimes\Bc$ be a $p$-summable quasihomomorphism. Thus $\varphi$ sends $q\Ac$ to the ideal $\ell^p\hotimes\Bc\subset\Lc\hotimes\Bc$. Since $\Lc$ is a Banach algebra, it is {\it tensoring} according to the terminology of \cite{Me}, which means that we can lift $\varphi$ to a bounded homomorphism $\varphi_*:\Tc Q\Ac\to \Lc\hotimes\Tc\Bc$. It is obtained first by passing to the analytic tensor algebras $\Tc\varphi: \Tc Q\Ac\to \Tc(\Lc\hotimes\Bc)$, and then composing by the obvious multiplication map $\Tc(\Lc\hotimes\Bc)\to\Lc\hotimes\Tc\Bc$ (this is the critical step where we use the fact that $\Lc$ is tensoring). Then $\varphi_*$ necessarily maps the ideal $\Ic=\ker(\Tc Q\Ac\to \Tc\Ac)$ to the ideal $\ell^p\hotimes\Tc\Bc$, whence a $X$-complex chain map compatible with the adic filtrations
\be
X(\varphi_*)\ :\ F^{n}_{\Ic}X(\Tc Q\Ac)\to F^{n}_{\ell^p\hotimes\Tc\Bc} X(\Lc\hotimes\Tc\Bc)\qquad \forall n\in\zz\ .
\ee
Consider the natural morphism of \cite{CQ1} obtained by the multiplication map on $\Lc$ followed by the projection $\Lc\to\Lc_{\nat}=\Lc/[\Lc,\Lc]$:
\be
\al: X(\Lc\hotimes\Tc\Bc)\to \Lc_{\nat}\hotimes X(\Tc\Bc)\ ,
\ee
and let $n$ be such that $2n\ge p$. Then the elements of $F^{4n}_{\ell^p\hotimes\Bc} X(\Lc\hotimes\Tc\Bc)$ contain at least $2n$ powers of $\ell^p$, so that their images under $\al$ lie in the complex $\ell^1_{\nat}\hotimes X(\Tc\Bc)$, where $\ell^1_{\nat}$ is the commutator quotient space $\ell^1/[\ell^1,\Lc]$. Thus composing with the canonical trace of operators $\Tr:\ell^1\to \cc$ yields a chain map
\be
\Tr\ :\ F^{4n}_{\ell^p\hotimes\Tc\Bc} X(\Lc\hotimes\Tc\Bc)\to X(\Tc\Bc)\qquad \forall 2n\ge p\ .
\ee
We define the bivariant Chern character of the quasihomomorphism $\varphi$ as the composite $\Tr\, X(\varphi_*)\gamma^{2n}_{\Ac}\in \hom(X(\Tc\Ac),X(\Tc\Bc))$.
\begin{proposition}
Let $\varphi: Q\Ac\to \Lc\hotimes\Bc$ be a $p$-summable quasihomomorphism. Then for any $2n\ge p$, the bivariant Chern character $\Tr\, X(\varphi_*)\gamma^{2n}_{\Ac}$ is a chain map of order $\le 2n$, thus defining a cyclic cohomology class
\be
\ch^{2n}(\varphi)\in HC_{2n}(\Ac,\Bc)\ .
\ee
$\ch^{2n}$ vanishes on degenerate quasihomomorphisms. Moreover, the image of $\ch^{2n}(\varphi)$ in the bivariant periodic theory $HP_0(\Ac,\Bc)$ is independent of $n$ and homotopy invariant. We denote by $\ch(\varphi)$ this canonical periodic class.
\end{proposition}
{\it Proof:} By proposition \ref{deg}, we already know that the chain map $\gamma^{2n}_{\Ac}$ is of order $\le 2n$. Since the homomorphism $\varphi_*:\Tc Q\Ac\to \Lc\hotimes\Tc\Bc$ clearly preserves the degree of differential forms, as well as the trace map $\Tr$, we deduce that $\Tr\, X(\varphi_*)\gamma^{2n}_{\Ac}$ is also of degree $\le 2n$. By refining the argument of proposition \ref{deg}, we can show that the periodic cyclic cohomology class of $\ch^{2n}(\varphi)$ is independent of $n$: by proposition \ref{pcycl0}, the difference $\chb^{2n+2}-\chb^{2n}$ is the coboundary of a chain in $\hom(X(\Tc\Tc\Ac), F^{4n}_{q\Tc\Ac}X(Q\Tc\Ac))$ vanishing on $F^k_{\Jc\Tc\Ac}X(\Tc\Tc\Ac)$ for $k$ sufficiently large, and this is compatible with taking the trace map $\Tr$. The homotopy invariance in periodic theory comes from the fact that the periodic cohomology class of the chain map $X(\varphi_*):X(\Tc Q\Ac) \to X(\Lc\hotimes\Tc\Bc)$ is always homotopy invariant with respect to the homomorphism $\varphi$ (see \cite{CQ2,Me}), in a way compatible with the filtration by the ideals $\Ic$ and $\ell^p\hotimes\Tc\Bc$ because the $p$-summable homotopy preserves the inclusion $\varphi_*(\Ic)\subset\ell^p\hotimes\Tc\Bc$. Finally, if $\varphi$ is degenerate, then the composite $\varphi_*\circ\phi: Q\Tc\Ac\to \Lc\hotimes\Tc\Bc$ sends both $\iota(\Tc\Ac)$ and $\iotab(\Tc\Ac)$ onto the same image, thus by the construction of $\chb^{2n}$, one sees that $\Tr\, X(\varphi_*)\gamma^{2n}_{\Ac}$ vanishes.\cqfd\\
\begin{remark}\textup{
If $\rho:\Ac\to\Bc$ is an homomorphism and $\varphi=\rho*0: Q\Ac\to \Bc$ the corresponding quasihomomorphism, there is no summability conditions and the Chern character $\ch^{2n}(\varphi)$ is defined in any degree. In particular, $\ch^0(\varphi)$ exactly corresponds to the class of the chain map $X(\Tc\rho):X(\Tc\Ac)\to X(\Tc\Bc)$ in $HC_0(\Ac,\Bc)$. Its periodic class $\ch(\varphi)$, accounting for the functoriality of $HP$, will simply be denoted by $\ch(\rho)$. More generally, the periodic Chern character of a quasihomomorphism $\rho_+*\rho_-:Q\Ac\to \Bc$ is the difference $\ch(\rho_+)-\ch(\rho_-)\in HP_0(\Ac,\Bc)$.}
\end{remark}
\begin{remark}\label{rem}\textup{
The periodic class of the cocycles $\gamma^{2n}_{\Ac}$ can be interpreted using excision for the exact sequence $0\to q\Ac\to Q\Ac\to \Ac\to 0$. Let $X(\Tc Q\Ac : \Tc\Ac)$ be the kernel of the chain map $X(\Tc(\id*\id)):X(\Tc Q\Ac)\to X(\Tc\Ac)$. Then excision shows \cite{CQ2,Me,Me2} that the morphism $X(\Tc q\Ac)\to X(\Tc Q\Ac : \Tc\Ac)$ induced by the injection $q\Ac\to Q\Ac$ is a homotopy equivalence. Hence $X(\Tc Q\Ac : \Tc\Ac)$ computes the cyclic homology of $q\Ac$. Moreover, $HP_0(\Ac,q\Ac)$ is a direct summand in $HP_0(\Ac,Q\Ac)$, and $\Ac$ is $HP$-equivalent to $q\Ac$. The class $\ch(\iota)-\ch(\iotab)\in HP_0(\Ac,q\Ac)$ of the chain map $X(\Tc\iota)-X(\Tc\iotab)$ sending $X(\Tc\Ac)$ to $X(\Tc Q\Ac: \Tc\Ac)$ is an invertible element realizing this equivalence. Its inverse in $HP_0(q\Ac,\Ac)$ is represented by the chain map $X(\Tc\pi_0)$ associated to the homomorphism $\pi_0=\id_{\Ac}*0: q\Ac\to\Ac$. Now, remark that our cocycles $\gamma^{2n}_{\Ac}$ are cohomologous in the complex $\hom(X(\Tc\Ac),X(\Tc Q\Ac: \Tc\Ac))$, and an easy computation shows that $\gamma^{0}_{\Ac}$ corresponds precisely to the difference $X(\Tc\iota)-X(\Tc\iotab)$. Hence, we have a realization of the $HP$-equivalence between $\Ac$ and $q\Ac$ by an infinite collection of {\it non-periodic} cocycles $[\gamma^{2n}_{\Ac}]\in HC_{2n}(\Ac,Q\Ac)$, whose periodic class coincide with $\ch(\iota)-\ch(\iotab)\in HP_0(\Ac,q\Ac)\subset HP_0(\Ac,Q\Ac)$. Of course, they all represent the bivariant Chern character of the universal Kasparov module $u_0\in\Esf_0(\Ac,\Qt\Ac)$ of section \ref{suniv}. The interesting feature of these cocycles is that they involve more powers of $q\Ac$ when $n$ increases, so that by selecting a sufficiently large $n$ we are able to construct a Chern character for any $p$-summable quasihomomorphism.}
\end{remark}
\begin{remark}\textup{
Excision also shows that the algebras $\ell^p\hotimes\Bc$ and $\Bc$ are $HP$-equivalent \cite{Cu2}. Once again, our bivariant Chern character realizes this Morita equivalence by two invertible elements in $HP_0(\Bc,\ell^p\hotimes\Bc)$ and $HP_0(\ell^p\hotimes\Bc,\Bc)$ obtained as follows. Let $i^{(p)}:\Bc\hookrightarrow \ell^p\hotimes\Bc$ be the canonical inclusion. It induces a chain map $X(\Tc i^{(p)}):X(\Tc\Bc)\to X(\Tc(\ell^p\hotimes\Bc))$ whose periodic cyclic cohomology class is $\ch(i^{(p)})\in HP_0(\Bc,\ell^p\hotimes\Bc)$. On the other hand, the canonical inclusion $j^{(p)}:\ell^p\hotimes\Bc\to \Lc\hotimes\Bc$ yields a $p$-summable quasihomomorphism $\varphi^{(p)}=j^{(p)}*0:Q(\ell^p\hotimes\Bc)\to \Lc\hotimes\Bc$ with Chern character $\ch(\varphi^{(p)})\in HP_0(\ell^p\hotimes\Bc,\Bc)$. Then $\ch(i^{(p)})$ and $\ch(\varphi^{(p)})$ are inverse to each other. Indeed, for $2n\ge p$, the composite $\ch^{2n}(\varphi^{(p)})\circ X(\Tc i^{(p)})\in\hom(X(\Tc\Bc),X(\Tc\Bc))$ exactly corresponds to the non-periodic Chern character $\ch^{2n}(\id_{\Bc}*0)$ of the quasihomomorphism $\id_{\Bc}*0: Q\Bc\to \Bc$ associated to the identity of $\Bc$. Therefore, the product in periodic cyclic cohomology $\ch(i^{(p)})\cdot\ch(\varphi^{(p)})=\ch(\id_{\Bc})$ is the unit of the ring $HP_0(\Bc,\Bc)$. Since we know by excision that $\ch(i^{(p)})$ is invertible \cite{Cu2}, we have also $\ch(\varphi^{(p)})\cdot\ch(i^{(p)})=1$ in $HP_0(\ell^p\hotimes\Bc,\ell^p\hotimes\Bc)$.}
\end{remark}
In fact all the relevant information of a $p$-summable quasihomomorphism $\varphi$ is contained in its restriction $\varphi:q\Ac\to \ell^p\hotimes\Bc$ to the ideal $q\Ac$ \cite{Cu1}. The remarks above show that up to homotopy, the periodic Chern character $\ch(\varphi)$ represented by any of the cocycles $\Tr\,X(\varphi_*)\gamma^{2n}_{\Ac}$ may be inserted in a commutative square
\be
\xymatrix{
X(\Tc (q\Ac)) \ar[r]^{X(\Tc\varphi)} & X(\Tc(\ell^p\hotimes\Bc)) \ar@{.>}[d] \\
X(\Tc\Ac) \ar@{.>}[u] \ar[r]^{\ch(\varphi)} & X(\Tc\Bc) }\label{square}
\ee
where the verical arrows realize the corresponding $HP$-equivalences, i.e. invert the chain maps induced respectively by the homomorphisms $\pi_0:q\Ac\to \Ac$ and $i^{(p)}:\Bc\to\ell^p\hotimes\Bc$. By the way, this property implies the compatibility of the Chern character with the Kasparov product whenever it is defined. To see this, let $\Ac,\Bc,\Cc$ be three algebras. For any $p,q\ge 1$, let $\mu: \ell^p\hotimes\ell^q\to \ell^r$, $r=\max(p,q)$ be the homomorphism induced by the identification of $\Hc=\ell^2(\nn)$ with $\ell^2(\nn^2)$. Given a $p$-summable quasihomomorphism $\varphi_1: q\Ac\to\ell^p\hotimes\Bc$ and a $q$-summable quasihomomorphism $\varphi_2:q\Bc\to \ell^q\hotimes\Cc$, we say that $\varphi_1$ and $\varphi_2$ are composable \cite{CQ2} if there exists a homomorphism $\varphi_1':q\Ac\to \ell^p\hotimes q\Bc$ such that $(\id\hotimes \pi_0)\circ \varphi_1'$ is homotopic to $\varphi_1$. Then the composition $\varphi_1\cdot \varphi_2= (\mu\hotimes\id)\circ(\id\hotimes \varphi_2)\circ \varphi_1'$ is a $r$-summable quasihomomorphism $ q\Ac\to \ell^r\hotimes\Cc$ defining a Kasparov product of $\varphi_1$ and $\varphi_2$. This may be depicted in the following diagram, which commutes up to homotopy:
\be
\xymatrix{
 &  \ell^p\hotimes q\Bc \ar[d]^{\id\hotimes \pi_0} \ar[r]^{\id\hotimes\varphi_2} & \ell^p\hotimes\ell^q\hotimes\Cc \ar[r]^{\mu\hotimes\id} & \ell^r\hotimes\Cc \\
q\Ac \ar[ur]^{\varphi_1'} \ar[r]_{\varphi_1} & \ell^p\hotimes\Bc &  &  }\label{Kasp}
\ee
The product $\varphi_1\cdot \varphi_2$ may not be well defined at the level of $KK$-groups. In particular its homotopy class in $KK_0(\Ac,\Cc)$ may depend on the choice of the lifting $\varphi_1'$. Nevertheless, the following lemma shows its compatibility with the Chern character:
\begin{lemma}
Let $\varphi_1: q\Ac\to\ell^p\hotimes\Bc$ and $\varphi_2:q\Bc\to \ell^q\hotimes\Cc$ be composable quasihomomorphisms. Then the bivariant Chern character is multiplicative, that is, if $\varphi_1\cdot\varphi_2: q\Ac\to \ell^r\hotimes\Cc$ is a Kasparov product, the equality
\be
\ch(\varphi_1\cdot\varphi_2)= \ch(\varphi_1)\cdot \ch(\varphi_2)
\ee
holds in the periodic theory $HP_0(\Ac,\Cc)$.
\end{lemma}
{\it Proof:} For any algebra $\Rc$ we let $\pi_0:q\Rc\to\Rc$ be the free product $\id_{\Rc}*0$ and $i^{(p)}:\Rc\to \ell^p\hotimes\Rc$ be the canonical inclusion. Then (\ref{Kasp}) may be enlarged to the following diagram commuting up to homotopy:
$$
\xymatrix{
 &  \ell^p\hotimes q\Bc \ar[d]|{\id\hotimes\pi_0} \ar[rr]^{\id\hotimes\varphi_2} & & \ell^p\hotimes\ell^q\hotimes\Cc \ar[r]^{\mu\hotimes\id} & \ell^r\hotimes\Cc \\
q\Ac \ar[d]_{\pi_0} \ar[ur]^{\varphi_1'} \ar[r]_{\varphi_1} & \ell^p\hotimes\Bc & q\Bc \ar[ul]_{i^{(p)}} \ar[dl]^{\pi_0} \ar[r]_{\varphi_2} & \ell^q\hotimes\Cc \ar[u]^{i^{(p)}} &  \\
\Ac & \Bc \ar[u]^{i^{(p)}} & & \Cc \ar[u]^{i^{(q)}} \ar[uur]_{i^{(r)}} &  }
$$
Applying the functor $X(\Tc\ \cdot\ )$, taking into account the fact that $i^{(p)}$ and $\pi_0$ induce homotopy equivalences and combining the result with (\ref{square}), we get the diagram
$$
\xymatrix{
X(\Tc (q\Ac)) \ar[rr]^{X(\Tc(\varphi_1\cdot\varphi_2))} & & X(\Tc(\ell^r\hotimes\Cc)) \ar@{.>}[d] \\
X(\Tc\Ac) \ar@{.>}[u] \ar[r]^{\ch(\varphi_1)} & X(\Tc\Bc) \ar[r]^{\ch(\varphi_2)} & X(\Tc\Cc) }
$$
commuting up to homotopy, whence $\ch(\varphi_1\cdot\varphi_2)= \ch(\varphi_1)\cdot \ch(\varphi_2)$. \cqfd\\

Note that according to the discussion at the end of section \ref{sretr}, if a $p$-summable quasihomomorphism $\varphi:Q\Ac\to\Lc\hotimes\Bc$ comes from a (sequentially) finitely summable unbounded module $(\Ec,\rho,D)\in\Psi_0(\Ac,\Bct)$, then the periodic Chern character $\ch(\varphi)$ is a retraction of the entire Chern character $\ch(\Ec,\rho,D)\in HE_0(\Ac,\Bc)$. We collect the preceding results in a theorem.
\begin{theorem}
The bivariant Chern character in periodic theory gives an additive map
\be
\ch: KK_0(\Ac,\Bc)\to HP_0(\Ac,\Bc)\ ,
\ee
compatible with the Kasparov product between quasihomomorphisms whenever the latter is defined.  \cqfd
\end{theorem}

We now deal with the odd finitely summable $KK$-group, which corresponds basically to invertible extensions of $\Ac$ by $\ell^p\hotimes\Bc$. We say that a $p$-summable bounded Fredholm bimodule $(\Ec,\rho,F)\in \Esf_1(\Ac,\Bct)$ is a $p$-summable invertible extension iff in matricial form one has
\beq
&&\Ec=\left(\begin{array}{c}
		(\Hc\oplus\Hc)\hotimes\Bct \\
		(\Hc\oplus\Hc)\hotimes\Bct
		\end{array} \right)\ ,\quad 
\rho(a)=\left(\begin{array}{cc}
		\al(a) & 0 \\
		0 & \al(a) 
		\end{array} \right)\in M_4(\Lc\hotimes\Bc)\ ,\\
&&F=\left(\begin{array}{cc}
		0 & X \\
		X & 0 
		\end{array} \right)\in M_4(\Lc\hotimes\Bct)\quad\mbox{with}\quad X=\left(\begin{array}{cc}
		1 & 0 \\
		0 & -1
		\end{array} \right)\in M_2(\Lc\hotimes\Bct)\ .\non
\eeq
Thus $\al(a)$ is a $2\times 2$ matrix over $\Lc\hotimes\Bc$ whose off-diagonal elements are in $\ell^p\hotimes\Bc$. The invertible extension is degenerate iff $[X,\rho(a)]=0$ for any $a\in\Ac$, in which case the matrix $\al(a)$ is diagonal. Now let $P=(1+X)/2$ be the projection onto the 1-eigenspace for $X$. The linear map $\si:\Ac\to \Lc\otimes\Bc$ sending an element $a$ to $P\al(a)P$ is almost multiplicative, in the sense that $\si(ab)-\si(a)\si(b)\in \ell^p\hotimes\Bc$ for any $a,b\in\Ac$. Then $\si$ yields an homomorphism $\pi\si:\Ac\to \Lc\hotimes\Bc/\ell^p\hotimes\Bc$, where $\pi$ is the projection onto the Calkin algebra $\Lc\hotimes\Bc/\ell^p\hotimes\Bc$. This homomorphism corresponds to the {\it Busby invariant} \cite{Bl} of the extension 
\be
\xymatrix{
0 \ar[r] & \ell^p\hotimes\Bc \ar[r] & (\ell^p\hotimes\Bc+ \si(\Ac)) \ar[r] & \Ac' \ar[r] & 0 \\
 & & & \Ac \ar[u] & }
\ee
where $\Ac'=\Ac/\ker(\pi\si)$. The inverse extension is determined by the almost multiplicative map $\si^{-1}(a)=(1-P)\al(a)(1-P)$.\\
Equivalently, the extension above is completely described by the homomorphism $\varphi:Q^s\Ac\to M_2(\Lc\hotimes\Bc)$ sending $\iota(a)=a+da$ to $\al(a)$ and $\iotab(a)=a-da$ to $X\al(a)X$. We may view this homomorphism as an even degree map between {\it $\zz_2$-graded algebras}, once $M_2(\Lc\hotimes\Bc)$ is endowed with its natural graduation given by diagonal/off-diagonal $2\times 2$ matrices. From now on we will denote by $M_2^s(\Lc\hotimes\Bc)$ this superalgebra. Then the element $q(a)=2da$ maps to $X[X,\al(a)]\in M_2^s(\ell^p\hotimes\Bc)$, hence $q^s\Ac$ goes to the $\zz_2$-graded ideal $M_2^s(\ell^p\hotimes\Bc)$. The extension is degenerate iff $q^s\Ac$ maps to 0. As in the even case, the set of invertible extensions is a semigroup for the direct sum upon the isomorphism $\Lc\cong M_2(\Lc)$. Homotopy is provided by interpolating $p$-summable extensions $Q^s\Ac\to M_2^s(\Lc\hotimes\Bc\hotimes \cinf[0,1])$.
\begin{definition}
Let $\Ac$ and $\Bc$ be complete bornological algebras. Then $KK_1(\Ac,\Bc)$ is the union, for all $p\ge 1$, of the classes of $p$-summable invertible extensions $\varphi:Q^s\Ac\to M_2^s(\Lc\hotimes\Bc)$ defined modulo homotopy and addition of degenerates. $KK_1(\Ac,\Bc)$ is an abelian group.
\end{definition}
The zero element of $KK_1(\Ac,\Bc)$ corresponds to the class of degenerates, and the inverse class of an extension $\varphi:Q^s\Ac\to M_2^s(\Lc\hotimes\Bc)$ is represented by the conjugate extension $Y^{-1}\varphi Y$, where $Y$ is the supersymmetric matrix $\left(\begin{array}{cc}
		0 & 1 \\
		1 & 0 
		\end{array} \right)\in M^s_2(\Lc\hotimes\Bct)$.\\

A bivariant Chern character on $KK_1(\Ac,\Bc)$ is constructed as in the even case, with the help of the universal cocycles $\chb^{2n+1}$ of proposition \ref{uni1}. First, one has a classifying map of superalgebras $\phi^s:Q^s\Tc\Ac\to \Tc Q^s\Ac$, sending $q^s\Tc\Ac$ to the ideal $\Ic^s:=\ker(\Tc Q^s\Ac\to\Tc\Ac)$. Then for any $n\in\nn$, we define the odd cocycle $\gamma^{2n+1}_{\Ac}$ as the composition of chain maps
\be
\begin{CD}
 X(\Tc\Ac) @>{X(v_{\Ac})}>> X(\Tc\Tc\Ac) @>{\chb^{2n+1}}>> F^{4n+2}_{q^s\Tc\Ac} X(Q^s\Tc\Ac) @>{X(\phi^s)}>> F^{4n+2}_{\Ic^s}X(\Tc Q^s\Ac)\ .
\end{CD}
\ee
The periodic cyclic cohomology class of $\gamma^{2n+1}_{\Ac}$ does not depend on $n$. Next, let $\varphi: Q^s\Ac\to M_2^s(\Lc\hotimes\Bc)$ be a $p$-summable invertible extension. Passing to analytic tensor algebras, it yields a bounded homomorphism of superalgebras $\varphi_*:\Tc Q^s\Ac\to M_2^s(\Lc\hotimes\Tc\Bc)$, mapping the ideal $\Ic^s$ to $M_2^s(\ell^p\hotimes\Tc\Bc)$, whence a chain map compatible with the adic filtrations
\be
X(\varphi_*)\ :\ F^{n}_{\Ic^s}X(\Tc Q^s\Ac)\to F^{n}_{M_2^s(\ell^p\hotimes\Tc\Bc)} X(M_2^s(\Lc\hotimes\Tc\Bc))\qquad \forall n\in\zz\ .
\ee
As in the non-supersymmetric case, there is a canonical morphism
\be
\al: X(M_2^s(\Lc\hotimes\Tc\Bc))\to M^s_2(\Lc)_{\nat}\hotimes X(\Tc\Bc)\ ,
\ee
where $M^s_2(\Lc)_{\nat}$ is the supercommutator quotient space. Since the subcomplex $F^{4n+2}_{M_2^s(\ell^p\hotimes\Bc)} X(M_2^s(\Lc\hotimes\Tc\Bc))$ involves elements of $M_2^s(\Lc\hotimes\Tc\Bc)$ containing at least $2n+1$ powers of $M_2^s(\ell^p)$, the canonical {\it supertrace} $\Tr_s:M_2^s(\ell^1)\to \cc$ yields a chain map for any $2n+1\ge p$:
\be
\Tr_s\ :\ F^{4n+2}_{M_2^s(\ell^p\hotimes\Tc\Bc)} X(M_2^s(\Lc\hotimes\Tc\Bc))\to X(\Tc\Bc)\qquad \forall 2n+1\ge p\ .
\ee
The bivariant Chern character of the invertible extension is the composite $\sqrt{2\pi i}\,\Tr_sX(\varphi_*)\gamma^{2n+1}_{\Ac}$ in $\hom(X(\Tc\Ac),X(\Tc\Bc))$. As in the even case, if the $p$-summable invertible extension $\varphi:Q^s\Ac\to M^s_2(\Lc\hotimes\Bc)$ comes from a (sequentially) finitely summable unbounded module $(\Ec,\rho,D)\in\Psi_1(\Ac,\Bct)$, then the periodic Chern character $\ch(\varphi)$ is a retraction of the entire Chern character $\ch(\Ec,\rho,D)\in HE_1(\Ac,\Bc)$.
\begin{theorem}
Let $\varphi:Q^s\Ac\to M_2^s(\Lc\hotimes\Bc)$ be a $p$-summable invertible extension. Then for any integer $n$ such that $2n+1\ge p$, the Chern character $\sqrt{2\pi i}\,\Tr_sX(\varphi_*)\gamma^{2n+1}_{\Ac}$ defines a cyclic cohomology class
\be
\ch^{2n+1}(\varphi)\in HC_{2n+3}(\Ac,\Bc)\ .
\ee 
$\ch^{2n+1}$ vanishes on degenerate extensions. The image of $\ch^{2n+1}(\varphi)$ in the periodic theory $HP_1(\Ac,\Bc)$ is independent of $n$ and homotopy invariant. Therefore, we get an additive map
\be
\ch: KK_1(\Ac,\Bc)\to HP_1(\Ac,\Bc)
\ee
defining the Chern character of invertible extensions in periodic theory.\cqfd
\end{theorem}
Remark that the degree of the non-periodic cocycle $\ch^{2n+1}(\varphi)$ is equal to $2n+3$ and not $2n+1$. This worse estimate is due to the choice of the parity of $n$ in proposition \ref{pfred}, which was only motivated by the simplicity of the corresponding formulas for the universal cocycles $\chb^{2n+1}$.

\vskip 0.5cm

\noindent {\bf Acknowledgements:} I would like to thank Dr. Ralf Meyer and Prof. Joachim Cuntz for several clarifications on this topic.


\begin{thebibliography}{9}


\bibitem{BGV} N. Berline, E. Getzler, M. Vergne, {\it Heat kernels and Dirac operators}, Grundlehren des Mathematischen Wissenschaft 298, Springer-Verlag (1992).

\bibitem{B} J. M. Bismut: The Atiyah-Singer index theorem for families of Dirac operators: two heat equation proofs, {\it Invent. Math.} {\bf 83} (1986) 91-151.

\bibitem{Bl} B. Blackadar: {\it $K$-theory for operator algebras}, Springer-Verlag, New-York (1986).

\bibitem{C0} A. Connes: Non-commutative differential geometry, {\it Publ. Math. IHES} {\bf 62} (1986) 41-144.


\bibitem{C1} A. Connes: {\it Non-commutative geometry}, Academic Press, New-York (1994).

\bibitem{C2} A. Connes: Entire cyclic cohomology of Banach algebras and characters of $\te$-summable Fredholm modules, {\it K-theory} {\bf 1} (1988) 519-548.

\bibitem{CM86} A. Connes, H. Moscovici: Transgression du caract\`ere de Chern et cohomologie cyclique, {\it C. R. Acad. Sci. Paris}, t. {\bf 303}, s\'erie I, {\bf 18} (1986) 913.

\bibitem{CM93} A. Connes, H. Moscovici: Transgression and the Chern character of finite-dimensional $K$-cycles, {\it Comm. Math. Phys.} {\bf 155} (1993) 103-122.

\bibitem{CM95} A. Connes, H. Moscovici: The local index formula in non-commutative geometry, GAFA {\bf 5} (1995) 174-243.

\bibitem{Cu1} J. Cuntz: A new look at $K\!K$-theory, {\it K-Theory} {\bf 1} (1987) 31-51.

\bibitem{Cu2} J. Cuntz: Cyclic theory and the bivariant Chern-Connes character, preprint SFB 478 (2000).

\bibitem{CQ0} J. Cuntz, D. Quillen: Algebra extensions and nonsingularity, JAMS {\bf 8} (1995) 251-289.

\bibitem{CQ1} J. Cuntz, D. Quillen: Cyclic homology and nonsingularity, JAMS {\bf 8} (1995) 373-442.

\bibitem{CQ2} J. Cuntz, D. Quillen: Excision in bivariant periodic cyclic cohomology, {\it Invent. Math.} {\bf 127} (1997) 67-98.

\bibitem{Gi} P. Gilkey: {\it Invariance theory, the heat equation, and the Atiyah-Singer index theorem}, 2nd ed., Studies in Advanced Mathematics, CRC Press (1995).

\bibitem{G} T. G. Goodwillie: Cyclic homology, derivations, and the free loopspace, {\it Topology} {\bf 24}(2) (1985) 187-215.

\bibitem{HN} H. Hogbe-Nlend: {\it Th\'eorie des bornologies et applications}, Springer-Verlag, Lecture Notes in Mathematics, Vol. {\bf 213} (1971).

\bibitem{JLO} A. Jaffe, A. Lesniewski, K. Osterwalder: Quantum $K$-theory, I. The Chern character, {\it Comm. Math. Phys.} {\bf 118} (1988) 1-14.

\bibitem{Kar} M. Karoubi: Homologie cyclique et $K$-th\'eorie, {\it Ast\'erisque} {\bf 149} (1987).

\bibitem{L} J. L. Loday: {\it Cyclic homology}, Grundlehren der mathematischen Wissenschaften 301, 2nd ed., Springer-Verlag (1998).

\bibitem{Me} R. Meyer: {\it Analytic cyclic cohomology}, Thesis, M\"unster (1999), math.KT/9906205.

\bibitem{Me2} R. Meyer: Excision in entire cyclic cohomology, {\it J. Eur. Math. Soc.} {\bf 3} (2001) 269-286.

\bibitem{Ni1} V. Nistor: A bivariant Chern character for $p$-summable quasihomomorphisms, {\it K-Theory} {\bf 5} (1991) 193-211.

\bibitem{Ni2} V. Nistor: A bivariant Chern-Connes character, {\it Ann. Math.} {\bf 138} (1993) 555-590.

\bibitem{P} D. Perrot: A bivariant Chern character for families of spectral triples, math-ph/0103016, to appear in {\it Comm. Math. Phys.}

\bibitem{Pu1} M. Puschnigg: Cyclic homology theories for topological algebras, {\it $K$-theory preprint archives} {\bf 292} (1998).

\bibitem{Pu2} M. Puschnigg: Excision in cyclic homology theories, {\it Invent. Math.} {\bf 143} (2001) 249-323.

\bibitem{Q1} D. Quillen: Superconnections and the Chern character, {\it Topology} {\bf 24} (1985) 89-95.

\bibitem{Q2} D. Quillen: Algebra cochains and cyclic cohomology, {\it Publ. Math. IHES} {\bf 68} (1989) 139-174.

\bibitem{Q3} D. Quillen:  Chern-Simons forms and cyclic cohomology, in: {\it The interface of mathematics and particle physics}, Oxford (1988) 117-134.

\bibitem{Z} R. Zekri: A new description of Kasparov's theory of $C^*$-algebra extensions, {\it J. Funct. Anal.} {\bf 84} (1989) 441-471.




\end{thebibliography}
\end{document}